\documentclass[aps,showpacs,superscriptaddress]{revtex4}
\usepackage{epsfig}
\usepackage{color}
\newcommand{\ben}{\begin{eqnarray}}
\newcommand{\een}{\end{eqnarray}}
\newcommand{\nnu}{\nonumber\\}
\newcommand{\bef}{\begin{figure}[htb]\centering}
\newcommand{\eef}{\end{figure}}
\newcommand{\bet}{\begin{table}[hbt]\centering}
\newcommand{\eet}{\end{table}}

\begin{document}
\title{Evolution of twist-3 multi-parton correlation functions relevant to single transverse-spin asymmetry}
\author{Zhong-Bo Kang}
\email{kangzb@iastate.edu}
\affiliation{Department of Physics and Astronomy, 
                 Iowa State University,  
                 Ames, IA 50011, USA}
\author{Jian-Wei Qiu}
\email{jwq@iastate.edu}
\affiliation{Department of Physics and Astronomy, 
                 Iowa State University,  
                 Ames, IA 50011, USA}                                     
\begin{abstract}
We constructed two sets of twist-3 correlation functions that 
are responsible for generating the novel single transverse-spin 
asymmetry in the QCD collinear factorization approach.
We derive evolution equations for these universal
three-parton correlation functions.  
We calculate evolution kernels 
relevant to the gluonic pole contributions to the asymmetry 
at the order of $\alpha_s$.  We find that  
all evolution kernels are infrared safe as they should be
and have a lot in common to the DGLAP evolution kernels
of unpolarized parton distributions.  
By solving the evolution equations, we explicitly demonstrate 
the factorization scale dependence of these twist-3 correlation 
functions.  
\end{abstract}                 
\date{\today}
\maketitle

\section{Introduction}
\label{introduction}

Large single transverse-spin asymmetries (SSAs),
$A_N \equiv (\sigma(s_T)-\sigma(-s_T))/(\sigma(s_T)+\sigma(-s_T))$, 
defined as the ratio of the difference and the sum of the 
cross sections when the spin $s_T$ is flipped,
have been consistently observed in various experiments 
involving one polarized hadron at different collision energies 
\cite{SSA-fixed-tgt,SSA-dis,SSA-rhic}. 
From the parity and time-reversal invariance of 
the strong interaction dynamics, the measured large asymmetries 
in high energy collisions should be directly connected to 
the transverse motion of partons inside a polarized hadron. 
Experimental data on SSAs provide excellent opportunities 
to probe QCD dynamics {\it beyond} what have been explored by 
the very successful leading power QCD collinear factorization 
formalism \cite{CSS-fac,CTEQPDF,MRSTPDF}.  

For high energy scattering cross sections with a large momentum 
transfer, $Q\gg \Lambda_{\rm QCD}$, QCD collinear factorization
approach is more relevant, and SSAs of these cross sections 
could be generated by twist-3 multi-parton correlation 
functions \cite{Efremov,qiu,Qiu_sterman,Qiu:1998ia,koike}.
With the parton's transverse momentum integrated, these 
multi-parton correlation functions represent a net (or integrated) 
spin dependence of parton's transverse motion 
inside a polarized hadron. 
On the other hand, SSAs of cross sections with two different 
momentum transfer scales, 
$Q_1 \gg Q_2 \gtrsim \Lambda_{\rm QCD}$, 
could be expressed in terms of the transverse momentum dependent
(TMD) parton distributions, which directly probe the spin dependence
of parton's transverse motion at the momentum scale $Q_2$ while
the larger scale $Q_1$ defines the hard collision 
\cite{Siv90,Brodsky:2002cx,MulTanBoe,JiMaYu04,mulders}.
These two approaches each have their own kinematic domain of validity 
and were shown to be consistent with each other in the kinematic 
regime where they both apply \cite{UnifySSA}.
Both approaches have been applied extensively in phenomenological studies 
\cite{Qiu:1998ia,Kanazawa:2000hz,mulders1,Boer:2003tx,Vogelsang:2005cs,Qiu:2007ar,BMVY,Bacch,Ans94,siverscompare,Ji:1992eu,Kang:2008qh,Yuan:2008it,anselminoD,muldrod,cedran}, 
and have had initial successes \cite{Kouvaris:2006zy,Anselmino:2008sga}.
However, all existing perturbative calculations are 
at the leading order (LO) in strong coupling constant, $\alpha_s(\mu)$,
and have a strong dependence on the choice of the renormalization scale 
$\mu$ as well as the factorization scale $\mu_F$, while the physically 
observed SSAs should not depend on the choice of the renormalization 
and/or the factorization scale.  The strong dependence on the choice 
of renormalization and factorization scale is an artifact of the 
lowest order perturbative calculation, and a significant cancellation
of the scale dependence between the leading and the next-to-leading
order (NLO) contribution is expected from the QCD factorization theorem
\cite{CSS-fac,CTEQPDF,MRSTPDF,Qiu_sterman}.  
In order to test QCD dynamics
for SSAs, it is necessary to calculate the evolution 
(or the scale dependence) of the universal long-distance distributions 
and to evaluate the perturbative short-distance contribution beyond
the lowest order.  

Since the 80's, a tremendous effort has been devoted to
derive evolution equations of twist-3 correlation functions
that contribute to the structure function $g_2$, which 
can be extracted from measurements of the double-spin asymmetry
for cross sections of inclusive deep inelastic scattering (DIS) 
between a longitudinally polarized lepton and a transversely 
polarized hadron \cite{g2_evo,sol_t3}.  Evolution equations for 
the chiral odd twist-3 correlation functions of an unpolarized
hadron $h_L$ and $e$, which are relevant to the SSAs in
connection with the twist-2 transversity distribution, 
were also derived by several authors \cite{hl_e}.  
Evolution of the moments of TMD parton distributions 
was discussed in Ref.~\cite{Henneman:2001ev}.  
However, a systematic study of the factorization 
scale dependence for the set of twist-3 correlation functions
that are responsible for the gluonic and fermionic pole contributions 
to the SSAs \cite{Efremov,qiu,Qiu:1998ia,koike}
is not available in the literature.  
It is important to note that this set of twist-3 correlation 
functions has a close connection to the TMD parton distributions
\cite{UnifySSA} and does not contribute to the inclusive 
structure function $g_2$.  
In most existing phenomenological studies of SSAs, 
the scale dependence of these correlation functions 
was often assumed to be the same as that of the unpolarized 
parton distribution functions (PDFs) 
\cite{qiu,Qiu:1998ia,koike,Vogelsang:2005cs,Kouvaris:2006zy,BMVY}.

In this paper, we construct two sets of twist-3 correlation 
functions that are responsible for generating the SSAs 
in the QCD collinear factorization approach, and 
derive a closed set of evolution equations 
for these universal twist-3 multi-parton correlation functions.
We then focus on the evolution of two types of 
twist-3 multi-parton correlation functions that give the 
leading gluonic contribution to SSAs \cite{Efremov,qiu}.
The first type covers the quark-gluon correlation functions 
defined as \cite{Efremov,qiu},
\ben
T_{q, F}(x, x,\mu_F)
=\int\frac{dy_1^-}{2\pi}e^{ixP^+y_1^-}
\langle P,s_T|\overline{\psi}_q(0)\frac{\gamma^+}{2}
\left[ \int dy_2^-\, \epsilon^{s_T\sigma n\bar{n}}
       F_\sigma^{~ +}(y_2^-)\right] 
\psi_q(y_1^-)|P,s_T\rangle \, .
\label{Tq}
\een
There is one quark-gluon correlation function, $T_{q,F}$, 
for each quark (anti-quark) flavor $q$ ($\bar{q}$).
The second type covers the correlation functions of three 
active gluons \cite{Ji:1992eu,Kang:2008qh},
\ben
T_{G,F}^{(f,d)}(x, x,\mu_F)
=\int\frac{dy_1^- }{2\pi}e^{ixP^+y_1^-}\frac{1}{xP^+}
\langle P,s_T|F^+_{~~\alpha}(0)
\left[ \int dy_2^-\, \epsilon^{s_T\sigma n\bar{n}}
       F_\sigma^{~ +}(y_2^-)\right] 
F^{\alpha+}(y_1^-)|P,s_T\rangle\, .
\label{Tg}
\een
There are two independent tri-gluon correlation functions, 
$T_{G,F}^{(f)}(x, x)$ and $T_{G,F}^{(d)}(x, x)$, 
because of the fact that the color of the three gluon field strengths 
in Eq.~(\ref{Tg}) can be neutralized by contracting with either the antisymmetric $if_{abc}$ or the symmetric $d_{abc}$ tensors 
with color indices, $a$, $b$, and $c$ \cite{Ji:1992eu,Kang:2008qh}.
In above equations, the proper gauge links that ensure the gauge 
invariance of these correlation functions have been suppressed 
\cite{qiu}. The $\mu_F$ is the factorization scale and
$\epsilon^{s_T\sigma n\bar{n}}=\epsilon^{\rho\sigma\mu\nu}
s_{T\rho} n_{\mu} \bar{n}_{\nu}$ with the light-cone vectors,
$n^\mu=(n^+,n^-,n_T)=(0,1,0_T)$ and $\bar{n}^{\nu}=(1,0,0_T)$,
which project out the light-cone components of any four-vector 
$V^\mu$ as $V\cdot n =V^+$ and $V\cdot\bar{n}=V^-$.
In Eqs.~(\ref{Tq}) and (\ref{Tg}), the subscript ``$F$'' 
refers to the fact that it is a field strength operator 
$F_\sigma^{~ +}$ (rather than a covariant derivative operator
$D_\sigma$) in the square brackets that represents 
the middle active parton.  In this paper, we also
calculate the evolution kernels for these two types of
correlation functions at the first non-trivial 
order in $\alpha_s(\mu)$ and study the factorization scale 
$\mu_F$-dependence of these correlation functions.  

There could be many different approaches to derive the 
evolution equations for the factorization scale dependence of 
these twist-3 correlation functions.  Since these correlation 
functions are universal, the evolution equations should not 
depend on how they were derived.  In this paper, we derive 
the evolution equations from the Feynman diagram representation 
of these correlation functions \cite{Collins:1988wj}.  We first
introduce the Feynman diagram representation for these 
twist-3 correlation functions that are relevant to the SSAs.  
From the operator definition of the twist-3 correlation functions, 
we then derive the cut vertices 
in momentum space to explicitly connect these correlation
functions to Feynman diagrams \cite{Mueller:1981sg}.  Following
the technique introduced in Ref.~\cite{Collins:1988wj}, 
we derive the evolution equations in two steps.  
First, we factorize, in terms of QCD collinear factorization 
approach \cite{CSS-fac,Ellis:1982cd,qiu_t4}, 
the perturbative modification to the twist-3 
correlation functions into a convolution of the short-distance 
evolution kernels with the twist-3 correlation functions.
Then, we calculate corresponding evolution kernels in the
light-cone gauge.  We also provide the prescription to calculate
the evolution kernels in a covariant gauge which should give 
the same results.  

The existence of evolution equations of these correlation functions
is an immediate consequence of the QCD collinear factorization of 
the single transverse-spin dependent cross sections \cite{CTEQHandbook,StermanTASI}.  The general 
form for a factorized hadronic inclusive cross section with 
one hadron transversely polarized and a large 
momentum transfer $Q$ may 
be written as \cite{qiu,Qiu_sterman}
\begin{equation}
\sigma(Q,s_T) = H_0\otimes f_2\otimes f_2 
+ (1/Q)\, H_1\otimes f_2 \otimes f_3 + {\cal O}(1/Q^2)\, .
\label{sigma_st}
\end{equation}
The $H_0$ and $H_1$ are perturbatively calculable 
coefficient functions expanded in power series of 
$\alpha_s$, and $f_n$ are nonperturbative matrix 
elements of the products of fields on the light cone and are
often loosely referred as ``twist-$n$'' parton distribution or 
correlation functions.  In Eq.~(\ref{sigma_st}), the ``$\otimes$''
represents the convolution over partons' momentum fractions. 
The first term in the right-hand-side (RHS) of 
Eq.~(\ref{sigma_st}), 
which is often referred as the leading twist term,
does not contribute to the single transverse-spin dependent 
cross section, defined as 
$\Delta\sigma(Q,s_T)\equiv [\sigma(Q,s_T)-\sigma(Q,-s_T)]/2$,
because of the parity and time-reversal invariance of the 
strong interaction.  Therefore, single transverse-spin
dependent cross section directly probe the twist-3 parton
distribution or correlation functions as \cite{qiu,Qiu:1998ia}
\begin{equation}
\Delta\sigma(Q,s_T) =
(1/Q) H_1 (Q/\mu_F,\alpha_s)\otimes f_2(\mu_F) \otimes f_3(\mu_F) 
+ {\cal O}(1/Q^2)\, ,
\label{Dsigma_st}
\end{equation}
where the summation over parton flavors and 
the dependence on renormalization scale have been suppressed.  
Since the physically measured cross section 
is independent of the choice
of renormalization and factorization scale, the factorization 
scale dependence in the RHS of the factorized formula 
in Eq.~(\ref{Dsigma_st}) should be cancelled between the 
$\mu_F$-dependence of the short-distance coefficient functions
and the $\mu_F$-dependence of the parton distribution and
correlation functions.  The $\mu_F$ dependence of the normal 
twist-2 PDFs satisfies the DGLAP evolution equation 
\cite{DGLAP},
\begin{equation}
\frac{\partial}{\partial\ln(\mu_F)}f_2(\mu_F) 
= P_2\otimes f_2(\mu_F)\, ,
\label{DGLAP}
\end{equation}
where the parton flavor dependence has been suppressed and
$P_2$ is the twist-2 evolution kernel, which can be calculated 
perturbatively and expressed in a power series of $\alpha_s$.
From $d\Delta\sigma(Q,s_T)/d\ln(\mu_F) = 0$, we 
derive the leading order generic evolution equation for $f_3$ as
\begin{equation}
\frac{\partial}{\partial\ln(\mu_F)}f_3
= \left(
\frac{\partial}{\partial\ln(\mu_F)}H_1^{(1)}-P_2^{(1)} \right)
\otimes f_3\, ,
\label{evolution_eq}
\end{equation}
by applying the factorized formula in Eq.~(\ref{Dsigma_st}) on
parton states and expanding it to the first non-trivial power of
$\alpha_s$.  In deriving Eq.~(\ref{evolution_eq}), we divided out  
the leading order coefficient function, $H_1^{(0)}$.  
Equation~(\ref{evolution_eq}) clearly shows that the evolution
equation is a consequence of QCD factorization and indicates 
that every perturbatively factorizable single transverse-spin 
dependent cross section could be used to derive the evolution 
kernels of twist-3 correlation functions.
For example, the order of $\alpha_s$ evolution kernels 
could be obtained by calculating $H_1^{(1)}$, the one-loop 
corrections to the short-distance partonic hard part of 
single transverse-spin dependent Drell-Yan 
cross section \cite{WV_spin08}.  The evolution kernels 
derived in this way should be the same as what we derived here 
directly from the Feynman diagram representation.

The quark-gluon and tri-gluon correlation functions in 
Eqs.~(\ref{Tq}) and (\ref{Tg}) give the leading soft
gluonic pole contribution to the transverse-spin 
dependent cross section with a single hard scale, 
$\Delta\sigma(Q,s_T)$ \cite{qiu,Kouvaris:2006zy,Kang:2008qh}.  
However, transverse-spin dependent cross sections 
with more than one physically observed hard scales 
could get additional hard pole contribution which is 
proportional to the off-diagonal part of the twist-3
correlation functions, $T_{q,F}(x,x',\mu_F)$ and 
$T_{G,F}^{(f,d)}(x,x',\mu_F)$ where $x'$ is not necessarily 
equal to $x$ \cite{UnifySSA,Qiu:2007ar}.
In addition to the gluonic pole, the SSAs or 
the transverse-spin dependent cross sections 
could obtain contributions from the fermionic pole of 
the partonic hard scattering \cite{Efremov,qiu}.  
The leading fermionic pole contribution is generated by
not only the off-diagonal part of the correlation
functions $T_{q,F}$ and $T_{G,F}^{(f,d)}$ but also
a new set of twist-3 correlation functions that 
have a vanishing diagonal contribution 
\cite{qiu,koike,Qiu:1998ia}.
In order to describe the phenomenon of SSAs for observables 
with more than one hard scale and evaluate the full 
perturbative contribution to SSAs beyond the lowest order 
in $\alpha_s$, it is necessary to study both the diagonal 
and off-diagonal twist-3 correlation functions that can 
generate the SSAs. 

The rest of this paper is organized as follows.  
In the next section, we construct two sets of 
twist-3 correlation functions that can generate the
SSAs.  In Sec.~\ref{evolution}, we introduce the 
Feynman diagram representation for these twist-3 
correlation functions.  To connect the Feynman diagrams
to the specific twist-3 correlation functions, we derive 
the cut vertices from the operator definition of these 
twist-3 correlation functions in momentum space.
From the perturbative modification to the correlation
functions, we derive the evolution equations.
In Sec.~\ref{kernels}, we calculate all evolution kernels 
at the order of $\alpha_s$ for the evolution equations 
of the quark-gluon and tri-gluon correlation functions
defined in Eqs.~(\ref{Tq}) and (\ref{Tg}).  
In Sec.~\ref{phenomenology}, we 
discuss the scale dependence of these correlation functions
by solving the evolution equations.  Finally, we give
our conclusions and a brief discussion of the impact of the 
calculated scale dependence of the correlation functions in 
Sec.~\ref{summary}.
 
\section{Twist-3 correlation functions relevant to SSAs}
\label{t3CFs}

In this section, we construct two sets of twist-3 
correlation functions that are responsible for generating 
the gluonic and fermionic pole contributions to the SSAs in 
the QCD collinear factorization approach \cite{Efremov,qiu}.  

We first introduce two twist-3 
correlation functions by generalizing the definition of the 
diagonal functions in Eqs.~(\ref{Tq}) and (\ref{Tg}),
\ben
\widetilde{\cal T}_{q,F,\sigma}(x, x+x_2,\mu_F,s_T)
\equiv 
\int\frac{dy_1^- dy_2^-}{(2\pi)^2}\,
e^{ixP^+y_1^-} e^{ix_2P^+y_2^-}
\langle P,s_T|\overline{\psi}_q(0)\,\frac{\gamma^+}{2}
\left[F_{\sigma}^{~ +}(y_2^-)\right] 
\psi_q(y_1^-)|P,s_T\rangle \, ,
\label{cTqVs}
\een
and 
\ben
\widetilde{\cal T}^{(f,d)}_{G,F,\sigma}(x, x+x_2,\mu_F,s_T)
\equiv
\int\frac{dy_1^- dy_2^-}{(2\pi)^2}\,
e^{ixP^+y_1^-} e^{ix_2P^+y_2^-}
\frac{1}{P^+}\langle P,s_T|F^{+\rho}(0)
\left[F_{\sigma}^{~ +}(y_2^-)\right] 
F^{+\lambda}(y_1^-)|P,s_T\rangle
\left(-g_{\rho\lambda}\right) \, ,
\label{cTgs}
\een
where the subscript ``$F$'' again 
indicates that a field strength
operator (not a covariant derivative operator
\cite{qiu,Qiu:1998ia}) is inserted in the middle of the 
bi-local operator that defines the twist-2 spin-averaged 
quark ($q$) or gluon ($G$) distribution function.  
The reality property of these two functions can be expressed 
as \cite{Qiu:1998ia},
\ben
\widetilde{\cal T}_{q,F,\sigma}(x, x+x_2,\mu_F,s_T)^*
&=&\widetilde{\cal T}_{q,F,\sigma}(x+x_2, x,\mu_F,s_T)\, ,
\nonumber\\
\widetilde{\cal T}^{(f,d)}_{G,F,\sigma}(x, x+x_2,\mu_F,s_T)^*
&=&
\widetilde{\cal T}^{(f,d)}_{G,F,\sigma}(x+x_2, x,\mu_F,s_T)\, .
\label{reality_1}
\een
That is, the real part of these two functions are symmetric
in the exchange of $x$ and $x+x_2$, while the imaginary part is
antisymmetric.
Similarly, from the parity and time-reversal invariance,
we find \cite{Qiu:1998ia}
\ben
\widetilde{\cal T}_{q,F,\sigma}(x, x+x_2,\mu_F,s_T)
&=&
- \widetilde{\cal T}_{q,F,\sigma}(x+x_2, x,\mu_F,-s_T)\, ,
\nonumber\\
\widetilde{\cal T}^{(f,d)}_{G,F,\sigma}(x, x+x_2,\mu_F,s_T)
&=&
-\widetilde{\cal T}^{(f,d)}_{G,F,\sigma}(x+x_2, x,\mu_F,-s_T)\, .
\label{PT_1}
\een
That is, these two functions are antisymmetric when the 
transverse spin vector $s_T$ reverses its direction.

From the definition in Eq.~(\ref{cTqVs}) and the symmetry 
properties in Eqs.~(\ref{reality_1}) and (\ref{PT_1}), 
we construct a twist-3 quark-gluon correlation 
function that is relevant to the SSA as follows,  
\ben
{\cal T}_{q,F}(x, x+x_2,\mu_F)
&\equiv &
\epsilon^{s_T\sigma n\bar{n}}\, \frac{1}{2} \left[
\widetilde{\cal T}_{q,F,\sigma}(x, x+x_2,\mu_F,s_T)
-\widetilde{\cal T}_{q,F,\sigma}(x, x+x_2,\mu_F,-s_T)
\right]
\nonumber \\
& = &
\epsilon^{s_T\sigma n\bar{n}}\, \frac{1}{2} \left[
\widetilde{\cal T}_{q,F,\sigma}(x, x+x_2,\mu_F,s_T)
+\widetilde{\cal T}_{q,F,\sigma}(x+x_2, x,\mu_F,s_T)
\right]
\nonumber\\
& \equiv &
\frac{1}{2}
\left[
\widetilde{\cal T}_{q,F}(x, x+x_2,\mu_F,s_T)
+\widetilde{\cal T}_{q,F}(x+x_2, x,\mu_F,s_T)
\right]
\nonumber \\
&=&
{\rm Re}\left[
\widetilde{\cal T}_{q,F}(x, x+x_2,\mu_F,s_T)
\right]\, ,
\label{Tqasy}
\een
where the spin-dependent twist-3 quark-gluon correlation
function is defined as
\ben
\widetilde{\cal T}_{q,F}(x, x+x_2,\mu_F,s_T)
&\equiv&
\int\frac{dy_1^- dy_2^-}{(2\pi)^2}\,
e^{ixP^+y_1^-} e^{ix_2P^+y_2^-}
\langle P,s_T|\overline{\psi}_q(0)\,\frac{\gamma^+}{2}
\left[ \epsilon^{s_T\sigma n\bar{n}}F_\sigma^{~ +}(y_2^-)\right] 
\psi_q(y_1^-)|P,s_T\rangle \, , 
\nonumber\\
&=&
\widetilde{\cal T}_{q,F}(x+x_2, x,\mu_F,-s_T)\, .
\label{Tqt}
\een
As shown in Eq.~(\ref{Tqasy}), 
the twist-3 quark-gluon correlation function
${\cal T}_{q,F}(x, x+x_2,\mu_F)$ is {\it real}\ and 
{\it symmetric}\ when the active momentum fraction
$x$ exchanges with $x+x_2$.

Similarly, we can construct the tri-gluon correlation function
relevant to the SSA as,
\ben
{\cal T}^{(f,d)}_{G,F}(x, x+x_2,\mu_F)
&\equiv &
\epsilon^{s_T\sigma n\bar{n}}\, \frac{1}{2} \left[
\widetilde{\cal T}^{(f,d)}_{G,F,\sigma}(x, x+x_2,\mu_F,s_T)
-\widetilde{\cal T}^{(f,d)}_{G,F,\sigma}(x, x+x_2,\mu_F,-s_T)
\right]
\nonumber \\
& = &
\epsilon^{s_T\sigma n\bar{n}}\, \frac{1}{2} \left[
\widetilde{\cal T}^{(f,d)}_{G,F,\sigma}(x, x+x_2,\mu_F,s_T)
+\widetilde{\cal T}^{(f,d)}_{G,F,\sigma}(x+x_2, x,\mu_F,s_T)
\right]
\nonumber\\
& \equiv &
\frac{1}{2}
\left[
\widetilde{\cal T}^{(f,d)}_{G,F}(x, x+x_2,\mu_F,s_T)
+\widetilde{\cal T}^{(f,d)}_{G,F}(x+x_2, x,\mu_F,s_T)
\right]
\nonumber \\
&=&
{\rm Re}\left[
\widetilde{\cal T}^{(f,d)}_{G,F}(x, x+x_2,\mu_F,s_T)
\right]\, ,
\label{Tgasy}
\een
where the spin-dependent twist-3 tri-gluon correlation
function is defined as
\ben
\widetilde{\cal T}^{(f,d)}_{G,F}(x, x+x_2,\mu_F,s_T)
&\equiv & 
\int\frac{dy_1^- dy_2^-}{(2\pi)^2}\,
e^{ixP^+y_1^-} e^{ix_2P^+y_2^-}
\frac{1}{P^+}\langle P,s_T|F^{+\rho}(0)
\left[ \epsilon^{s_T\sigma n\bar{n}}F_\sigma^{~ +}(y_2^-)\right] 
F^{+\lambda}(y_1^-)|P,s_T\rangle (-g_{\rho\lambda}) 
\nonumber\\
&=&
\widetilde{\cal T}^{(f,d)}_{G,F}(x+x_2, x,\mu_F,-s_T)\, .
\label{Tgt}
\een
The tri-gluon correlation function 
${\cal T}^{(f,d)}_{G,F}(x, x+x_2,\mu_F)$ is also {\it real}
and {\it symmetric} in the exchange of $x$ and $x+x_2$.

The newly defined twist-3 correlation functions, 
${\cal T}_{q,F}(x, x+x_2,\mu_F)$ and 
${\cal T}^{(f,d)}_{G,F}(x, x+x_2,\mu_F)$, 
are related to the diagonal correlation functions
in Eqs.~(\ref{Tq}) and (\ref{Tg}) as
\ben
T_{q,F}(x,x,\mu_F) &=& 
\int dx_2\, \left[2\pi\,\delta(x_2)\right]\, 
            {\cal T}_{q,F}(x,x+x_2,\mu_F)
            \, ,
\nnu
T_{G,F}^{(f,d)}(x,x,\mu_F) &=& 
\int dx_2\, \left[2\pi\,\delta(x_2)\right]\, 
\left(\frac{1}{x}\right)
            {\cal T}^{(f,d)}_{G,F}(x,x+x_2,\mu_F)\, .
\label{T_diag}
\een
Notice that the new tri-gluon correlation function, 
${\cal T}_{G,F}(x, x',\mu_F)$, is symmetric in the exchange of  
$x$ and $x'$, while a direct generalization of 
the diagonal tri-gluon correlation function in Eq.~(\ref{Tg}),
$T_{G,F}(x,x',\mu_F)\equiv{\cal T}_{G,F}(x, x',\mu_F)/x$ 
is not symmetric in exchanging $x$ and $x'$.

In addition to the gluonic pole contribution, 
the SSA could also be generated by the fermionic 
pole of partonic hard scattering \cite{Efremov,qiu}.  
The fermionic pole contribution at twist-3 is proportional to 
the off-diagonal part of the correlation functions ${\cal T}_{q,F}$
and ${\cal T}_{G,F}$, as well as a new set of twist-3 correlation 
functions which vanishes when $x_2=0$ \cite{qiu,Qiu:1998ia,koike}.
To construct this new set of twist-3 correlation functions, 
we introduce two new twist-3 correlation functions,
\ben
\widetilde{\cal T}_{\Delta q,F,\sigma}(x, x+x_2,\mu_F,s_T)
\equiv 
\int\frac{dy_1^- dy_2^-}{(2\pi)^2}\,
e^{ixP^+y_1^-} e^{ix_2P^+y_2^-}
\langle P,s_T|\overline{\psi}_q(0)\,
\frac{\gamma^+\gamma^5}{2}
\left[ i\, F_{\sigma}^{~ +}(y_2^-)\right] 
\psi_q(y_1^-)|P,s_T\rangle \, ,
\label{cTqAs}
\een
and 
\ben
\widetilde{\cal T}^{(f,d)}_{\Delta G,F,\sigma}(x, x+x_2,\mu_F,s_T)
\equiv
\int\frac{dy_1^- dy_2^-}{(2\pi)^2}\,
e^{ixP^+y_1^-} e^{ix_2P^+y_2^-}
\frac{1}{P^+}\langle P,s_T|F^{+\rho}(0)
\left[ i\, F_{\sigma}^{~ +}(y_2^-)\right] 
F^{+\lambda}(y_1^-)|P,s_T\rangle
\left(i\epsilon_{\perp\rho\lambda}\right) \, ,
\label{cTDgs}
\een
where the antisymmetric tensor
$\epsilon_{\perp\rho\lambda}=\epsilon_{\perp}^{\rho\lambda}
=-\epsilon^{\rho\lambda n\bar{n}}$ and subscript ``$\Delta q$''
and ``$\Delta G$'' indicate that the field strength 
operator is inserted in the middle of the bi-local field 
operators that define the twist-2 quark helicity distribution  
$\Delta q$ and the gluon helicity distribution $\Delta G$,
respectively.  Similar to Eq.~(\ref{reality_1}),
the reality property of these two new twist-3 correlation 
functions can be expressed as,
\ben
\widetilde{\cal T}_{\Delta q,F,\sigma}(x, x+x_2,\mu_F,s_T)^*
&=&
-\widetilde{\cal T}_{\Delta q,F,\sigma}(x+x_2, x,\mu_F,s_T)\, ,
\nonumber\\
\widetilde{\cal T}^{(f,d)}_{\Delta G,F,\sigma}(x, x+x_2,\mu_F,s_T)^*
&=&
-\widetilde{\cal T}^{(f,d)}_{\Delta G,F,\sigma}(x+x_2, x,\mu_F,s_T)\, .
\label{reality_2}
\een
That is, the real part of these two new functions are 
{\it antisymmetric} in the exchange of $x$ and $x+x_2$, 
while the imaginary part is symmetric.  This reality property
is different from that of the functions ${\cal T}_{q,F,\sigma}$
and ${\cal T}_{G,F,\sigma}$. 
Similarly, from the parity and time-reversal invariance,
we find,
\ben
\widetilde{\cal T}_{\Delta q,F,\sigma}(x, x+x_2,\mu_F,s_T)
&=&
\widetilde{\cal T}_{\Delta q,F,\sigma}(x+x_2, x,\mu_F,-s_T)\, ,
\nonumber\\
\widetilde{\cal T}^{(f,d)}_{\Delta G,F,\sigma}(x, x+x_2,\mu_F,s_T)
&=&
\widetilde{\cal T}^{(f,d)}_{\Delta G,F,\sigma}(x+x_2, x,\mu_F,-s_T)\, .
\label{PT_2}
\een
That is, these two functions are {\it symmetric} when the 
transverse spin vector $s_T$ reverses its direction.  

From the definition of these new correlation functions 
in Eqs.~(\ref{cTqAs}) and (\ref{cTDgs}) and their properties
in Eqs.~(\ref{reality_2}) and (\ref{PT_2}), 
we construct the second set of twist-3 quark-gluon 
and tri-gluon correlation functions 
that could also contribute to the SSAs.  The new quark-gluon
correlation function is defined as,
\ben
{\cal T}_{\Delta q,F}(x, x+x_2,\mu_F)
&\equiv &
s_T^\sigma\, \frac{1}{2} \left[
\widetilde{\cal T}_{\Delta q,F,\sigma}(x, x+x_2,\mu_F,s_T)
-\widetilde{\cal T}_{\Delta q,F,\sigma}(x, x+x_2,\mu_F,-s_T)
\right]
\nonumber \\
& = &
s_T^\sigma\, \frac{1}{2} \left[
\widetilde{\cal T}_{\Delta q,F,\sigma}(x, x+x_2,\mu_F,s_T)
-\widetilde{\cal T}_{\Delta q,F,\sigma}(x+x_2, x,\mu_F,s_T)
\right]
\nonumber\\
& \equiv &
\frac{1}{2}
\left[
\widetilde{\cal T}_{\Delta q,F}(x, x+x_2,\mu_F,s_T)
-\widetilde{\cal T}_{\Delta q,F}(x+x_2, x,\mu_F,s_T)
\right] 
\nonumber \\
&=& 
{\rm Re}\left[
\widetilde{\cal T}_{\Delta q,F}(x, x+x_2,\mu_F,s_T)
\right]\, ,
\label{TDqasy}
\een
where the spin-dependent new twist-3 quark-gluon correlation
function is defined as
\ben
\widetilde{\cal T}_{\Delta q,F}(x, x+x_2,\mu_F,s_T)
&\equiv &
\int\frac{dy_1^- dy_2^-}{(2\pi)^2}\,
e^{ixP^+y_1^-} e^{ix_2P^+y_2^-}
\langle P,s_T|\overline{\psi}_q(0)\,
\frac{\gamma^+\gamma^5}{2}
\left[i\, s_T^\sigma \, F_\sigma^{~ +}(y_2^-)\right] 
\psi_q(y_1^-)|P,s_T\rangle \, , 
\nonumber \\
& = &
-\widetilde{\cal T}_{\Delta q,F}(x+x_2, x,\mu_F,-s_T)\, 
\label{TDqt}
\een
which was also discussed in Ref.~\cite{koike}.  
As shown in Eq.~(\ref{TDqasy}), this new twist-3 quark-gluon
correlation function ${\cal T}_{\Delta q,F}(x, x+x_2,\mu_F)$
that is relevant to the SSA is also {\it real}, but, is
{\it antisymmetric} in the exchange of $x$ and $x+x_2$.
Similarly, the new tri-gluon correlation function is 
defined as,
\ben
{\cal T}^{(f,d)}_{\Delta G,F}(x, x+x_2,\mu_F)
&\equiv &
s_T^\sigma\, \frac{1}{2} \left[
\widetilde{\cal T}^{(f,d)}_{\Delta G,F,\sigma}(x, x+x_2,\mu_F,s_T)
-\widetilde{\cal T}^{(f,d)}_{\Delta G,F,\sigma}(x, x+x_2,\mu_F,-s_T)
\right]
\nonumber \\
& = &
s_T^\sigma\, \frac{1}{2} \left[
\widetilde{\cal T}^{(f,d)}_{\Delta G,F,\sigma}(x, x+x_2,\mu_F,s_T)
-\widetilde{\cal T}^{(f,d)}_{\Delta G,F,\sigma}(x+x_2, x,\mu_F,s_T)
\right]
\nonumber\\
& \equiv &
\frac{1}{2}
\left[
\widetilde{\cal T}^{(f,d)}_{\Delta G,F}(x, x+x_2,\mu_F,s_T)
-\widetilde{\cal T}^{(f,d)}_{\Delta G,F}(x+x_2, x,\mu_F,s_T)
\right] 
\nonumber \\
&=& 
{\rm Re}\left[
\widetilde{\cal T}^{(f,d)}_{\Delta G,F}(x, x+x_2,\mu_F,s_T)
\right]\, ,
\label{TDgasy}
\een
where the spin-dependent new twist-3 tri-gluon correlation
function is defined as
\ben
\widetilde{\cal T}^{(f,d)}_{\Delta G,F}(x, x+x_2,\mu_F,s_T)
&\equiv & 
\int\frac{dy_1^- dy_2^-}{(2\pi)^2}\,
e^{ixP^+y_1^-} e^{ix_2P^+y_2^-}
\frac{1}{P^+}\langle P,s_T|F^{+\rho}(0)
\left[ i\, s_T^\sigma\, F_\sigma^{~ +}(y_2^-)\right] 
F^{+\lambda}(y_1^-)|P,s_T\rangle 
\left(i\epsilon_{\perp\rho\lambda}\right)\, .
\nonumber \\
&=&
-\widetilde{\cal T}^{(f,d)}_{\Delta G,F}(x+x_2, x,\mu_F,-s_T)
\label{TDgt}
\een
From Eq.~(\ref{TDgasy}), it is clear that the 
new twist-3 tri-gluon correlation function
${\cal T}^{(f,d)}_{\Delta G,F}(x, x+x_2,\mu_F)$
is also {\it real}, but, {\it antisymmetric} in 
the exchange of  
$x$ and $x+x_2$.  Consequently, the diagonal part of 
these two new correlation functions vanishes, 
\ben
T_{\Delta q,F}(x,x,\mu_F) &\equiv & 
\int dx_2\, \left[2\pi\,\delta(x_2)\right]\, 
     {\cal T}_{\Delta q,F}(x,x+x_2,\mu_F)
=0            \, ,
\nnu
T_{\Delta G,F}^{(f,d)}(x,x,\mu_F) &\equiv& 
\int dx_2\, \left[2\pi\,\delta(x_2)\right]\, 
\left(\frac{1}{x}\right)
     {\cal T}^{(f,d)}_{\Delta G}(x,x+x_2,\mu_F)
=0            \, .
\label{TA_diag}
\een
That is, this set of twist-3 correlation functions
does not directly generate soft gluonic pole contribution to 
the SSAs \cite{qiu,koike}.

To complete this subsection, we summarize the key properties
of these twist-3 correlation functions that are responsible for
generating the SSAs from the unpinched gluonic and fermionic 
poles of partonic scattering in the QCD collinear 
factorization approach.  From their operator structure, these 
correlation functions can be grouped into two sets.  One 
set is for the ${\cal T}_{q,F}$ and ${\cal T}^{(f,d)}_{G,F}$, 
and the other includes ${\cal T}_{\Delta q,F}$ and 
${\cal T}^{(f,d)}_{\Delta G,F}$. The operators for the first
set of correlation functions, ${\cal T}_{q,F}$ and 
${\cal T}^{(f,d)}_{G,F}$, are constructed from the bi-local
operators that define the twist-2 {\it spin-averaged} PDFs 
with an insertion of the following operator, 
\ben
\int \frac{dy_2^-}{2\pi}\, e^{ix_2 P^+ y_2^-}    
\left[\epsilon^{s_T\sigma n\bar{n}}\,
      F_{\sigma}^{\ +}(y_2^-)\right]
=i\int \frac{dy_2^-}{2\pi}\, e^{ix_2 P^+ y_2^-}
  \left[i\,\epsilon_\perp^{\rho\sigma}\,
      s_{T\rho}\, F_{\sigma}^{\ +}(y_2^-)\right]\, ;
\label{Lorentz}
\een
and the operators for the second set of correlation functions,
${\cal T}_{\Delta q,F}$ and ${\cal T}^{(f,d)}_{\Delta G,F}$,
are constructed from the bi-local operators that define the 
twist-2 {\it spin-dependent} parton helicity distributions
with an insertion of a slightly different operator, 
\ben
i\int \frac{dy_2^-}{2\pi}\, e^{ix_2 P^+ y_2^-}  
\left[s_T^{\sigma}\, F_{\sigma}^{\ +}(y_2^-)\right]\, .
\label{magmoment}
\een
The $i\epsilon_\perp^{\rho\sigma}$ in Eq.~(\ref{Lorentz})
takes care of the parity invariance of the spin asymmetry
for the first set of correlation functions,
while the same property was taken care of naturally by
the $\gamma^5$ or $i\epsilon_{\perp\rho\lambda}$ in the 
operator definition of the spin-dependent helicity 
distributions.  The extra ``$i$'' in both Eq.~(\ref{Lorentz})
and Eq.~(\ref{magmoment}) provides the necessary phase for
the SSAs and is a result of taking the contribution from the 
gluonic or fermionic pole of partonic scattering \cite{qiu}.
It is this phase that both sets of twist-3 correlation 
functions relevant to the SSAs {\it do not} contribute to 
the long-distance correlation functions 
extracted from any measurement of parity conserving
{\it double-spin} asymmetries.  For example, none of these 
twist-3 correlation functions,
${\cal T}_{q,F}$, ${\cal T}^{(f,d)}_{G,F}$, 
${\cal T}_{\Delta q,F}$, and ${\cal T}^{(f,d)}_{\Delta G,F}$,
directly contributes to the DIS structure function $g_2$.

\section{Evolution Equations}
\label{evolution}

In this section we introduce the Feynman diagram representation 
of the twist-3 quark-gluon and tri-gluon correlation functions 
defined in the last section.  We derive cut vertices in momentum space
to connect the Feynman diagrams to specific twist-3 correlation
functions \cite{Mueller:1981sg}.  
From the Feynman diagram representation, 
we derive the evolution equations for the scale dependence of 
these twist-3 correlation functions.

\bef
\psfig{file=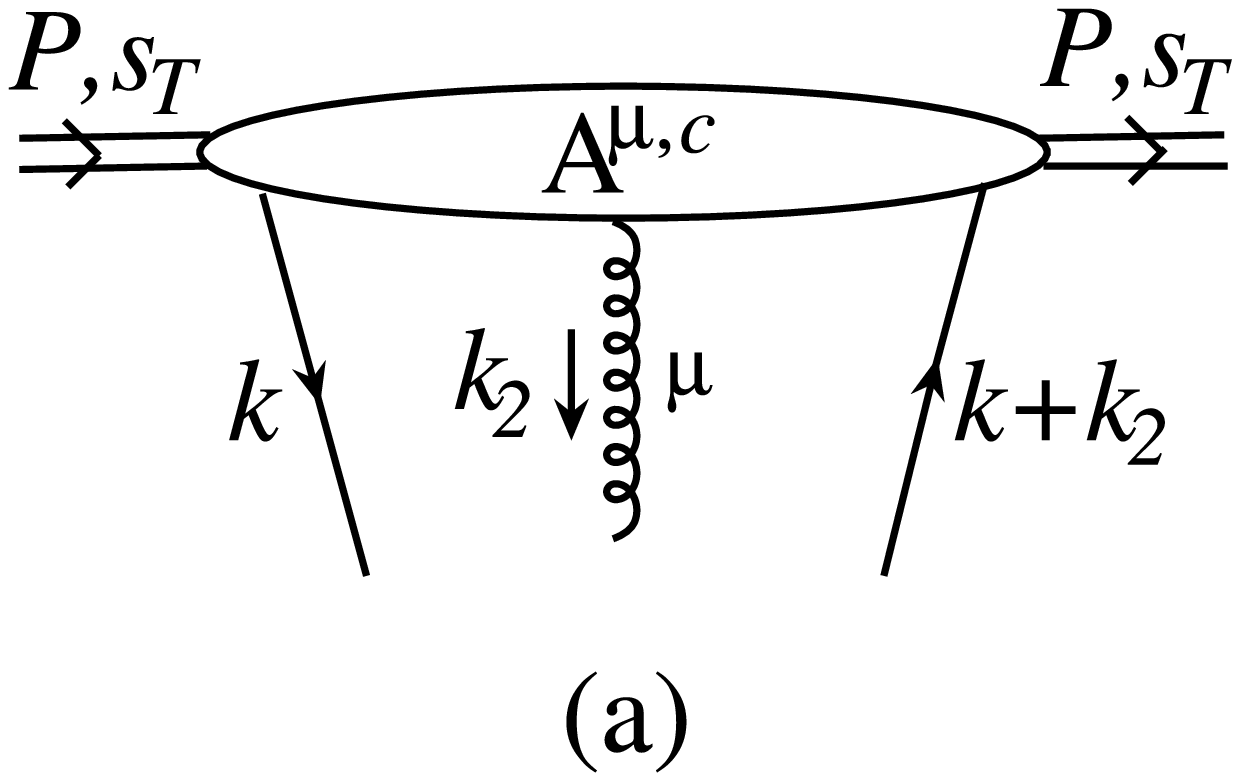,width=1.5in}
\hskip 0.5in
\psfig{file=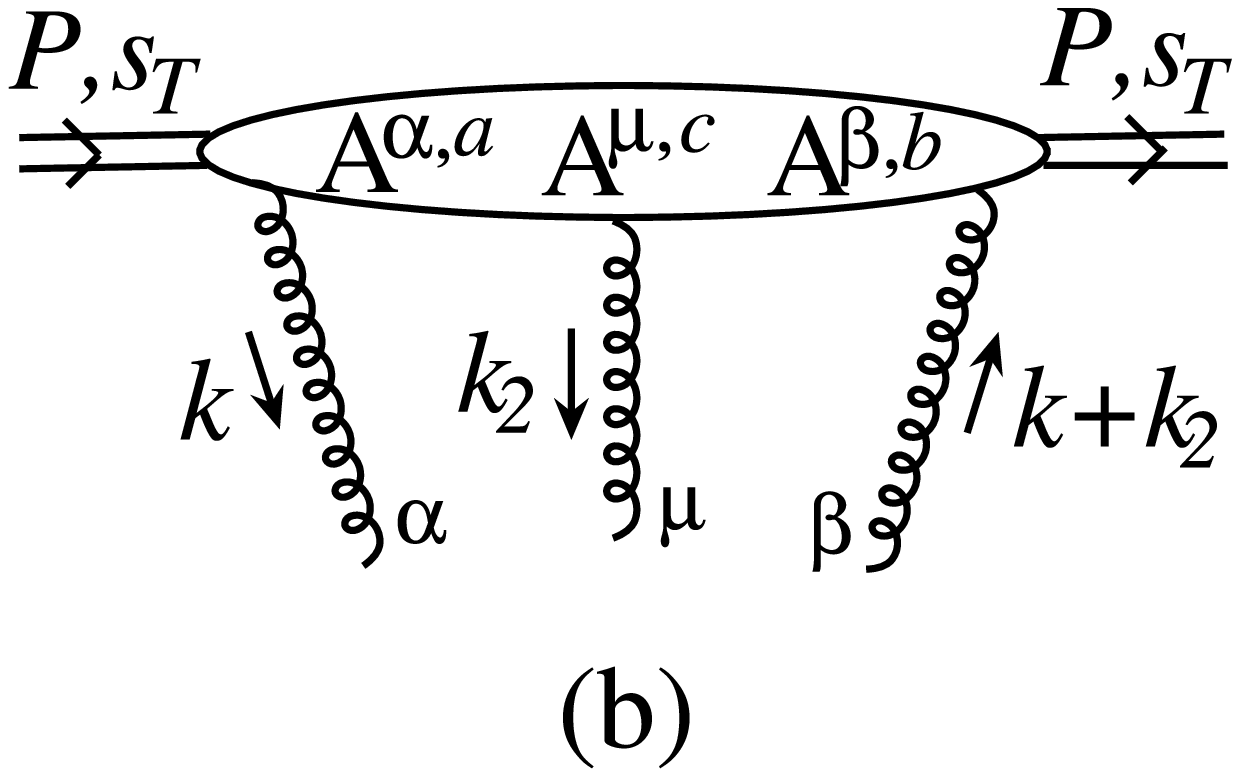,width=1.5in}
\caption{Feynman diagrams that contribute to the twist-3
quark-gluon (a) and tri-gluon (b) correlation functions.  
$\alpha,\beta,\mu$ and $a,b,c$ are Lorentz and color indices 
of gluon field operators, respectively.}
\label{fig1}
\eef

\subsection{Feynman diagram representation and cut vertex}

In QCD collinear factorization approach to SSAs, the twist-3
three-parton correlation functions measure the net effect of 
the quantum interference between two scattering amplitudes of 
the transversely polarized hadron: one with single active 
parton and the other with two active partons, participating in
the short-distance hard scattering \cite{qiu}.  
Like the normal PDFs,
the quark-gluon and tri-gluon correlation functions, as defined
in last section, could be represented by the cut forward 
scattering diagrams as sketched in Figs.~\ref{fig1}(a) and 
\ref{fig1}(b), respectively.  The cut represents a particular 
final-state.  The Feynman diagrams in Fig.~\ref{fig1} should
include all possible cuts to sum over all possible final 
states.  We suppress the explicit cuts for the diagrams in
Fig.~\ref{fig1} since the matrix element of the three-parton 
correlation functions with the middle gluon field strength 
in the left side of the cut is equal to the matrix element 
with the gluon field strength in the right side of the 
final-state cut.  This is because the field operators of 
hadronic matrix elements commute on the light-cone 
\cite{Qiu_sterman,Jaffe:1983hp}.  
Because of the odd number of active fields defining
the twist-3 correlation functions, unlike the normal PDFs,
these correlation functions do not have the probability 
interpretation.

As discussed in the last section, one set of twist-3 
correlation functions is expressed in terms of a 
{\it sum} of two spin-dependent twist-3 correlation functions, 
as in Eqs.~(\ref{Tqasy}) and (\ref{Tgasy}), and the other
by a {\it difference} of two spin-dependent twist-3 correlation 
functions, as in Eqs.~(\ref{TDqasy}) and (\ref{TDgasy}).  
These spin-dependent twist-3 correlation
functions are given in terms of explicit matrix elements
of the transverse-spin dependent hadronic state and 
could be represented by Feynman diagrams.
However, since all gluon lines in Feynman diagrams
are connected to gluon fields,  
calculating the Feynman diagrams in Fig.~\ref{fig1}
does not immediately give the twist-3 correlation functions
whose gluonic degree of freedom is represented by 
the field strength, $F^{+\mu}$, not the gluon
field, $A^\mu$.  Therefore, in order to fully define 
the Feynman diagram representation of the spin-dependent
twist-3 correlation functions, we need to derive the cut vertex
\cite{Mueller:1981sg} to connect the operator definition of 
the spin-dependent twist-3 correlation functions 
to the cut forward scattering Feynman diagrams 
in Fig.~\ref{fig1}.  With different cut vertices, 
the same diagrams in Fig.~\ref{fig1}
can represent both sets of the spin-dependent 
twist-3 correlation functions.

The cut vertex can be derived by rewriting the operator 
definition of the correlation functions in terms of 
hadronic matrix elements of quark and gluon operators 
in momentum space \cite{Mueller:1981sg}.  

To derive the cut vertex to connect the spin-dependent
quark-gluon correlation function in Eq.~(\ref{Tqt}) 
to the Feynman diagram in Fig.~\ref{fig1}(a), 
we reexpress the operator definition of the
correlation function in Eq.~(\ref{Tqt}) in terms of 
hadronic matrix elements of quark and gluon operators 
in momentum space and find,
\ben
\widetilde{\cal T}_{q,F}(x, x+x_2,\mu_F,s_T)
&=&
\int \frac{d^4k}{(2\pi)^4}\, \frac{d^4k_2}{(2\pi)^4} \,
\langle P,s_T|\,\widetilde{\overline{\psi}}_{q,i}(-k-k_2)
\nonumber \\
&\ & \times
\left[
  \frac{\gamma^+}{2P^+}\,\delta\left(x-\frac{k^+}{P^+}\right)
  x_2\, \delta\left(x_2-\frac{k^+_2}{P^+}\right)
  \left(i\, \epsilon^{s_T\sigma n \bar{n}}\right)
  \left(-g_{\sigma\mu}+\frac{k_{2\sigma} n_\mu}{k_2^+}\right)
\right] \left({\cal C}_q\right)^c_{ij}
\nonumber\\
&\ & \hskip 2.2in \times
\widetilde{A}^{\mu,c}(k_2)\, \widetilde{\psi}_{q,j}(k)\,
|P,s_T\rangle \, ,
\label{Tqm}
\een
where the fermionic color contraction factor ${\cal C}_q$
is given by
\ben
\left({\cal C}_q\right)^c_{ij} 
= \left(t^c\right)_{ij} \, ,
\label{color_qc}
\een
with quark and gluon color indices,
$i,j=1,2,3=N_c$ and $c=1,2,...,8=N_c^2-1$, respectively, and 
$t^c$ are the generators of the fundamental representation 
of SU(3) color.  The field operators listed with ``$\sim$'' in
Eq.~(\ref{Tqm}) represent the momentum space field operators 
of those in Eq.~(\ref{Tqt}).  
The matrix element,
$\langle P,s_T|\,\widetilde{\overline{\psi}}_{q,i}(-k-k_2)
\widetilde{A}^{\mu,c}(k_2)\, \widetilde{\psi}_{q,j}(k)\,
|P,s_T\rangle$, in Eq.~(\ref{Tqm}) can be represented by 
the Feynman diagram in Fig.~\ref{fig1}(a).  
By comparing the definition of quark-gluon correlation function
in Eq.~(\ref{Tqm}) and the Feynman diagram in Fig.~\ref{fig1}(a), 
it is clear that we can derive the quark-gluon 
correlation function, $\widetilde{\cal T}_{q,F}$, 
from the Feynman diagram in Fig.~\ref{fig1}(a) by contracting
the quark and gluon lines with the expression in the 
square brackets and the color contraction factor $(t^c)_{ij}$,
plus the integration over the loop momenta in Eq.~(\ref{Tqm}). 
The expression in the square brackets 
plus the color contraction factor ${\cal C}_q$
defines the cut vertex that connects 
the Feynman diagram in Fig.~\ref{fig1}(a) to the quark-gluon 
correlation function $\widetilde{\cal T}_{q,F}$ in Eq.~(\ref{Tqt}),
\ben
{\cal V}_{q,F}\equiv
\frac{\gamma^+}{2P^+}\,\delta\left(x-\frac{k^+}{P^+}\right)
\left(i\,\epsilon^{s_T\sigma n\bar{n}}\right)\,
x_2\, \delta\left(x_2-\frac{k_2^+}{P^+}\right)
\left[-g_{\sigma\mu}\, 
      + \frac{k_{2\sigma} n_\mu}{k_2^+}\,   \right]\, {\cal C}_q\, .
\label{cv_q}
\een
Similarly, we can rewrite the tri-gluon correlation function
in Eq.~(\ref{Tgt}) as
\ben
\widetilde{\cal T}_{G}^{(f,d)}(x, x+x_2,\mu_F,s_T)
&=&
\int \frac{d^4k}{(2\pi)^4}\, \frac{d^4k_2}{(2\pi)^4} \,
\langle P,s_T|\,
\widetilde{A}^{\beta,b}(-k-k_2)\,
\widetilde{A}^{\mu,c}(k_2)\,
\widetilde{A}^{\alpha,a}(k)\, |P,s_T\rangle
\nonumber \\
&\ & \times
\left[
  \,x(x+x_2)
  \left(-g_{\alpha\beta} 
        + \frac{(k+k_2)_\alpha n_\beta}{(k+k_2)^+}
        + \frac{k_\beta\, n_\alpha}{k^+}
        - \frac{k\cdot(k+k_2)\, n_\alpha n_\beta}
               {k^+\, (k+k_2)^+}
  \right) \delta\left(x-\frac{k^+}{P^+}\right)
\right.
\nonumber \\
&\ & \ \ \times
\left.
  x_2\, \delta\left(x_2-\frac{k^+_2}{P^+}\right)
  \left(i\, \epsilon^{s_T\sigma n \bar{n}}\right)
  \left(-g_{\sigma\mu}+\frac{k_{2\sigma}\, n_\mu}{k_2^+}\right) 
\right] \left({\cal C}_g\right)^{(f,d)}_{bca} \, ,
\label{Tgm}
\een
where the gluonic color contraction factor ${\cal C}_g$ 
is given by 
\ben
\left({\cal C}_g\right)^{(f)}_{bca} 
= i f_{bca}=({\cal F}^c)_{ba}\, ,
\quad\mbox{and}\quad
\left({\cal C}_g\right)^{(d)}_{bca} = d_{bca}\, ,
\label{color_gc}
\een
where ${\cal F}^c$ are the generators of adjoint representation 
of SU(3) color.  In Eq.~(\ref{Tgm}), 
the matrix element
$\langle P,s_T|\,
\widetilde{A}^{\beta,b}(-k-k_2)\,
\widetilde{A}^{\mu,c}(k_2)\,
\widetilde{A}^{\alpha,a}(k)\, |P,s_T\rangle$
can be represented by the Feynman diagram in Fig.~\ref{fig1}(b).
Similar to the situation of quark-gluon correlation, 
the expression in the square brackets plus 
the color contraction factor ${\cal C}^{(f,d)}_{bca}$
defines the cut vertex for calculating the 
tri-gluon correlation function $\widetilde{\cal T}_{G,F}^{(f,d)}$
from the diagram in Fig.~\ref{fig1}(b),
\ben
{\cal V}_{G,F} &\equiv& 
x(x+x_2)
  \left(-g_{\alpha\beta} 
        + \frac{(k+k_2)_\alpha n_\beta}{(k+k_2)^+}
        + \frac{k_\beta\, n_\alpha}{k^+}
        - \frac{k\cdot(k+k_2)\, n_\alpha n_\beta}
               {k^+\, (k+k_2)^+}
  \right) \delta\left(x-\frac{k^+}{P^+}\right)
\nonumber \\
&\ & \times\,
  x_2\, \delta\left(x_2-\frac{k^+_2}{P^+}\right)
  \left(i\, \epsilon^{s_T\sigma n \bar{n}}\right)
  \left[-g_{\sigma\mu}+\frac{k_{2\sigma}\, n_\mu}{k_2^+}\right] 
\left({\cal C}_g\right)^{(f,d)}_{bca} \, .
\label{cv_g}
\een

Similarly, by rewriting the operator definitions of the second
set of spin-dependent twist-3 correlation functions in terms of 
quark and gluon field operators in momentum space, we derive 
the following cut vertices,
\ben
{\cal V}_{\Delta q,F}\equiv
\frac{\gamma^+\gamma^5}{2P^+}\,\delta\left(x-\frac{k^+}{P^+}\right)
\left(-s_T^\sigma \right)\,
x_2\, \delta\left(x_2-\frac{k_2^+}{P^+}\right)
\left[-g_{\sigma\mu}\, 
      + \frac{k_{2\sigma} n_\mu}{k_2^+}\,   \right]\, {\cal C}_q
\label{cv_Dq}
\een
for connecting the same Feynman diagram in Fig.~\ref{fig1}(a) to 
the second set quark-gluon correlation function 
$\widetilde{\cal T}_{\Delta q,F}$ in Eq.~(\ref{TDqt}), and 
\ben
{\cal V}_{\Delta G,F} &\equiv& 
x(x+x_2) \left(i\epsilon_{\perp\rho\lambda}\right)
  \left[-g^{\rho\beta} 
        + \frac{(k+k_2)^\rho n^\beta}{(k+k_2)^+}\right]
  \left[-g^{\lambda\alpha} 
        + \frac{k_\lambda\, n_\alpha}{k^+}\right]
  \delta\left(x-\frac{k^+}{P^+}\right)
\nonumber \\
&\ & \times\,
  x_2\, \delta\left(x_2-\frac{k^+_2}{P^+}\right)\left(-s_T^\sigma\right)\,
  \left[-g_{\sigma\mu}+\frac{k_{2\sigma}\, n_\mu}{k_2^+}\right] \,
  {\cal C}_g^{(f,d)} \, 
\label{cv_Dg}
\een
for connecting the same Feynman diagram in Fig.~\ref{fig1}(b) to 
the second set tri-gluon correlation function 
$\widetilde{\cal T}_{\Delta G,F}$ in Eq.~(\ref{TDgt}).
The color factors in Eqs.~(\ref{cv_Dq}) and Eqs.~(\ref{cv_Dg})
are the same as those in Eqs.~(\ref{cv_q}) and Eqs.~(\ref{cv_g}),
respectively.

For our calculation of the evolution kernels in the next section
in the light-cone gauge, $n\cdot A=0$, the cut vertices 
are simplified as,
\ben
{\cal V}_{q,F}^{\rm LC}
&=&
\frac{\gamma^+}{2P^+}\,\delta\left(x-\frac{k^+}{P^+}\right)
x_2\,\delta\left(x_2-\frac{k_2^+}{P^+}\right)
\left(i\,\epsilon^{s_T\sigma n\bar{n}}\right)
\left[-g_{\sigma\mu}\right]\,
{\cal C}_q \, ,
\label{cv_q_lc}\\
{\cal V}_{G,F}^{\rm LC}
&=&
x(x+x_2)
  \left(-g_{\alpha\beta} \right)
  \delta\left(x-\frac{k^+}{P^+}\right)
  x_2\, \delta\left(x_2-\frac{k^+_2}{P^+}\right)
  \left(i\, \epsilon^{s_T\sigma n \bar{n}}\right)
  \left[-g_{\sigma\mu}\right] \,
  {\cal C}_g^{(f,d)} \, ,
\label{cv_g_lc}\\
{\cal V}_{\Delta q,F}^{\rm LC}
&=&
\frac{\gamma^+\gamma^5}{2P^+}\,\delta\left(x-\frac{k^+}{P^+}\right)
x_2\, \delta\left(x_2-\frac{k_2^+}{P^+}\right)
\left(-s_T^\sigma \right)\,
\left[-g_{\sigma\mu}\right]\, {\cal C}_q
\label{cv_Dq_lc}\\
{\cal V}_{\Delta G,F}^{\rm LC}
&=&
x(x+x_2) \left(i\epsilon_{\perp}^{\beta\alpha}\right)
  \delta\left(x-\frac{k^+}{P^+}\right)
  x_2\, \delta\left(x_2-\frac{k^+_2}{P^+}\right)
    \left(-s_T^\sigma\right)\, 
    \left[-g_{\sigma\mu}\right]\,
    {\cal C}_g^{(f,d)}
\label{cv_Dg_lc}
\een
for the tri-gluon correlation functions, 
$\widetilde{\cal T}_{q,F}$, 
$\widetilde{\cal T}^{(f,d)}_{G,F}$,
$\widetilde{\cal T}_{\Delta q,F}$, 
$\widetilde{\cal T}^{(f,d)}_{\Delta G,F}$,
respectively.

\subsection{Factorization and evolution equations}

In order to derive the evolution equations and evolution kernels 
from the definition of the twist-3 correlation functions, 
we need to compute the perturbative modification to these 
correlation functions caused by the quark-gluon 
interaction in QCD \cite{Collins:1988wj}.  For example,
we need to calculate the diagram in Fig.~\ref{fig2} for 
extracting the flavor non-singlet evolution kernel of the
quark-gluon correlation function.    

\bef
\psfig{file=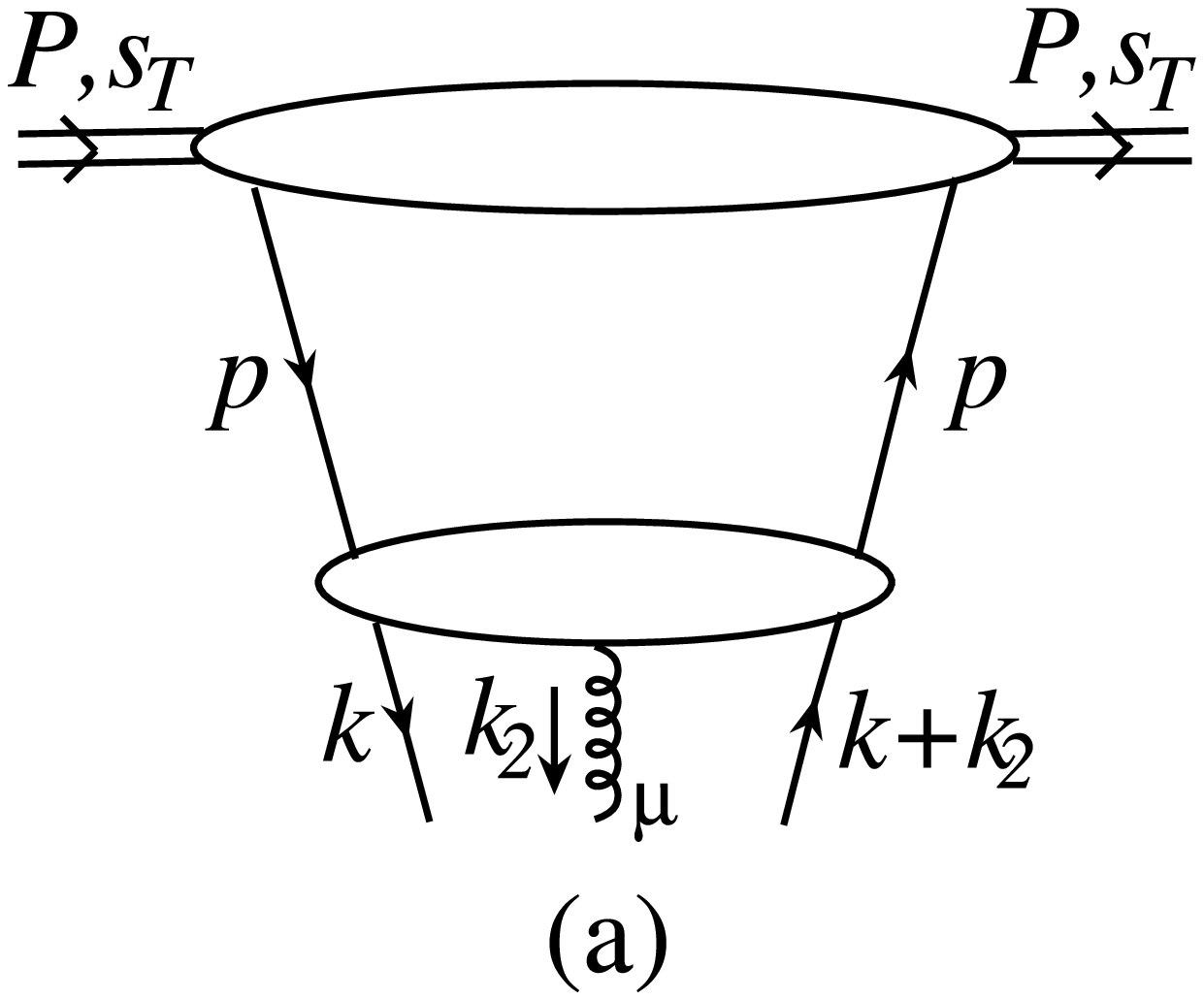,width=1.5in}
\hskip 0.5in
\psfig{file=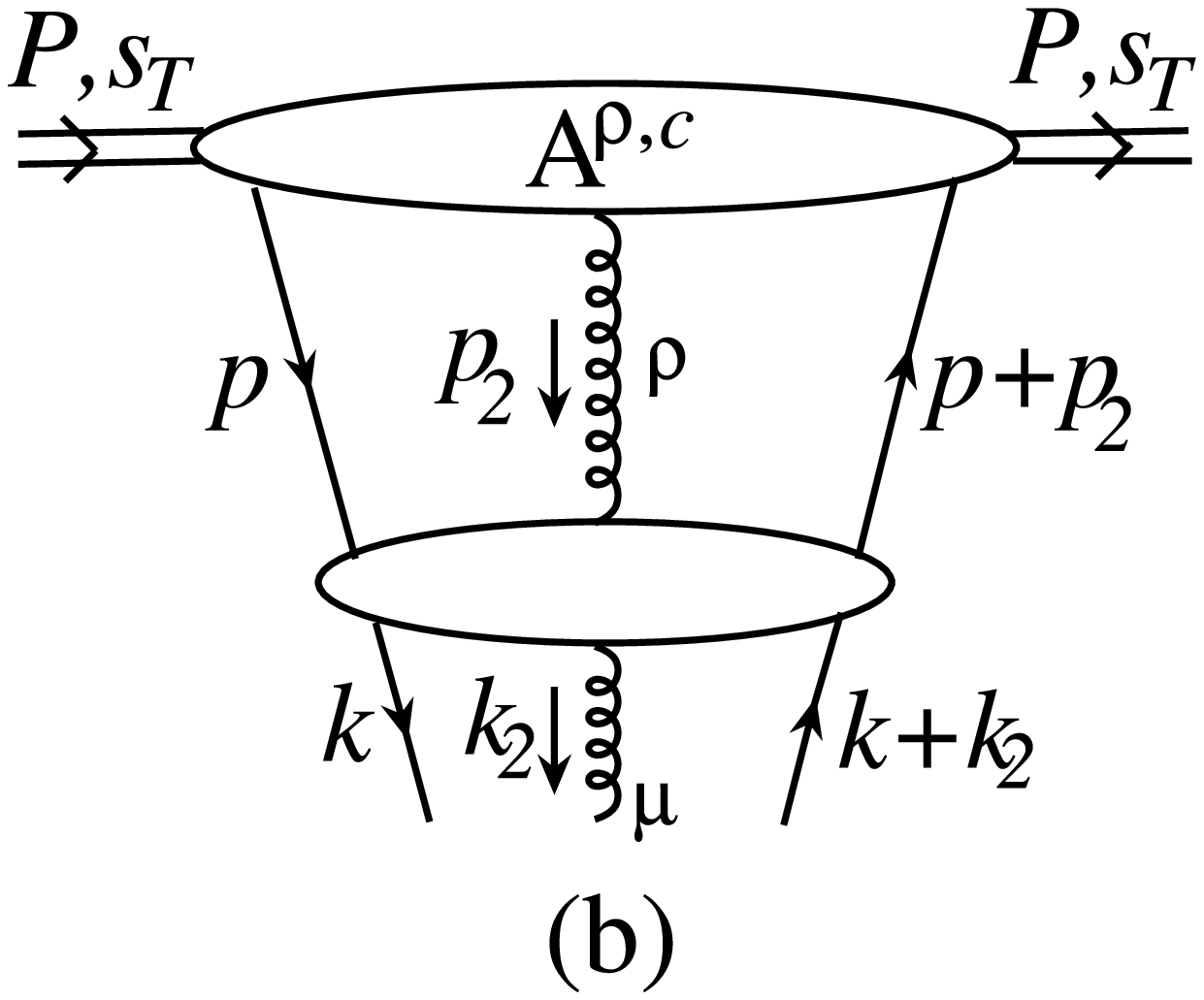,width=1.5in}
\caption{Feynman diagrams that contribute to the flavor non-singlet 
change of the twist-3 quark-gluon correlation function where
$\mu,\rho$ and $c$ are Lorentz and color indices 
of gluon field operators, respectively.
The lower part of quark and gluon lines are contracted to 
the cut vertex that define the quark-gluon correlation function.}
\label{fig2}
\eef

We first evaluate the perturbative change to all spin-dependent 
correlation functions,
$\widetilde{\cal T}_{q,F}$, $\widetilde{\cal T}^{(f,d)}_{G,F}$,
$\widetilde{\cal T}_{\Delta q,F}$, and
$\widetilde{\cal T}^{(f,d)}_{\Delta G,F}$, because
they are defined in terms of hadronic matrix elements and 
represented by the Feynman diagrams with proper cut vertices.
We follow the standard QCD collinear factorization approach 
to factorize the perturbative change to these spin-dependent
correlation functions into short-distance evolution kernels
convoluted with corresponding gauge invariant long-distance 
matrix elements or the correlation functions
\cite{Collins:1988wj,Ellis:1982cd,qiu_t4}.
From Eqs.~(\ref{Tqasy}), (\ref{Tgasy}), (\ref{TDqasy}),
and (\ref{TDgasy}), we then derive the evolution equations
for the two sets of twist-3 correlation functions that
are responsible for generating the SSAs in QCD collinear 
factorization approach.

We start with the flavor non-singlet change to 
the quark-gluon correlation function, $\widetilde{\cal T}_{q,F}$,
as represented by the diagrams in Fig.~\ref{fig2}
with the cut vertex in Eq.~(\ref{cv_q}). 
For the twist-3 correlation functions relevant to the SSAs,
we are interested in the difference of the diagrams 
in Fig.~\ref{fig2} with hadron spin $s_T$ and that with $-s_T$.
Only survival leading twist matrix element from the top of 
the diagram in Fig.~\ref{fig2}(a) after taking the difference 
is the transversity distribution that 
does not contribute to the change of the quark-gluon correlation
function of massless quark due to the symmetry of time-reversal 
or simply due to an odd number of gamma matrices in the spinor 
trace.  Following the same derivation and steps presented in 
Refs.~\cite{Ellis:1982cd,qiu_t4}, we find that 
the twist-3 contribution from the diagram in 
Fig.~\ref{fig2}(a) can be combined with the leading 
contribution of the diagram in Fig.~\ref{fig2}(b) 
due to color gauge invariance \cite{qiu,koike}.  
Long-distance physics of the combined contribution from 
two diagrams in Fig.~\ref{fig2} could be expressed in terms
of four twist-3 long-distance matrix elements or 
correlation functions, $T_{(D,F)}^{(V,A)}$,
as defined in Ref.~\cite{Qiu:1998ia}, where superscripts 
$V$ and $A$ represent the vector and axial vector current, 
respectively, and subscripts, $D$ and $F$, refer to the 
standard QCD covariant derivative and field strength, respectively. 
The correlation functions, $T_{F}^{V}$ and $T_{F}^{A}$, 
correspond to our spin-dependent quark-gluon correlation functions,
$\widetilde{\cal T}_{q,F}$ and $\widetilde{\cal T}_{\Delta q,F}$,
respectively, while the other two functions could be obtained 
by replacing the field strength operator $F_{\sigma}^{~ +}$ 
by the covariant derivative operator $D_{\sigma}$. 
As explained in Ref.~\cite{Qiu:1998ia}, the two correlation
functions with the covariant derivative operator do not
contribute to the SSAs.

We now provide a detailed derivation of the projection
operator for extracting the flavor non-singlet evolution
kernels or the short-distance contribution from both diagrams 
in Fig.~\ref{fig2}.  There are two sources of twist-3
or subleading power contribution from the diagram 
in Fig.~\ref{fig2}(a) \cite{qiu,Ellis:1982cd,qiu_t4}.  
One is from the transverse momentum expansion of the 
parton momenta entering the bottom part of the diagram
and the other is from the spinor trace decomposition when
the bottom part of the diagram is contracted by $\gamma\cdot n$
instead of the leading $\gamma\cdot P$ 
\cite{Ellis:1982cd,qiu_t4}.  The part from the 
transverse momentum expansion contributes to the 
matrix elements with a covariant derivative, $T_D^{V,A}$, 
which do not contribute to the SSAs as discussed in 
Ref.~\cite{Qiu:1998ia}.  The other subleading contribution 
of the diagram in Fig.~\ref{fig2}(a) due to the spinor 
decomposition could contribute. 
Although the matrix element of the subleading term from 
the spinor decomposition of the diagram in Fig.~\ref{fig2}(a) 
has only two quark field operators, it can be expressed in 
terms of a matrix element of two quark fields and a gluon 
field by applying the equation of motion \cite{Ellis:1982cd}.
That is, this part of subleading contribution from the 
diagram in Fig.~\ref{fig2}(a) can be represented by the 
same diagram in Fig.~\ref{fig2}(b) except that the partonic 
Feynman diagrams in the bottom part of the diagram
are given by the diagrams with the contact interaction
\cite{qiu,qiu_t4}.  Therefore, we can derive the full 
flavor non-singlet evolution kernels from the diagram in 
Fig.~\ref{fig2}(b) with the understanding that the bottom
part of the diagram also includes those with the contact
interaction or the special propagator \cite{qiu,qiu_t4}.

We represent the perturbative change to 
$\widetilde{\cal T}_{q,F}$ from the diagram in 
Fig.~\ref{fig2}(b) as
\ben
d\widetilde{\cal T}_{q,F}(x,x+x_2,\mu_F,s_T) 
\equiv \int \frac{d^4p\,d^4p_2}{(2\pi)^8}\
{\rm Tr}\left[\hat{T}^\rho(p,p_2,P,s_T)
              \hat{H}_{\rho}(p,p_2,x,x_2,\mu_F)\right]\, ,
\label{Iq}
\een
where the ``Tr'' represents the trace over the fermion
fields' spinor indices, and $\hat{T}$ and $\hat{H}$ represent 
the top part and the bottom part of the Feynman diagram, 
respectively.  In the momentum space, the $\hat{T}$ is given by 
the matrix element,
\ben
\hat{T}^\rho(p,p_2,P,s_T)
= \langle P,s_T|\,\widetilde{\overline{\psi}}_{q,i}(-p-p_2)
\widetilde{A}^{\rho,c}(p_2)\, \widetilde{\psi}_{q,j}(p)\,
|P,s_T\rangle\, ,
\label{T12}
\een
where $i,j$ are color indices of the quark fields and $\rho, c$ 
are Lorentz and color indices of the gluon field, respectively.  
The $\hat{H}$ represents the bottom blob that includes all cut 
Feynman diagrams for the given external quark and gluon lines.
The list of all cut diagrams at order of $\alpha_s$ will be
given in the next section when we present the calculation of
evolution kernels.  
The bottom quark and gluon lines of these diagrams are contracted 
by the cut vertex that defines the correlation function.  
The dependence of $x$ and $x_2$ in the argument of $\hat{H}$ 
in Eq.~(\ref{Iq}) is from the cut vertex, and the scale 
$\mu_F$ represents the hardness or the off-shellness of the 
parton momenta, $k$ and $k_2$.  To pick up the leading power
contribution from the perturbative modification to the 
quark-gluon correlation function, 
we first separate the spinor trace for the case 
of massless partons by \cite{qiu_t4}
\ben
\hat{H}_\rho(p,p_2,x,x_2,\mu_F) \approx
H_{\rho,\alpha}(p,p_2,x,x_2,\mu_F) \left(
\frac{1}{2}\gamma^\alpha \right) 
+ \widetilde{H}_{\rho,\alpha}(p,p_2,x,x_2,\mu_F) \left(
\frac{1}{2}\gamma^\alpha\left(i\gamma^5\right) \right) 
+ \dots \, ,
\label{Tqspin}
\een
where ``$\dots$'' represents terms with even number of 
$\gamma$-matrices and subleading, and 
\ben
H_{\rho,\alpha}(p,p_2,x,x_2,\mu_F)
=\frac{1}{2} {\rm Tr}\left[
\hat{H}_\rho(p,p_2,x,x_2,\mu_F)\gamma_\alpha \right]\, ,
\label{Hma}
\een
and
\ben
\widetilde{H}_{\rho,\alpha}(p,p_2,x,x_2,\mu_F)
=\frac{1}{2} {\rm Tr}\left[
\hat{H}_\rho(p,p_2,x,x_2,\mu_F)\gamma_\alpha 
\left(i\gamma^5\right)\right]\, .
\label{Hma5}
\een
In order to derive the contribution from the first term in
Eq.~(\ref{Tqspin}) in details, we introduce
\ben
I_q\equiv \int \frac{d^4p\,d^4p_2}{(2\pi)^8}\
T^{\rho,\alpha}(p,p_2,P,s_T)
H_{\rho,\alpha}(p,p_2,x,x_2,\mu_F)\, ,
\label{Iqspin}
\een
with 
\ben
T^{\rho,\alpha}(p,p_2,P,s_T)
= \frac{1}{2}{\rm Tr}\left[
\hat{T}^\rho(p,p_2,P,s_T)\gamma^\alpha \right]\, .
\label{Tma}
\een
We then apply the strong ordering 
in the off-shellness of active partons, $|p^2|\ll \mu_F^2$ and
$|p_2^2|\ll \mu_F^2$, and make the collinear approximation to 
expand the parton momenta entering into the $\hat{H}$ 
in Fig.~\ref{fig2}(b) around $p=\xi P$ and $p_2=\xi_2 P$ as
\ben
H_{\rho,\alpha}(p,p_2,x,x_2,\mu_F) 
&\approx &
H_{\rho,\alpha}(\xi P,\xi_2 P,x,x_2,\mu_F)
+ \frac{\partial H_{\rho,\alpha}(\xi P,\xi_2 P,x,x_2,\mu_F)}
       {\partial p^\beta} \,
  (p-\xi P)^\beta
\nonumber \\
&\ &
+\frac{\partial H_{\rho,\alpha}(\xi P,\xi_2 P,x,x_2,\mu_F)}
      {\partial p_2^\beta}\, 
  (p_2-\xi_2 P)^\beta
+ \dots \, .
\label{Hma_co}
\een
By substituting Eq.~(\ref{Hma_co}) into Eq.~(\ref{Iqspin}), 
we can rewrite the $I_q$ as 
\ben
I_q & \approx &
\int d\xi\, d\xi_2\, 
T^{\rho,\alpha}(\xi,\xi+\xi_2)\, 
H_{\rho,\alpha}(\xi,\xi_2,x,x_2,\mu_F)
+ \int d\xi\, d\xi_2\, 
T_1^{\rho,\alpha,\beta}(\xi,\xi+\xi_2)\, 
\frac{\partial H_{\rho,\alpha}(\xi,\xi_2,x,x_2,\mu_F)}
     {\partial p^\beta}
\nonumber \\
&\ &
+ \int d\xi\, d\xi_2\, 
T_2^{\rho,\alpha,\beta}(\xi,\xi+\xi_2)\, 
\frac{\partial H_{\rho,\alpha}(\xi,\xi_2,x,x_2,\mu_F)}
     {\partial p_2^\beta}
+ \dots\, ,
\label{Iq_co}
\een
where the explicit $P$ dependence in $H_{\rho,\alpha}$ is
suppressed.  The correlation functions in Eq.~(\ref{Iq_co}) 
are given by
\ben
T^{\rho,\alpha}(\xi,\xi+\xi_2)
&=& 
\int \frac{d^4p\, d^4 l_2}{(2\pi)^8}\,
\delta\left(\xi-\frac{p^+}{P^+}\right)
\delta\left(\xi_2-\frac{p_2^+}{P^+}\right)
\langle P,s_T|\,\widetilde{\overline{\psi}}_{q,i}(-p-p_2)
\frac{\gamma^\alpha}{2}
\widetilde{A}^{\rho,c}(p_2)\, \widetilde{\psi}_{q,j}(p)\,
|P,s_T\rangle\, ;
\nonumber \\
T_1^{\rho,\alpha,\beta}(\xi,\xi+\xi_2)
&=& 
\int \frac{d^4p\, d^4 l_2}{(2\pi)^8}\,
\delta\left(\xi-\frac{p^+}{P^+}\right)
\delta\left(\xi_2-\frac{p_2^+}{P^+}\right)(p-\xi P)^\beta
\nonumber\\
&\ & \hskip 0.5in \times
\langle P,s_T|\,\widetilde{\overline{\psi}}_{q,i}(-p-p_2)
\frac{\gamma^\alpha}{2}
\widetilde{A}^{\rho,c}(p_2)\, \widetilde{\psi}_{q,j}(p)\,
|P,s_T\rangle\, ;
\nonumber \\
T_2^{\rho,\alpha,\beta}(\xi,\xi+\xi_2)
&=& 
\int \frac{d^4p\, d^4 l_2}{(2\pi)^8}\,
\delta\left(\xi-\frac{p^+}{P^+}\right)
\delta\left(\xi_2-\frac{p_2^+}{P^+}\right)(p_2-\xi_2 P)^\beta
\nonumber \\
&\ & \hskip 0.5in \times
\langle P,s_T|\,\widetilde{\overline{\psi}}_{q,i}(-p-p_2)
\frac{\gamma^\alpha}{2}
\widetilde{A}^{\rho,c}(p_2)\, \widetilde{\psi}_{q,j}(p)\,
|P,s_T\rangle\, .
\label{T_co} 
\een
Finally, we decouple the contraction of Lorentz indices in 
the RHS of Eq.~(\ref{Iq_co}) to express the 
quark-gluon correlation functions in terms of the 
$\widetilde{\cal T}_{q,F}$, defined in Eq.~(\ref{Tqm}), 
so that we can factorize 
the leading term of the RHS of Eq.~(\ref{Iq_co}) 
into a convolution of the $\widetilde{\cal T}_{q,F}$ and 
corresponding evolution kernel.  We find
\ben
T^{\rho,\alpha}(\xi,\xi+\xi_2)
\approx \left[\, \widetilde{\cal C}_q 
\left(\frac{-1}{\xi_2}\right) 
\left(i\, \epsilon^{s_T\rho n \bar{n}}\right) P^\alpha\,\right]
\widetilde{\cal T}_{q,F}^{\rm (LC)}(\xi,\xi+\xi_2,s_T) + \dots \,
\label{Tma_Tq}
\een
where the factorization scale dependence is suppressed and
the fermionic color projection operator
$\widetilde{\cal C}_q$ is given by
\ben 
(\widetilde{\cal C}_q)^c_{ji}=2/(N_c^2-1) (t^c)_{ji}\, ,
\label{color_qp}
\een
with the quark and gluon color indices $ij$ and $c$ 
as labeled in Fig.~\ref{fig2}(a), so that
$\widetilde{\cal C}_q\, {\cal C}_q = 1$.  
In Eq.~(\ref{Tma_Tq}), the quark-gluon correlation function 
$\widetilde{\cal T}_{q,F}^{\rm (LC)}$ has the same definition as 
that of $\widetilde{\cal T}_{q,F}$ in Eq.~(\ref{Tqt}), except
the cut vertex in the square brackets is replaced by 
the cut vertex in the light-cone gauge in Eq.~(\ref{cv_q_lc}).
The superscript ``LC'' indicates that this quark-gluon
correlation function is calculated by using the 
light-cone gauge cut vertex instead of the full cut vertex. 
We find that the term proportional to 
$T_1^{\rho,\alpha,\beta}(\xi,\xi+\xi_2)$ in 
Eq.~(\ref{Iq_co}) does not give the leading power 
contribution.  For the third term in Eq.~(\ref{Iq_co}), 
we have
\ben
T_2^{\rho,\alpha,\beta}(\xi,\xi+\xi_2)
\approx \left[ \, \widetilde{\cal C}_q 
\left(i\, \epsilon^{s_T\beta n \bar{n}}\right)
P^\rho\, P^\alpha\,\right]
\widetilde{\cal T}_{q,F}^{\rm (CO)}(\xi,\xi+\xi_2,s_T) + \dots \,
\label{T2ma_Tq}
\een
where the long-distance quark-gluon correlation function, 
$\widetilde{\cal T}_{q,F}^{\rm (CO)}$,  
has the same definition as that of $\widetilde{\cal T}_{q,F}$ 
in Eq.~(\ref{Tqt}), except the cut vertex in the square 
brackets is replaced by 
\ben
\frac{\gamma^+}{2P^+}\,\delta\left(\xi-\frac{p^+}{P^+}\right)
\left(i\,\epsilon^{s_T\sigma n\bar{n}}\right)
\xi_2\, \delta\left(\xi_2-\frac{p_2^+}{P^+}\right)
\frac{p_{2\sigma}\, n_\rho}{p_2^+}\, 
\left({\cal C}_q\right)^c_{ij}\, ,
\label{cv_q_co}
\een
which corresponds to the second term in the square brackets  
in Eq.~(\ref{cv_q}).  The superscript ``CO'' indicates that
this term provides the leading contribution in a covariant 
gauge calculation of the correlation functions 
\cite{qiu,Luo:1993ui}.  
From the factorized expression for the first and 
the third term, we find the leading contribution from the RHS 
of Eq.~(\ref{Iq_co}) can be factorized as
\ben
I_q & \approx & 
\int d\xi\, d\xi_2\, 
\Bigg\{
\widetilde{\cal T}_{q,F}^{\rm (LC)}(\xi,\xi+\xi_2,s_T)\, 
\left[\widetilde{\cal C}_q \left(\frac{-1}{\xi_2}\right) 
\left(i\, \epsilon^{s_T\rho\, n \bar{n}}\right) P^\alpha\, 
H_{\rho,\alpha}(\xi,\xi_2,x,x_2,\mu_F)\right]
\nonumber\\
&\ & \hskip 0.6in 
+ \widetilde{\cal T}_{q,F}^{\rm (CO)}(\xi,\xi+\xi_2,s_T)\, 
\left[\widetilde{\cal C}_q\, 
\left(i\, \epsilon^{s_T\beta\, n \bar{n}}\right) 
P^\rho\, P^\alpha\, \left.
\frac{\partial H_{\rho,\alpha}(\xi,\xi_2,x,x_2,\mu_F)}
     {\partial p_2^\beta}\right|_{p_2=\xi_2P}
\right]
\Bigg\} + \dots\, ,
\label{Iq_fac}
\een
where the ``$\dots$'' again represents the subleading term which 
includes the contribution from the $T_1^{\rho,\alpha,\beta}$ in
Eq.~(\ref{Iq_co}).  From the definitions of 
$\widetilde{\cal T}_{q,F}^{\rm (LC)}(\xi,\xi+\xi_2,s_T)$ and 
$\widetilde{\cal T}_{q,F}^{\rm (CO)}(\xi,\xi+\xi_2,s_T)$, we have 
\ben
\widetilde{\cal T}_{q,F}(\xi,\xi+\xi_2,s_T)
=
\widetilde{\cal T}_{q,F}^{\rm (LC)}(\xi,\xi+\xi_2,s_T)
+
\widetilde{\cal T}_{q,F}^{\rm (CO)}(\xi,\xi+\xi_2,s_T)\, .
\label{Tq_gauge}
\een
Therefore, QCD color gauge invariance requires
\ben
\widetilde{\cal C}_q \left(\frac{-1}{\xi_2}\right) 
\left(i\, \epsilon^{s_T\rho\, n \bar{n}}\right) P^\alpha\, 
H_{\rho,\alpha}^{\rm (LC)}(\xi,\xi_2,x,x_2,\mu_F)
= 
\widetilde{\cal C}_q\, 
\left( i\, \epsilon^{s_T\beta\, n \bar{n}}\right) 
P^\rho\, P^\alpha\, \left.
\frac{\partial H_{\rho,\alpha}^{\rm (CO)}(\xi,\xi_2,x,x_2,\mu_F)}
     {\partial p_2^\beta}\right|_{p_2=\xi_2P}\, ,
\label{gauge_eq}
\een
when the LHS is evaluated in the light-cone gauge and the RHS 
is evaluated in a covariant gauge.  Then, 
the two terms in Eq.~(\ref{Iq_fac}) can be combined
into one term proportional to the quark-gluon correlation
function, $\widetilde{\cal T}_{q,F}(\xi,\xi+\xi_2,s_T)$.
Since $\widetilde{\cal T}_{q,F}^{\rm (CO)}(\xi,\xi+\xi_2,s_T)$ 
vanishes in the light-cone gauge, the left-hand-side (LHS)
of the equality in Eq.~(\ref{gauge_eq}) represents
the short-distance partonic part calculated in the 
light-cone gauge.  On the other hand, 
the RHS of the equality in Eq.~(\ref{gauge_eq}) represents
the short-distance partonic part calculated in a covariant
gauge \cite{qiu}.  This is because the matrix element 
$\widetilde{\cal T}_{q,F}^{\rm (CO)}(\xi,\xi+\xi_2,s_T)$ dominates over 
$\widetilde{\cal T}_{q,F}^{\rm (LC)}(\xi,\xi+\xi_2,s_T)$ 
in a covariant gauge calculation \cite{qiu,Luo:1993ui}.  
That is, the equality in Eq.~(\ref{gauge_eq}) 
provides an excellent consistency test for the perturbative
modification of the quark-gluon correlation functions evaluated
in different gauges.

By using Eqs.~(\ref{Tq_gauge}) and (\ref{gauge_eq}), we 
can combine the two factorized terms in Eq.~(\ref{Iq_fac})
into one factorized term as 
\ben
I_q \approx
\int d\xi\, d\xi_2\, 
\widetilde{\cal T}_{q,F}(\xi,\xi+\xi_2,s_T)\,
dK_{qq}(\xi,\xi+\xi_2,x,x+x_2,\mu_F) + \dots\, ,
\label{devo_ns_q}
\een
where the perturbative modification to the correlation 
function, $dK_{qq}(\xi,\xi+\xi_2,x,x+x_2,\mu_F)$,
can be calculated by using either side of the equality in 
Eq.~(\ref{gauge_eq}) depending on the gauge used for the
calculation.  For the light-cone gauge calculation,
\ben
dK_{qq}(\xi,\xi+\xi_2,x,x+x_2,\mu_F)
=\widetilde{\cal C}_q \left(\frac{-1}{\xi_2}\right) 
\left(i\, \epsilon^{s_T\rho\, n \bar{n}}\right) P^\alpha\, 
H_{\rho,\alpha}^{\rm (LC)}(\xi,\xi_2,x,x_2,\mu_F)\, .
\label{dKqq}
\een
From Eqs.~(\ref{Hma}) and Eq.~(\ref{dKqq}), we 
derive the projection operator in the light-cone gauge,
\ben
{\cal P}_{q,F}^{(\rm LC)}
=\frac{1}{2}\, \gamma\cdot P \left(\frac{-1}{\xi_2}\right) 
\left(i\, \epsilon^{s_T\rho\, n \bar{n}}\right) \, 
\widetilde{\cal C}_q\, ,
\label{proj_q_lc}
\een
for extracting the perturbative modification $dK_{qq}$ from 
the partonic diagram in Fig.~\ref{fig3}(a), which is equal 
to the lower blob of the diagram in Fig.~\ref{fig2}(b) plus 
all diagrams with the contact interaction.  From the RHS
of Eq.~(\ref{gauge_eq}), we have the projection operator
for the covariant gauge calculation
\ben
{\cal P}_{q,F}^{(\rm CO)}
=\frac{1}{2}\, \gamma\cdot P \ P^\rho \,
\left(i\, \epsilon^{s_T\beta\, n \bar{n}}\right) \,
\widetilde{\cal C}_q\
\frac{\partial}{\partial p_2^\beta}\, ,
\label{proj_q_co}
\een
where the $p_2$ is set to $\xi_2 P$ following the derivative
\cite{qiu}.

\bef
\psfig{file=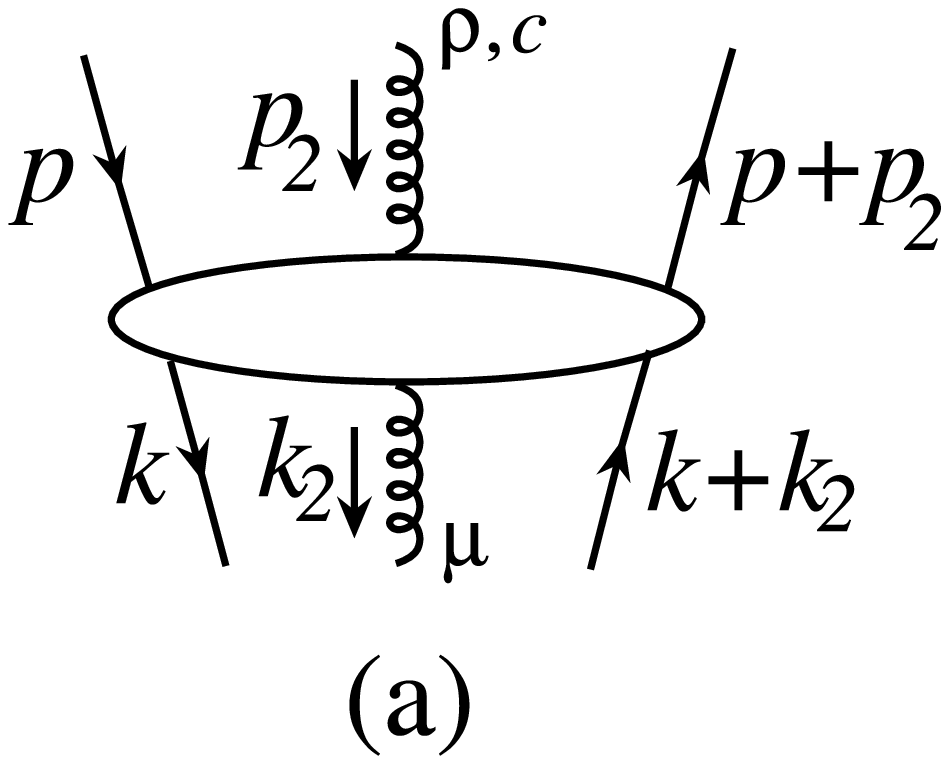,width=1.2in}
\hskip 0.5in
\psfig{file=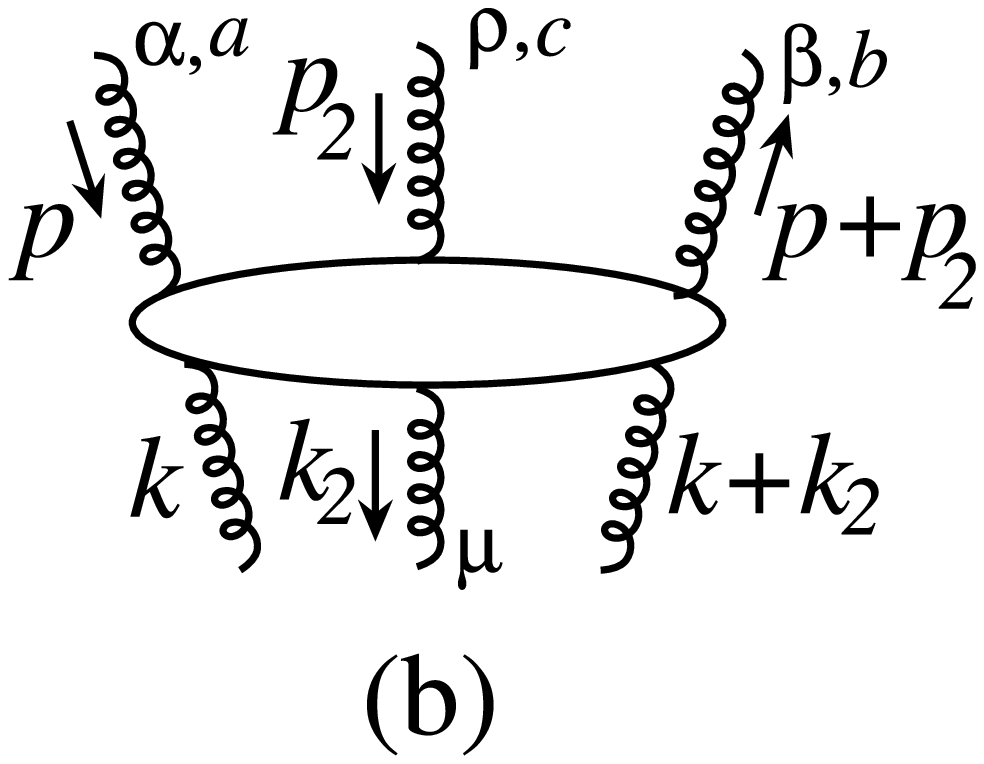,width=1.2in}
\hskip 0.5in
\psfig{file=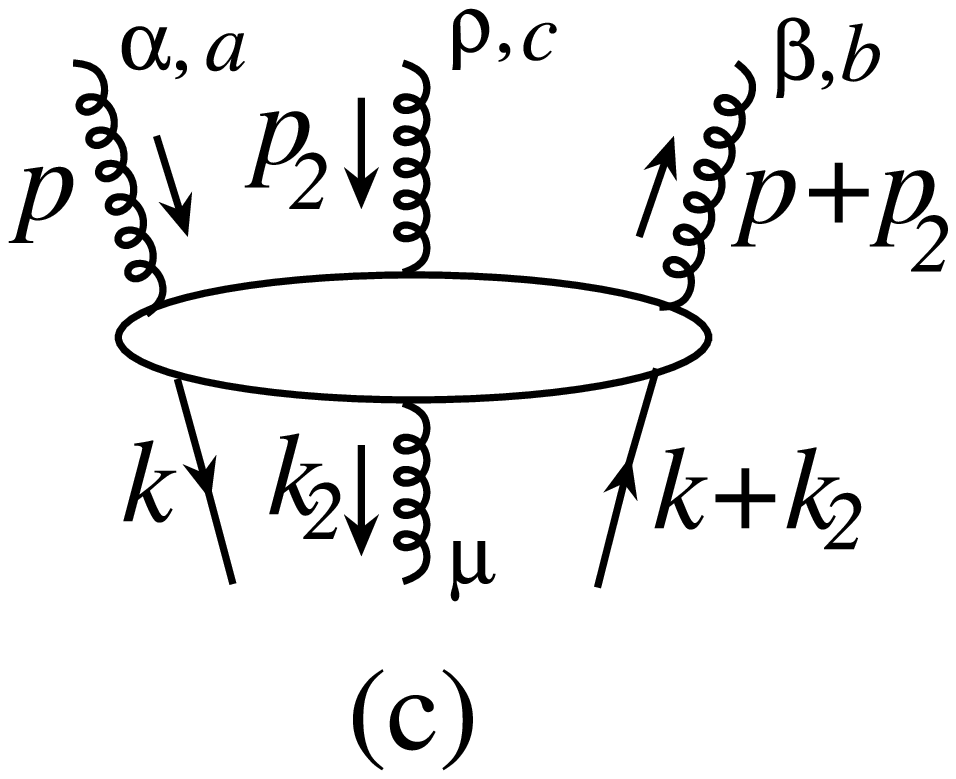,width=1.2in}
\hskip 0.5in
\psfig{file=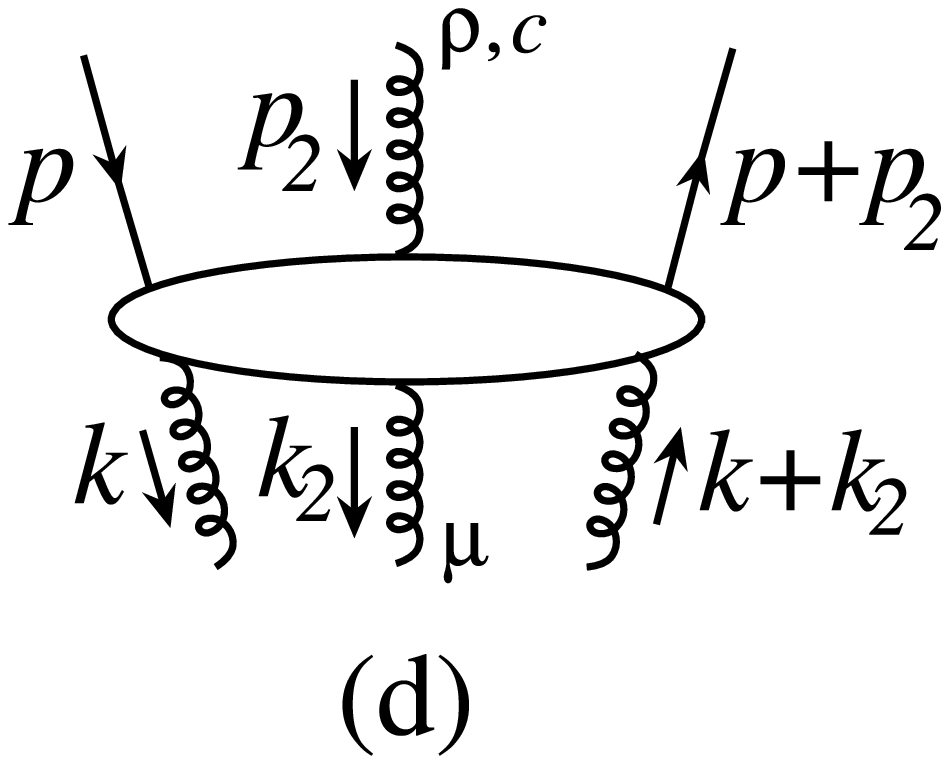,width=1.2in}
\caption{Partonic Feynman diagrams that contribute to 
the evolution kernels of 
the twist-3 correlation functions. }
\label{fig3}
\eef

In order to derive the leading contribution from the second term in
Eq.~(\ref{Tqspin}), we introduce 
\ben
I_{\Delta q}\equiv \int \frac{d^4p\,d^4p_2}{(2\pi)^8}\
\widetilde{T}^{\rho,\alpha}(p,p_2,P,s_T)
\widetilde{H}_{\rho,\alpha}(p,p_2,x,x_2,\mu_F)\, ,
\label{IDqspin}
\een
with 
\ben
\widetilde{T}^{\rho,\alpha}(p,p_2,P,s_T)
= \frac{1}{2}{\rm Tr}\left[
\hat{T}^\rho(p,p_2,P,s_T)\gamma^\alpha 
\left(i\gamma^5\right)\right]\, .
\label{Tma5}
\een
Following the same derivation as that for $I_q$, we find
\ben
I_{\Delta q} & \approx & 
\int d\xi\, d\xi_2\, 
\Bigg\{
\widetilde{\cal T}_{\Delta q,F}^{\rm (LC)}(\xi,\xi+\xi_2,s_T)\, 
\left[\widetilde{\cal C}_q \left(\frac{-1}{\xi_2}\right) 
\left(i\,s_T^\rho \right) P^\alpha\, 
\widetilde{H}_{\rho,\alpha}(\xi,\xi_2,x,x_2,\mu_F)\right]
\nonumber\\
&\ & \hskip 0.6in 
+ \widetilde{\cal T}_{\Delta q,F}^{\rm (CO)}(\xi,\xi+\xi_2,s_T)\, 
\left[\widetilde{\cal C}_q\, 
\left(i\,s_T^\beta\right) 
P^\rho\, P^\alpha\, \left.
\frac{\partial \widetilde{H}_{\rho,\alpha}(\xi,\xi_2,x,x_2,\mu_F)}
     {\partial p_2^\beta}\right|_{p_2=\xi_2P}
     \right]
\Bigg\} + \dots
\label{IDq_fac}\\
&\equiv &
\int d\xi\, d\xi_2\,
\widetilde{\cal T}_{\Delta q,F}(\xi,\xi+\xi_2,s_T)\,
dK_{q\Delta q}(\xi,\xi+\xi_2,x,x+x_2,\mu_F) + \dots\, ,
\label{devo_ns_Dq}
\een
where the perturbative modification to $\widetilde{\cal T}_{q,F}$ 
from $\widetilde{\cal T}_{\Delta q,F}$ is given by
\ben
dK_{q\Delta q}(\xi,\xi+\xi_2,x,x+x_2,\mu_F) 
&=& 
\widetilde{\cal C}_q \left(\frac{-1}{\xi_2}\right) 
\left(i\,s_T^\rho \right) P^\alpha\, 
\widetilde{H}_{\rho,\alpha}^{\rm (LC)}(\xi,\xi_2,x,x_2,\mu_F)
\label{dKqDq}\\
&=& \left.
\widetilde{\cal C}_q\, 
\left(i\,s_T^\beta\right) 
P^\rho\, P^\alpha\, 
\frac{\partial 
      \widetilde{H}_{\rho,\alpha}^{\rm (CO)}(\xi,\xi_2,x,x_2,\mu_F)}
     {\partial p_2^\beta}\right|_{p_2=\xi_2P}\, ,
\nonumber
\een
where the subscript ``LC'' (``CO'') again indicates 
the light-cone (covariant) gauge calculation.
From Eq.~(\ref{dKqDq}), we obtain the projection operator,
\ben
{\cal P}_{\Delta q,F}^{(\rm LC)}
=\frac{1}{2}\, \gamma\cdot P\, \gamma^5 
\left(\frac{-1}{\xi_2}\right) 
\left(-s_T^\rho\right) \, 
\widetilde{\cal C}_q\, ,
\label{proj_Dq_lc}
\een
for extracting $dK_{q\Delta q}$ from the same diagram in 
Fig.~\ref{fig3}(a) in the light-cone gauge.  Similarly,
one can easily derive the projection operator for 
the covariant gauge calculation from Eq.~(\ref{dKqDq}).

By adding contributions from Eqs.~(\ref{devo_ns_q}) and
(\ref{devo_ns_Dq}), we obtain the factorized perturbative 
modification to $\widetilde{\cal T}_{q,F}$,
\ben
d\widetilde{\cal T}_{q,F}(x,x+x_2,\mu_F,s_T)
&\approx &
\int d\xi\, d\xi_2\,\left[
\widetilde{\cal T}_{q,F}(\xi,\xi+\xi_2,s_T)\,
dK_{qq}(\xi,\xi+\xi_2,x,x+x_2,\mu_F) 
\right.
\nonumber \\ 
&\ & \hskip 0.5in \left.
+\widetilde{\cal T}_{\Delta q,F}(\xi,\xi+\xi_2,s_T)\,
dK_{q\Delta q}(\xi,\xi+\xi_2,x,x+x_2,\mu_F) \right] \, .
\label{devo_ns}
\een
As shown in the next section, 
the leading power perturbative modification, $dK_{ij}$
with $i,j=q,\Delta q,g,\Delta g$, can be expressed as 
\ben
dK_{ij}(\xi,\xi+\xi_2,x,x+x_2,\mu_F) 
= \int^{\mu_F^2} \frac{dk_T^2}{k_T^2}\, 
  K_{ij}(\xi,\xi+\xi_2,x,x+x_2,\alpha_s) + \dots\, ,
\label{dk_ij}
\een
where $K_{ij}(\xi,\xi+\xi_2,x,x+x_2,\alpha_s)$ is referred as
the short-distance perturbative evolution kernel.  
Substituting Eq.~(\ref{dk_ij})
into Eq.~(\ref{devo_ns}) and taking the derivative 
with respect to the factorization scale $\mu_F$ in both
sides in Eq.~(\ref{devo_ns}), we derive the leading order 
flavor non-singlet evolution equation for 
the quark-gluon correlation function, 
\ben
\mu_F^2 \frac{\partial}{\partial \mu_F^2} 
\widetilde{\cal T}_{q,F}(x,x+x_2,\mu_F,s_T) 
&=& 
\int d\xi\, d\xi_2\, \left[
\widetilde{\cal T}_{q,F}(\xi,\xi+\xi_2,\mu_F,s_T)\,
K_{qq}(\xi,\xi+\xi_2,x,x+x_2,\alpha_s)
\right.
\nonumber\\
&\ & \hskip 0.5in \left.
+\widetilde{\cal T}_{\Delta q,F}(\xi,\xi+\xi_2,\mu_F,s_T)\,
K_{q\Delta q}(\xi,\xi+\xi_2,x,x+x_2,\alpha_s) \right]
\, ,
\label{evo_ns}
\een
which is consistent with that in Eq.~(\ref{evolution_eq}) and 
has the generic homogeneous differential-integral form
of the typical evolution equation, such as the DGLAP 
evolution equation of PDFs \cite{Collins:1988wj,DGLAP}.

\bef
\psfig{file=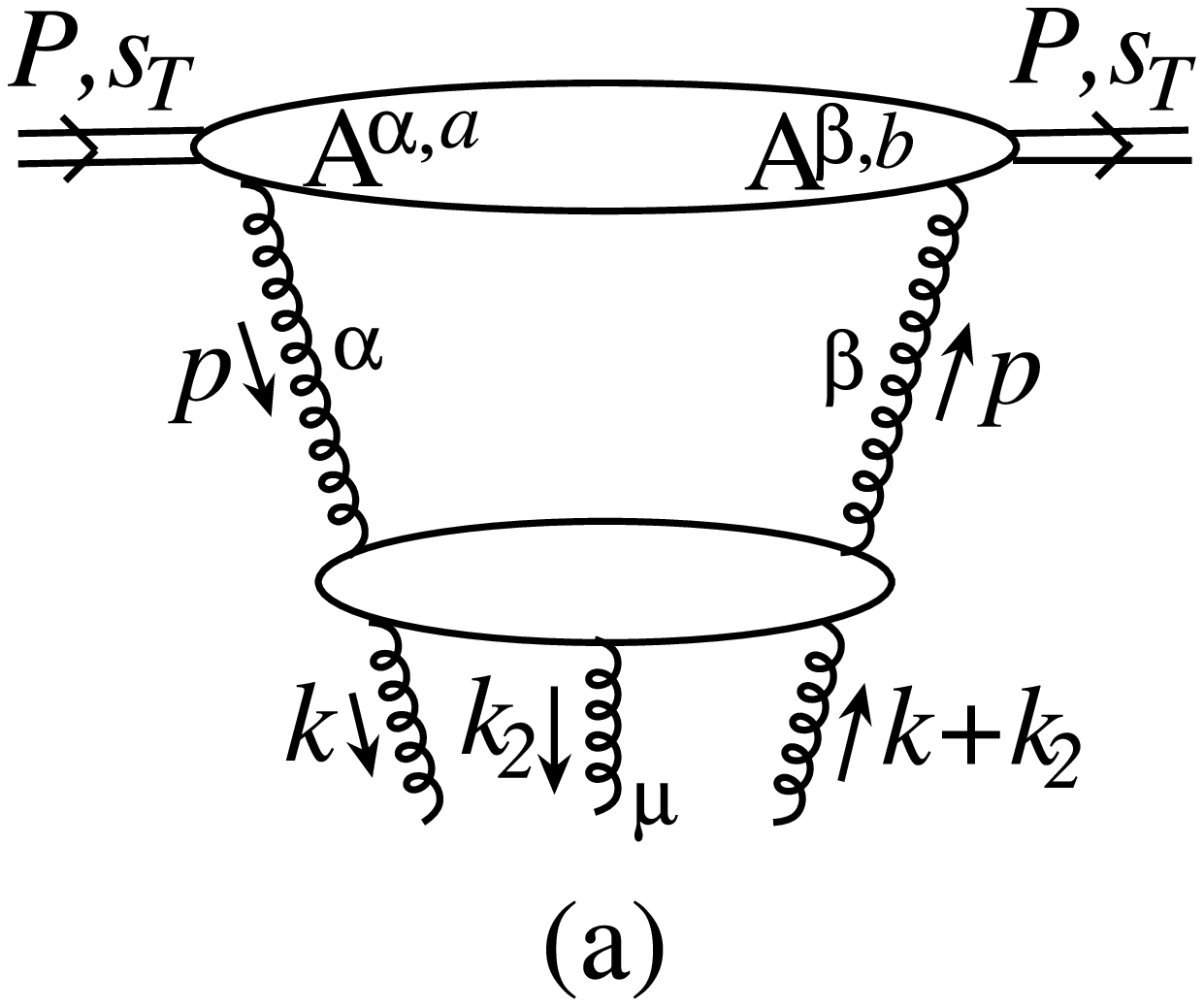,width=1.5in}
\hskip 0.5in
\psfig{file=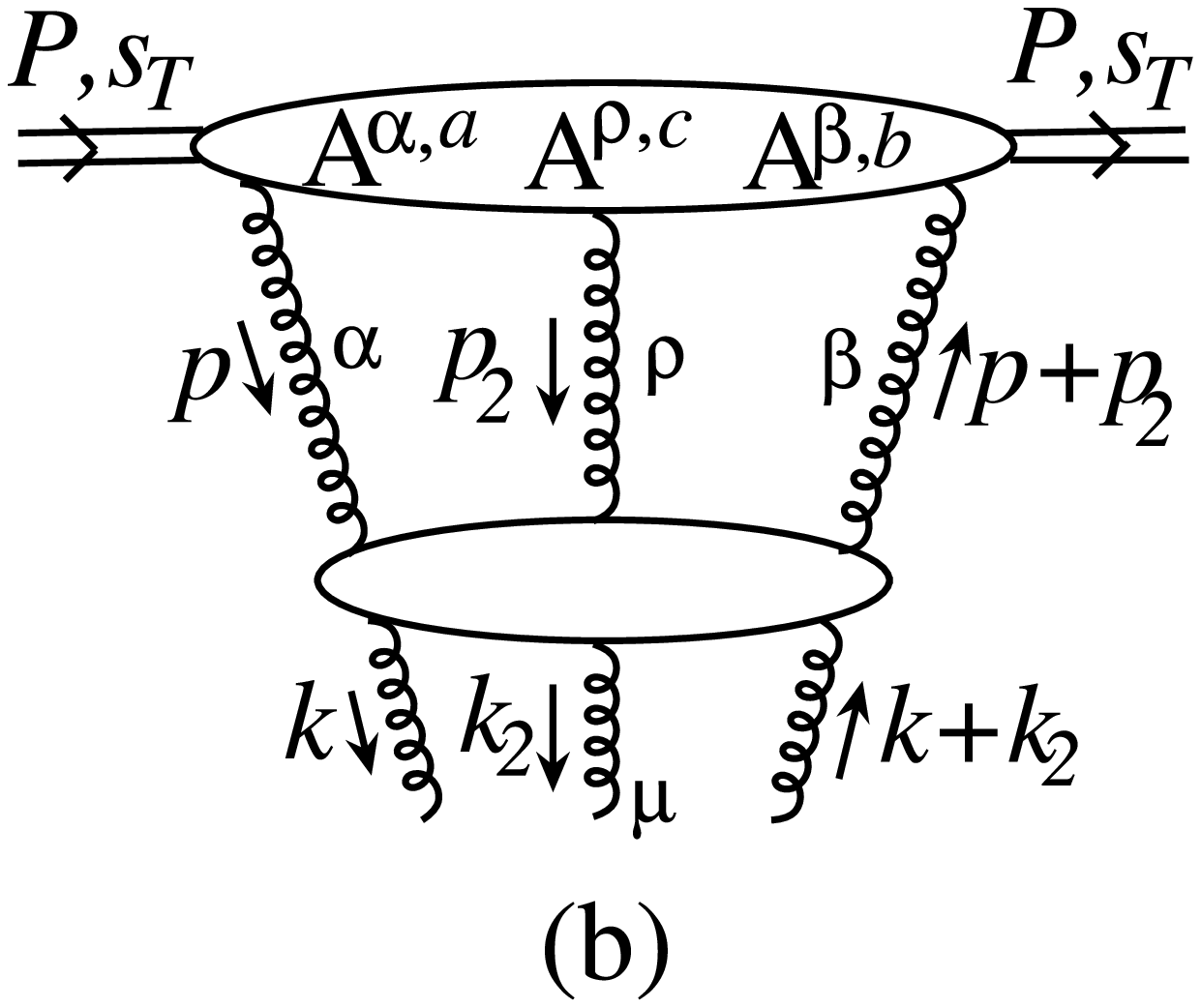,width=1.5in}
\caption{Feynman diagrams that contribute to the change of 
the twist-3 tri-gluon correlation functions where
$\alpha,\beta,\mu,\rho$ and $a,b,c$ are Lorentz and color indices 
of gluon field operators, respectively.
The lower part of gluon lines are contracted to 
the cut vertices that define the tri-gluon correlation functions.}
\label{fig4}
\eef

Next, we derive the perturbative change of 
tri-gluon correlation function
$\widetilde{\cal T}_{G,F}^{(f,d)}$ from the diagrams in 
Fig.~\ref{fig4}.  Since gluon transversity distribution
vanishes \cite{h_g}, there is no leading twist or leading
power contribution to the evolution of the tri-gluon
correlation functions from Fig.~\ref{fig4}(a).
Similar to the case of the flavor non-singlet change of 
quark-gluon correlation function discussed above, 
the subleading power contribution from the diagram in 
Fig.~\ref{fig4}(a) can be combined with the leading 
contribution of the diagram in Fig.~\ref{fig4}(b) 
\cite{qiu_t4}.  We can then derive the projection 
operator for calculating the gluonic evolution kernel by 
factorizing the diagram in Fig.~\ref{fig4}(b).
 
We express the diagram in Fig.~\ref{fig2}(b) as
\ben
d\widetilde{\cal T}_{G,F}^{(i)}(x,x+x_2,\mu_F,s_T)
\equiv 
\int \frac{d^4p\, d^4p_2}{(2\pi)^8}\,
\sum_{\rho,\alpha,\beta}\left[
T^{\rho,\alpha,\beta}(p,p_2,P,s_T)\,
H^{(i)}_{\rho,\alpha,\beta}(p,p_2,x,x_2,\mu_F)\right]
\equiv I_G^{(i)} + I_{\Delta G}^{(i)}\, ,
\label{Ig}
\een
where the superscript $i=f,d$ from the cut vertex and 
$I_G$ ($I_{\Delta G}$) represents the part
of the perturbative change that is symmetric (antisymmetric) 
in the exchange of the Lorentz indices $\alpha$ and $\beta$. 
In Eq.~(\ref{Ig}), the partonic part
$H_{\rho,\alpha,\beta}$ is given by the bottom part
of the diagram in Fig.~\ref{fig4}(b) plus 
diagrams with the contact interaction from the subleading
contribution of the diagram in Fig.~\ref{fig4}(a).  All
partonic diagrams are contracted by the cut vertex 
in Eq.~(\ref{cv_g}).
The tri-gluon matrix element $T^{\rho,\alpha,\beta}$ 
in Eq.~(\ref{Ig}) is defined as
\ben
T^{\rho,\alpha,\beta}(p,p_2,P,s_T)=
\langle P,s_T|\,
\widetilde{A}^{\beta,b}(-p-p_2)\,
\widetilde{A}^{\rho,c}(p_2)\,
\widetilde{A}^{\alpha,a}(p)\, |P,s_T\rangle
\label{H3g}
\een
with the gluon color indices, $b,c,a$ and is represented by
the top part of the Feynman diagram in Fig.~\ref{fig4}(b).
Following the same steps used to factorize the diagram in 
Fig.~\ref{fig2}(b), we can factorize the leading power 
contribution to the part that is symmetric in
$\alpha$ and $\beta$ as
\ben
I_G^{(i)} 
& \approx & 
\int d\xi\, d\xi_2\, 
\Bigg\{
\left[\frac{\widetilde{\cal T}^{(j){\rm (LC)}}_{G,F}
            (\xi,\xi+\xi_2,s_T)}
     {\xi(\xi+\xi_2)}\right] 
\left[\,\widetilde{\cal C}_g^{(j)}\ 
      \frac{1}{2}\, d^{\alpha\beta}
      \left(\frac{-1}{\xi_2}\right) 
      \left(i\, \epsilon^{s_T\rho\, n \bar{n}}\right) 
      H^{(i)}_{\rho,\alpha,\beta}(\xi,\xi_2,x,x_2,\mu_F)\right]
\label{Ig_fac}\\
&\ & \hskip 0.5in 
+ 
\left[\frac{\widetilde{\cal T}^{(j){\rm (CO)}}_{G,F}
            (\xi,\xi+\xi_2,s_T)}
     {\xi(\xi+\xi_2)}\right] 
\left[\,\widetilde{\cal C}_g^{(j)}\ 
      \frac{1}{2} d^{\alpha\beta}\,P^\rho\,
      \left(i\, \epsilon^{s_T\sigma\, n \bar{n}}\right) 
      \left. 
\frac{\partial H^{(i)}_{\rho,\alpha,\beta}(\xi,\xi_2,x,x_2,\mu_F)}
     {\partial p_2^\sigma}\right|_{p_2=\xi_2 P}
     \right]
\Bigg\} + \dots\, ,
\nonumber
\een
where the transverse polarization tensor\
$d^{\alpha\beta}\equiv -g^{\alpha\beta} + n^\alpha \bar{n}^\beta
+\bar{n}^\alpha n^\beta$\ and 
the gluonic color projection operators $\widetilde{\cal C}_g^{(i)}$
with $j=f,d$ are given by
\ben 
(\widetilde{\cal C}_g^{(f)})_{acb}
   =\frac{1}{N_c(N_c^2-1)}\, i f_{acb}\, ,
\quad
\mbox{and}
\quad
(\widetilde{\cal C}_g^{(d)})_{acb}
   =\frac{N_c}{(N_c^2-4)(N_c^2-1)}\, d_{acb}\, ,
\label{color_gp}
\een
for color indices labeled in Fig.~\ref{fig4}(b), 
so that $\widetilde{\cal C}_g^{(j)}\,{\cal C}_g^{(j)} = 1$ 
for $j=f,d$.  In Eq.~(\ref{Ig_fac}), 
$\widetilde{\cal T}_{G,F}^{(j){\rm (LC)}}$ and 
$\widetilde{\cal T}_{G,F}^{(j){\rm (CO)}}$ with $j=f,d$
are tri-gluon correlation functions that have the same definition
as that of $\widetilde{\cal T}_{G,F}^{(f,d)}$ 
in Eq.~(\ref{Tgm}), except 
that the cut vertex is now replaced by the corresponding one 
in the light-cone gauge and the one in a covariant gauge, 
respectively, and they satisfy
\ben
\widetilde{\cal T}_{G,F}^{(f,d)}(\xi,\xi+\xi_2,s_T)
=
\widetilde{\cal T}_{G,F}^{(f,d){\rm (LC)}}(\xi,\xi+\xi_2,s_T)
+
\widetilde{\cal T}_{G,F}^{(f,d){\rm (CO)}}(\xi,\xi+\xi_2,s_T)\, .
\label{Tg_gauge}
\een
Again, the color gauge invariance requires 
\ben
 \frac{1}{2}\, d^{\alpha\beta}
 \left(\frac{-1}{\xi_2}\right) 
 \left(i\, \epsilon^{s_T\rho\, n \bar{n}}\right) 
 H_{\rho,\alpha,\beta}^{(i){\rm (LC)}}(\xi,\xi_2,x,x_2,\mu_F)
=\frac{1}{2} d^{\alpha\beta}\,P^\rho\,
 \left(i\, \epsilon^{s_T\sigma\, n \bar{n}}\right)
 \left. 
\frac{\partial 
      H_{\rho,\alpha,\beta}^{(i){\rm (CO)}}(\xi,\xi_2,x,x_2,\mu_F)}
     {\partial p_2^\sigma}\right|_{p_2=\xi_2P}\, ,
\label{gauge_eq_g}
\een
when the LHS is evaluated in the light-cone gauge and the
RHS in a covariant gauge.  Therefore, the two leading terms 
in Eq.~(\ref{Ig_fac}) can be combined together as
\ben
I_G^{(i)} \approx 
\int d\xi\, d\xi_2\, 
\widetilde{\cal T}^{(j)}_{G,F}(\xi,\xi+\xi_2,s_T)\,
dK_{gg}^{(ij)}(\xi,\xi+\xi_2,x,x+x_2,\mu_F)
\label{Ig_dK}
\een
with
\ben
dK_{gg}^{(ij)}(\xi,\xi+\xi_2,x,x+x_2,\mu_F)
=
 \widetilde{\cal C}_g^{(j)}\
 \frac{1}{2}\, d^{\alpha\beta}
 \frac{1}{\xi(\xi+\xi_2)}\,
 \left(\frac{-1}{\xi_2}\right) 
 \left(i\, \epsilon^{s_T\rho\, n \bar{n}}\right) 
 H_{\rho,\alpha,\beta}^{(i){\rm (LC)}}(\xi,\xi_2,x,x_2,\mu_F)
\label{dKgg_lc}
\een
in the light-cone gauge.  The $dK_{gg}^{(ij)}$ can also
be derived in a covariant gauge from the RHS of 
Eq.~(\ref{gauge_eq_g}).  
The equality in Eq.~(\ref{gauge_eq_g}) provides an independent 
check of perturbative calculation done in different gauges. 
From Eq.~(\ref{dKgg_lc}), we
obtain the light-cone gauge projection operator,
\ben
{\cal P}_{G,F}^{(f,d){\rm (LC)}}
= \frac{1}{2}\, d^{\alpha\beta}\,
 \frac{1}{\xi(\xi+\xi_2)}
 \left(\frac{-1}{\xi_2}\right) 
 \left(i\, \epsilon^{s_T\rho\, n \bar{n}}\right) \,
 \widetilde{\cal C}_g^{(f,d)}\, ;
\label{proj_g_lc}
\een
for calculating the perturbative modification to the 
tri-gluon correlation function 
$\widetilde{\cal T}_{G,F}^{(f,d)}$ from the diagrams
in Fig.~\ref{fig3}(b), which includes all the partonic
Feynman diagrams from the lower
blob of the diagram in Fig.~\ref{fig4}(b) plus 
corresponding twist-3 contribution from the diagram in 
Fig.~\ref{fig4}(a) expressed in terms of diagrams 
with the contact interaction \cite{qiu_t4}.
Similar projection operator can be derived from the RHS of 
Eq.~(\ref{gauge_eq_g}) for the covariant gauge calculation.

Similarly, we derive the perturbative modification to
$\widetilde{\cal T}_{G,F}^{(f,d)}$ from the tri-gluon
correlation function $\widetilde{\cal T}_{\Delta G,F}^{(f,d)}$,
\ben
I_{\Delta G}^{(i)} \approx 
\int d\xi\, d\xi_2\, 
\widetilde{\cal T}^{(j)}_{\Delta G,F}(\xi,\xi+\xi_2,s_T)\,
dK_{g\Delta g}^{(ij)}(\xi,\xi+\xi_2,x,x+x_2,\mu_F)
\label{IDg_dK}
\een
with
\ben
dK_{g\Delta g}^{(ij)}(\xi,\xi+\xi_2,x,x+x_2,\mu_F)
=
 \widetilde{\cal C}_g^{(j)}\
 \frac{1}{2}\, \left(i\epsilon_\perp^{\alpha\beta}\right)\,
 \frac{1}{\xi(\xi+\xi_2)}\,
 \left(\frac{-1}{\xi_2}\right) 
 \left(-s_T^\rho \right) 
 H_{\rho,\alpha,\beta}^{(i){\rm (LC)}}(\xi,\xi_2,x,x_2,\mu_F)
\label{dKgDg_lc}
\een
in the light-cone gauge.  One can easily derive the 
expression for $dK_{g\Delta g}^{(ij)}$ in a covariant gauge 
as well.  From Eq.~(\ref{dKgDg_lc}), we
obtain the light-cone gauge projection operator,
\ben
{\cal P}_{\Delta G,F}^{(f,d){\rm (LC)}}
= \frac{1}{2}\, 
 \left(i\epsilon_\perp^{\alpha\beta}\right)\,
 \frac{1}{\xi(\xi+\xi_2)}\,
 \left(\frac{-1}{\xi_2}\right) 
 \left(-s_T^\rho \right)\, 
  \widetilde{\cal C}_g^{(f,d)}\, ;
\label{proj_Dg_lc}
\een
for calculating the perturbative modification from the 
tri-gluon correlation function 
$\widetilde{\cal T}_{\Delta G,F}^{(f,d)}$ from the same 
diagrams in Fig.~\ref{fig3}(b).

Using the generic expression of the leading power 
contribution to the perturbative modification factor
$dK_{ij}$ in Eq.~(\ref{dk_ij}), we derive the 
evolution equation for the factorization scale dependence
of the tri-gluon correlation function 
$\widetilde{\cal T}_{G,F}^{(f,d)}$ by factorizing the 
perturbative correction from the diagrams 
in Fig.~\ref{fig4},
\ben
\mu_F^2 \frac{\partial}{\partial \mu_F^2} 
\widetilde{\cal T}_{G,F}^{(i)}(x,x+x_2,\mu_F,s_T) 
&=& 
\sum_{j=f,d} \int d\xi\, d\xi_2\, \left[
\widetilde{\cal T}_{G,F}^{(j)}(\xi,\xi+\xi_2,\mu_F,s_T)\,
K_{gg}^{(ji)}(\xi,\xi+\xi_2,x,x+x_2,\alpha_s)
\right.
\nonumber\\
&\ & \hskip 0.8in
+ \left.
\widetilde{\cal T}_{\Delta{G},F}^{(j)}
      (\xi,\xi+\xi_2,\mu_F,s_T)\,
K_{g\Delta{g}}^{(ji)}(\xi,\xi+\xi_2,x,x+x_2,\alpha_s)
\right]
\label{evo_gg}
\een
where the superscript $i,j=f,d$, $K_{gg}^{(ji)}$
and $K_{g\Delta{g}}^{(ji)}$ are evolution kernels that
can be perturbatively calculated from the diagram 
in Fig.~\ref{fig3}(b) with proper projection operators
as discussed above.
 
\bef
\psfig{file=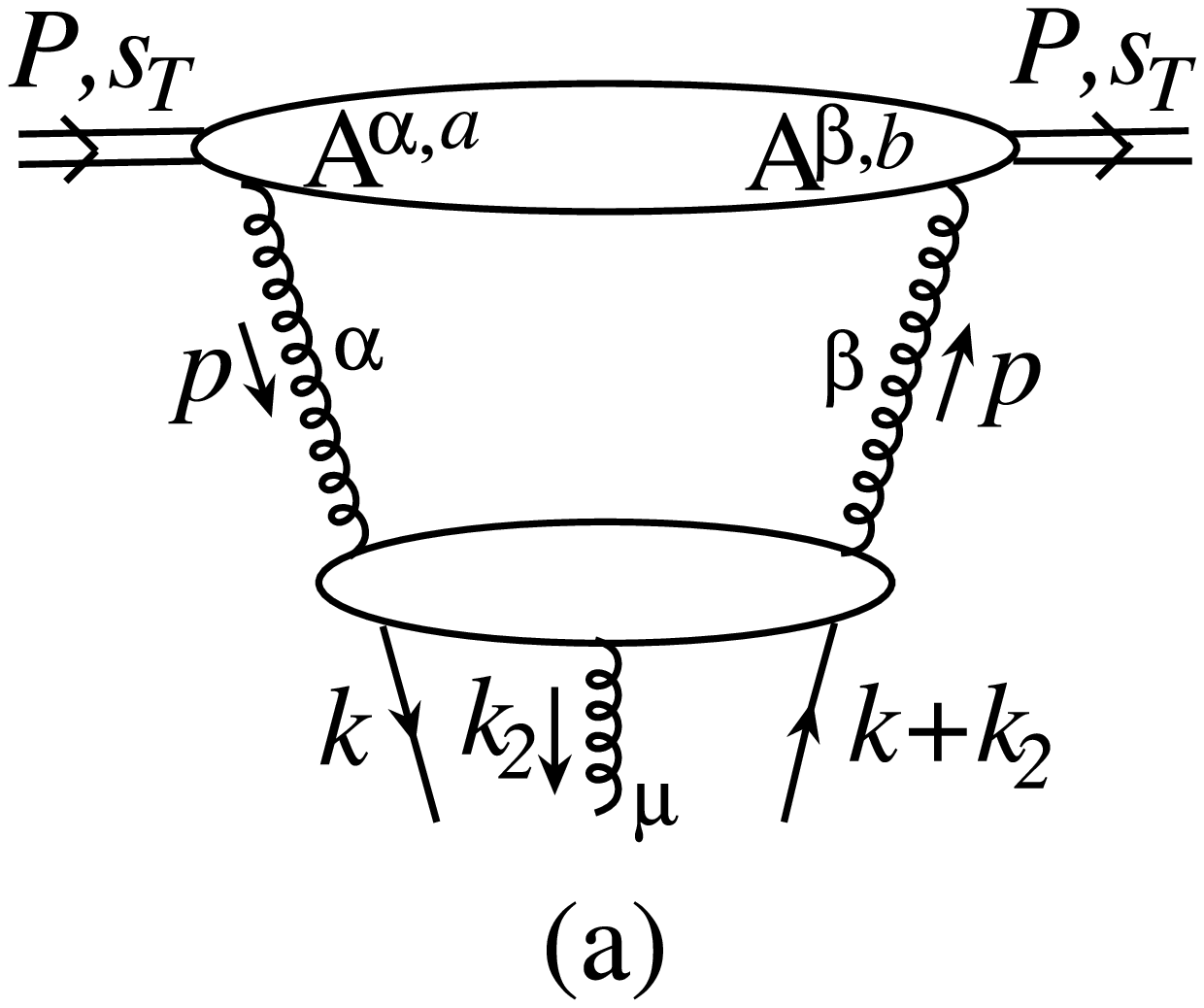,width=1.5in}
\hskip 0.5in
\psfig{file=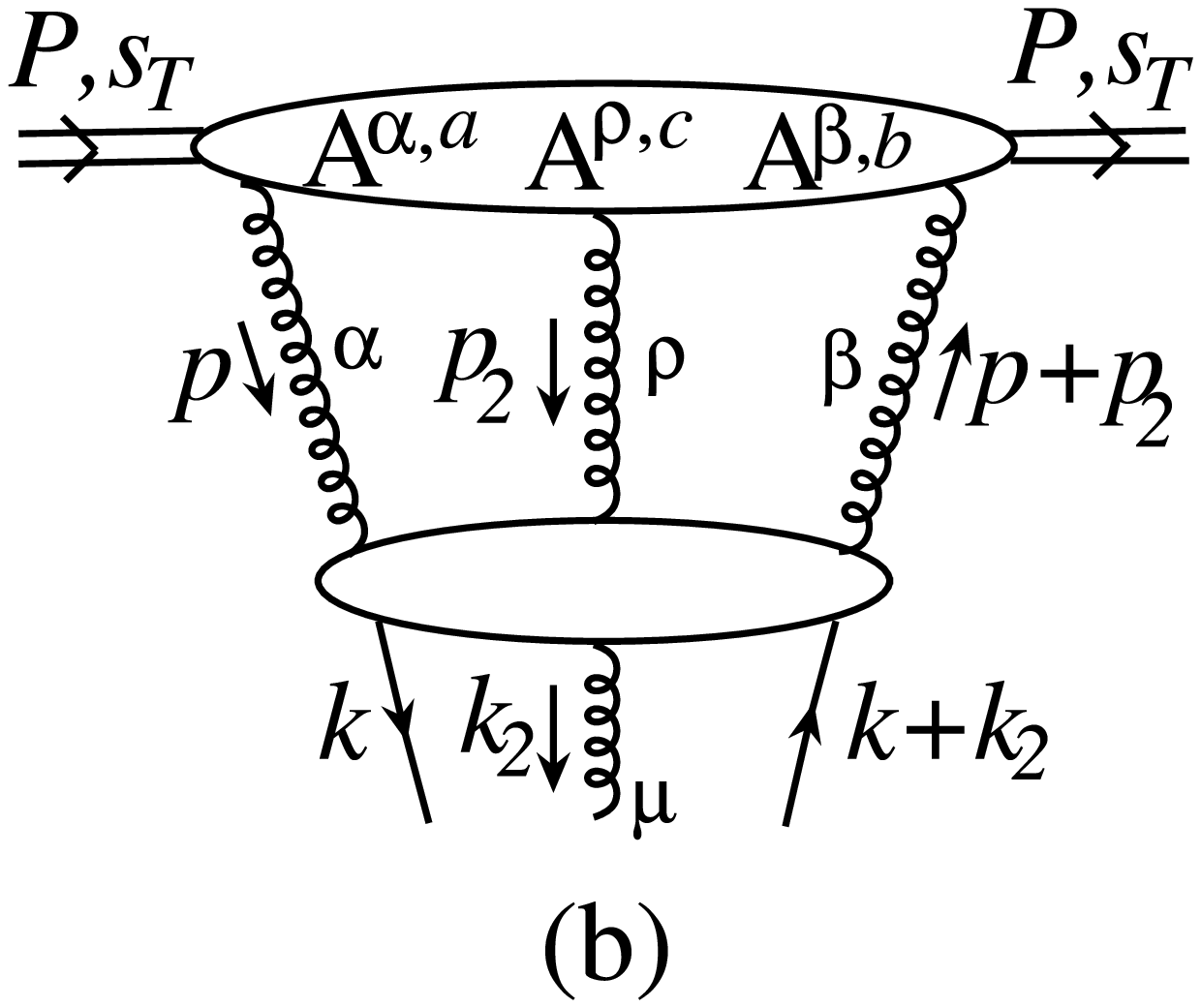,width=1.5in}
\caption{Feynman diagrams that contribute to the change of 
the twist-3 quark-gluon correlation function where
$\alpha,\beta,\mu,\rho$ and $a,b,c$ are Lorentz and color indices 
of gluon field operators, respectively.
The lower part of quark and gluon lines are contracted to 
the cut vertex that defines the quark-gluon correlation function.}
\label{fig5}
\eef

\bef
\psfig{file=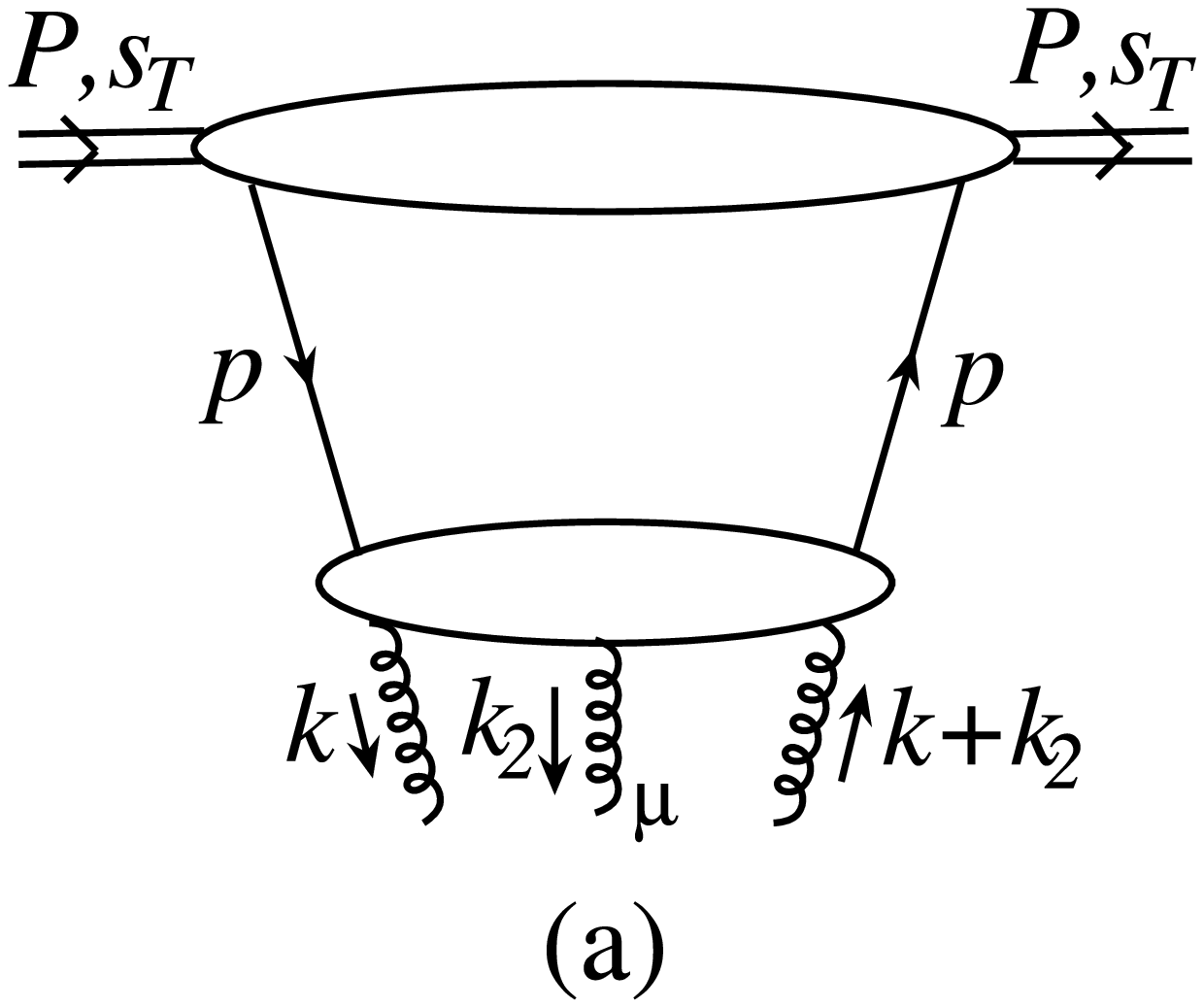,width=1.5in}
\hskip 0.5in
\psfig{file=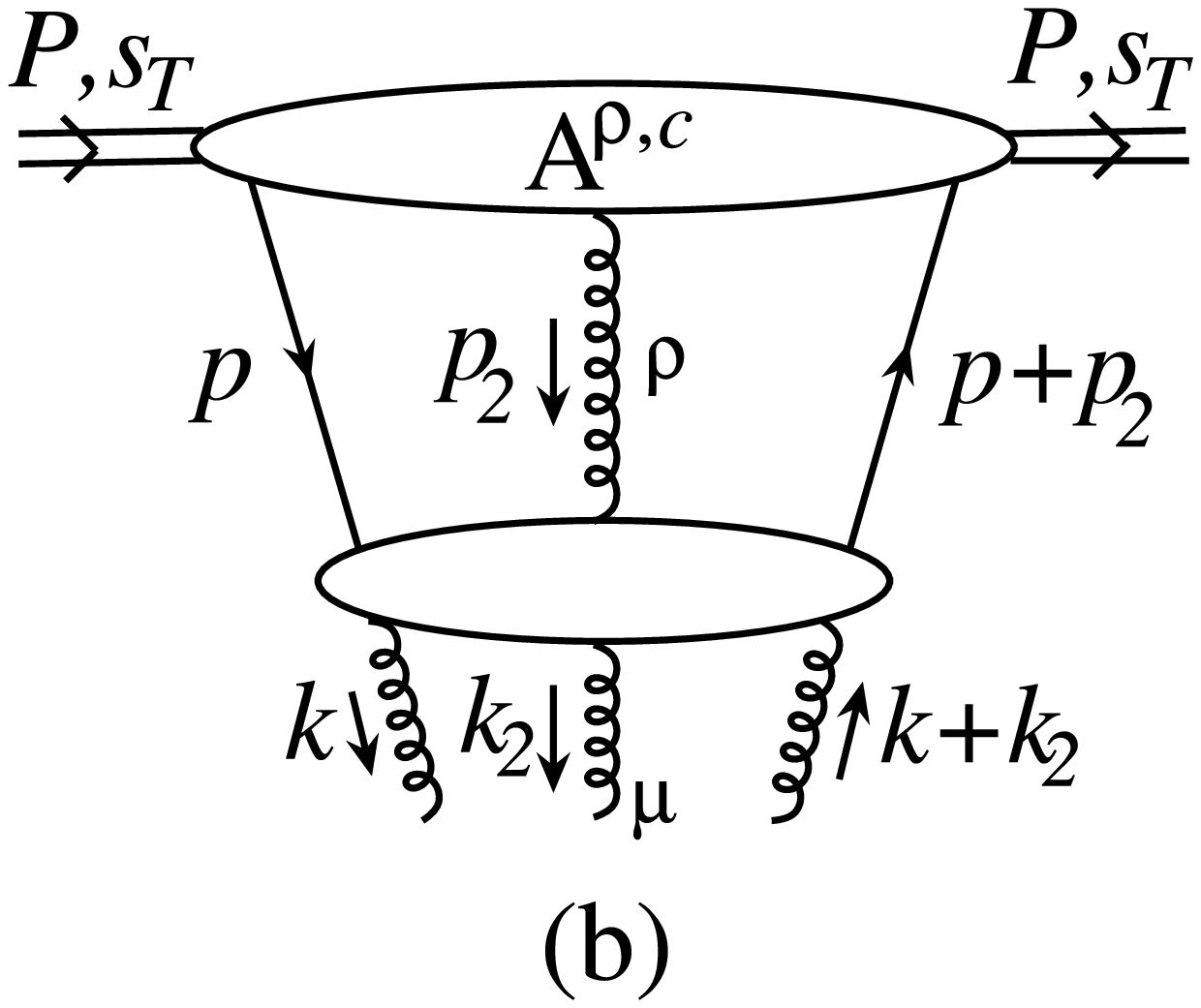,width=1.5in}
\caption{Feynman diagrams that contribute to the change of 
the twist-3 tri-gluon correlation functions from the interaction
initiated from the quark-gluon correlation functions.  
The lower part of gluon lines are contracted to 
the cut vertices that define the tri-gluon correlation functions.}
\label{fig6}
\eef

The evolution equation for the scale dependence of the 
quark-gluon correlation function in Eq.~(\ref{evo_ns}) 
can also get 
contribution from the tri-gluon correlation functions via
the diagrams in Fig.~\ref{fig5}.  Similarly, 
the evolution equation for the tri-gluon correlation functions 
in Eq.~(\ref{evo_gg}) can get additional contribution from the 
quark-gluon correlation function via the diagrams in 
Fig.~\ref{fig6}. 

Following the same procedure to factorize the diagrams
in Fig.~\ref{fig4}, we derive the additional contribution
to the evolution of the quark-gluon correlation function from 
the tri-gluon correlation functions and have,
\ben
\mu_F^2 \frac{\partial}{\partial \mu_F^2} 
\widetilde{\cal T}_{q,F}(x,x+x_2,\mu_F,s_T) 
&=& 
\int d\xi\, d\xi_2\, \left[
\widetilde{\cal T}_{q,F}(\xi,\xi+\xi_2,\mu_F,s_T)\,
K_{qq}(\xi,\xi+\xi_2,x,x+x_2,\alpha_s)
\right.
\nonumber\\
&\ & \hskip 0.5in \left.
+\widetilde{\cal T}_{\Delta q,F}(\xi,\xi+\xi_2,\mu_F,s_T)\,
K_{q\Delta q}(\xi,\xi+\xi_2,x,x+x_2,\alpha_s) \right]
\nonumber \\
&+&
\sum_{i=f,d} \int d\xi\, d\xi_2\, \left[
\widetilde{\cal T}_{G,F}^{(i)}(\xi,\xi+\xi_2,\mu_F,s_T)\,
K_{qg}^{(i)}(\xi,\xi+\xi_2,x,x+x_2,\alpha_s)
\right.
\nonumber\\
&\ & \hskip 0.75in
+ \left.
\widetilde{\cal T}_{\Delta{G},F}^{(i)}
      (\xi,\xi+\xi_2,\mu_F,s_T)\,
K_{q\Delta{g}}^{(i)}(\xi,\xi+\xi_2,x,x+x_2,\alpha_s)
\right]
\, .
\label{evo_q}
\een
The evolution kernels from the tri-gluon correlation functions 
to the quark-gluon correlation function, $K_{qg}^{(f,d)}$
and $K_{q\Delta g}^{(f,d)}$, 
can be obtained by calculating the diagram in Fig.~\ref{fig3}(c)
with proper projection operators.  
If the kernels are evaluated in the 
light-cone gauge, the three gluon lines on the top of
the diagram are contracted by the projection operator in
Eqs.~(\ref{proj_g_lc}) and (\ref{proj_Dg_lc}), respectively. 
The diagram in Fig.~\ref{fig3}(c)
includes all Feynman diagrams from the bottom part of 
the diagram in Fig.~\ref{fig5}(b) plus diagrams from the 
subleading contribution of the diagram in Fig.~\ref{fig5}(a),
which can be effectively expressed in terms of the diagrams
with the contact interaction and  
the same external lines as those in Fig.~\ref{fig5}(b).
The combination of these diagrams forms a gauge invariant
set \cite{qiu_t4}.  

Similarly, following the same procedure to factorize the 
diagrams in Fig.~\ref{fig2}, we derive the additional 
contribution to the evolution equation of the tri-gluon  
correlation functions from the quark-gluon correlation functions and have,
\ben
\mu_F^2 \frac{\partial}{\partial \mu_F^2} 
\widetilde{\cal T}_{G,F}^{(i)}(x,x+x_2,\mu_F,s_T) 
&=& 
\sum_{j=f,d} \int d\xi\, d\xi_2\, \left[
\widetilde{\cal T}_{G,F}^{(j)}(\xi,\xi+\xi_2,\mu_F,s_T)\,
K_{gg}^{(ji)}(\xi,\xi+\xi_2,x,x+x_2,\alpha_s)
\right.
\nonumber\\
&\ & \hskip 0.8in
+ \left.
\widetilde{\cal T}_{\Delta{G},F}^{(j)}
      (\xi,\xi+\xi_2,\mu_F,s_T)\,
K_{g\Delta{g}}^{(ji)}(\xi,\xi+\xi_2,x,x+x_2,\alpha_s)
\right]
\nonumber\\
&+& 
\sum_q
\int d\xi\, d\xi_2\, \left[
\widetilde{\cal T}_{q,F}(\xi,\xi+\xi_2,\mu_F,s_T)\,
K_{gq}^{(i)}(\xi,\xi+\xi_2,x,x+x_2,\alpha_s)
\right.
\nonumber\\
&\ & \hskip 0.7in \left.
+\widetilde{\cal T}_{\Delta q,F}(\xi,\xi+\xi_2,\mu_F,s_T)\,
K_{g\Delta q}^{(i)}(\xi,\xi+\xi_2,x,x+x_2,\alpha_s) 
\right]\, .
\label{evo_g}
\een
where $\sum_q$ runs over all quark and antiquark flavors,
the superscript, $i,j=f,d$.  The evolution kernels from
the quark-gluon correlation functions to the tri-gluon 
correlation functions, $K_{gq}^{(f,d)}$ and 
$K_{g\Delta q}^{(f,d)}$, can be obtained by
calculating the diagram in Fig.~\ref{fig3}(d)
with proper projection operators.  
If the evolution kernels are evaluated in the
light-cone gauge, the quark and gluon lines on the top of
the diagram are contracted by the projection operator in
Eqs.~(\ref{proj_q_lc}) and (\ref{proj_Dq_lc}), respectively.
The diagram in Fig.~\ref{fig3}(d) includes
all partonic Feynman diagrams from the bottom part of the diagram 
in Fig.~\ref{fig6}(b) plus the diagrams with the contact
interaction representing the subleading contribution of 
the diagram in Fig.~\ref{fig6}(a).  

Following the same derivation for the perturbative corrections
to the first set twist-3 correlation functions, we derive
the evolution equations for the second set of twist-3
correlation functions,
\ben
\mu_F^2 \frac{\partial}{\partial \mu_F^2} 
\widetilde{\cal T}_{\Delta q,F}(x,x+x_2,\mu_F,s_T) 
&=& 
\int d\xi\, d\xi_2\, \left[
\widetilde{\cal T}_{\Delta q,F}(\xi,\xi+\xi_2,\mu_F,s_T)\,
K_{\Delta q\Delta q}(\xi,\xi+\xi_2,x,x+x_2,\alpha_s)
\right.
\nonumber\\
&\ & \hskip 0.5in \left.
+\widetilde{\cal T}_{q,F}(\xi,\xi+\xi_2,\mu_F,s_T)\,
K_{\Delta q\,q}(\xi,\xi+\xi_2,x,x+x_2,\alpha_s) \right]
\nonumber \\
&+&
\sum_{i=f,d} \int d\xi\, d\xi_2\left[
\widetilde{\cal T}_{G,F}^{(i)}(\xi,\xi+\xi_2,\mu_F,s_T)\,
K_{\Delta q\,g}^{(i)}(\xi,\xi+\xi_2,x,x+x_2,\alpha_s)
\right.
\nonumber\\
&\ & \hskip 0.75in
+ \left.
\widetilde{\cal T}_{\Delta{G},F}^{(i)}
      (\xi,\xi+\xi_2,\mu_F,s_T)\,
K_{\Delta q\Delta{g}}^{(i)}(\xi,\xi+\xi_2,x,x+x_2,\alpha_s)
\right]
\, ,
\label{evo_Dq}
\een
and 
\ben
\mu_F^2 \frac{\partial}{\partial \mu_F^2} 
\widetilde{\cal T}_{\Delta G,F}^{(i)}(x,x+x_2,\mu_F,s_T) 
&=& 
\sum_{j=f,d} \int d\xi\, d\xi_2\left[
\widetilde{\cal T}_{\Delta G,F}^{(j)}(\xi,\xi+\xi_2,\mu_F,s_T)\,
K_{\Delta g\Delta g}^{(ji)}(\xi,\xi+\xi_2,x,x+x_2,\alpha_s)
\right.
\nonumber\\
&\ & \hskip 0.8in
+ \left.
\widetilde{\cal T}_{G,F}^{(j)}
      (\xi,\xi+\xi_2,\mu_F,s_T)\,
K_{\Delta{g}\, g}^{(ji)}(\xi,\xi+\xi_2,x,x+x_2,\alpha_s)
\right]
\nonumber\\
&+& 
\sum_q
\int d\xi\, d\xi_2\, \left[
\widetilde{\cal T}_{q,F}(\xi,\xi+\xi_2,\mu_F,s_T)\,
K_{\Delta g\,q}^{(i)}(\xi,\xi+\xi_2,x,x+x_2,\alpha_s)
\right.
\nonumber\\
&\ & \hskip 0.7in \left.
+\widetilde{\cal T}_{\Delta q,F}(\xi,\xi+\xi_2,\mu_F,s_T)\,
K_{\Delta g\Delta q}^{(i)}(\xi,\xi+\xi_2,x,x+x_2,\alpha_s) 
\right]\, .
\label{evo_Dg}
\een
All evolution kernels in Eqs.~(\ref{evo_Dq}) and (\ref{evo_Dg})
can be derived by calculating diagrams in Fig.~\ref{fig3}
with proper projection operators discussed in this section.

Equations~(\ref{evo_q}), (\ref{evo_g}), (\ref{evo_Dq}), 
and (\ref{evo_Dg}) form a closed set of evolution equations 
for the scale dependence of the two sets of twist-3 
quark-gluon and tri-gluon correlation functions defined in
the last section.  From these evolution equations, we can 
construct the evolution equations of twist-3 correlation
functions that are responsible for the SSAs as,
\ben
\mu_F^2 \frac{\partial}{\partial \mu_F^2} 
{\cal T}_{q,F}(x,x+x_2,\mu_F) 
&=&
\frac{1}{2}\left[
\mu_F^2 \frac{\partial}{\partial \mu_F^2} 
\widetilde{\cal T}_{q,F}(x,x+x_2,\mu_F,s_T) 
+ 
\mu_F^2 \frac{\partial}{\partial \mu_F^2} 
\widetilde{\cal T}_{q,F}(x+x_2,x,\mu_F,s_T) 
\right] \, ,
\label{ssa_q}\\
\mu_F^2 \frac{\partial}{\partial \mu_F^2} 
{\cal T}_{G,F}^{(i)}(x,x+x_2,\mu_F) 
&=&
\frac{1}{2}\left[
\mu_F^2 \frac{\partial}{\partial \mu_F^2} 
\widetilde{\cal T}_{G,F}^{(i)}(x,x+x_2,\mu_F,s_T)
+
\mu_F^2 \frac{\partial}{\partial \mu_F^2} 
\widetilde{\cal T}_{G,F}^{(i)}(x+x_2,x,\mu_F,s_T) 
\right] \, ,
\label{ssa_g} \\
\mu_F^2 \frac{\partial}{\partial \mu_F^2} 
{\cal T}_{\Delta q,F}(x,x+x_2,\mu_F) 
&=&
\frac{1}{2}\left[
\mu_F^2 \frac{\partial}{\partial \mu_F^2} 
\widetilde{\cal T}_{\Delta q,F}(x,x+x_2,\mu_F,s_T) 
-
\mu_F^2 \frac{\partial}{\partial \mu_F^2} 
\widetilde{\cal T}_{\Delta q,F}(x+x_2,x,\mu_F,s_T) 
\right]\, ,
\label{ssa_Dq}\\
\mu_F^2 \frac{\partial}{\partial \mu_F^2} 
{\cal T}_{\Delta G,F}^{(i)}(x,x+x_2,\mu_F) 
&=&
\frac{1}{2}\left[
\mu_F^2 \frac{\partial}{\partial \mu_F^2} 
\widetilde{\cal T}_{\Delta G,F}^{(i)}(x,x+x_2,\mu_F,s_T)
-
\mu_F^2 \frac{\partial}{\partial \mu_F^2} 
\widetilde{\cal T}_{\Delta G,F}^{(i)}(x+x_2,x,\mu_F,s_T)
\right], . 
\label{ssa_Dg} 
\een
As we show in the next section, the sum or the difference 
in the RHS of above equations determines the symmetry property
of these correlation functions when the active momentum
fractions $x$ and $x+x_2$ are switched.

\section{Evolution Kernels}
\label{kernels}

We present in this section our calculation of the order of 
$\alpha_s$ evolution kernels for the evolution equations 
that are derived in the last section at $x_2=0$.  More
precisely, we derive the order of $\alpha_s$ 
evolution equations for the diagonal twist-3 correlation 
functions defined in Eq.~(\ref{T_diag}).
We will present the complete evolution kernels at the 
order of $\alpha_s$ in a future publication.

The evolution kernels can be derived from the order of 
$\alpha_s$ diagrams in Fig.~\ref{fig3} after setting 
$x_2=0$ or integrating over $x_2$ weighted by $\delta(x_2)$.
We use the light-cone gauge {\it cut vertices} and 
{\it projection operators} derived in the last section 
to contract the quark and gluon lines at the 
{\it bottom} and the {\it top} of these diagrams, respectively.
Since the cut vertices with the middle gluon in the LHS of 
the cut are the same as that with the gluon in the RHS of 
the cut, we only need to calculate the cut Feynman diagrams 
in Fig.~\ref{fig3} that have the middle gluon at the 
{\it bottom} part of the diagrams in {\it one} side of the cut.
On the other hand, the sum of the all final-state cuts 
requires us to calculate all diagrams with the middle gluon 
on the {\it top} part of the diagrams in {\it both} sides 
of the cut.  In addition, we need to calculate the same
diagrams in Fig.~\ref{fig3} with the active momentum fractions
$x$ and $x+x_2$ switched, as indicated by the equations 
in Eq.~(\ref{ssa_q}) to Eq.~(\ref{ssa_Dg}).

\bef
\psfig{file=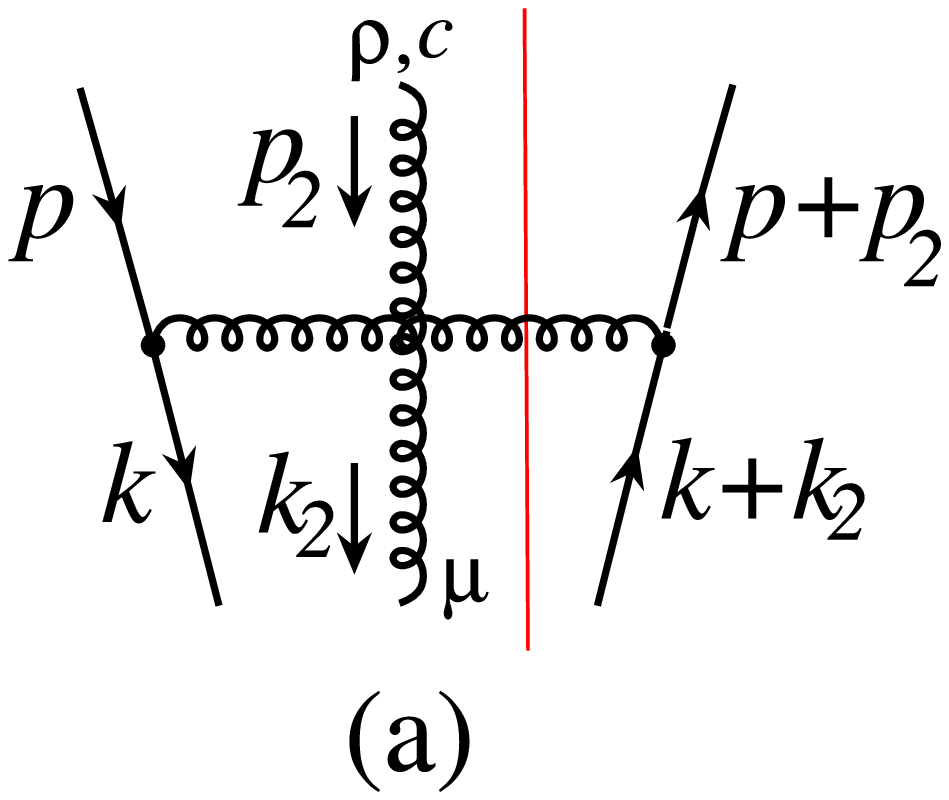,width=1.0in}
\hskip 0.3in
\psfig{file=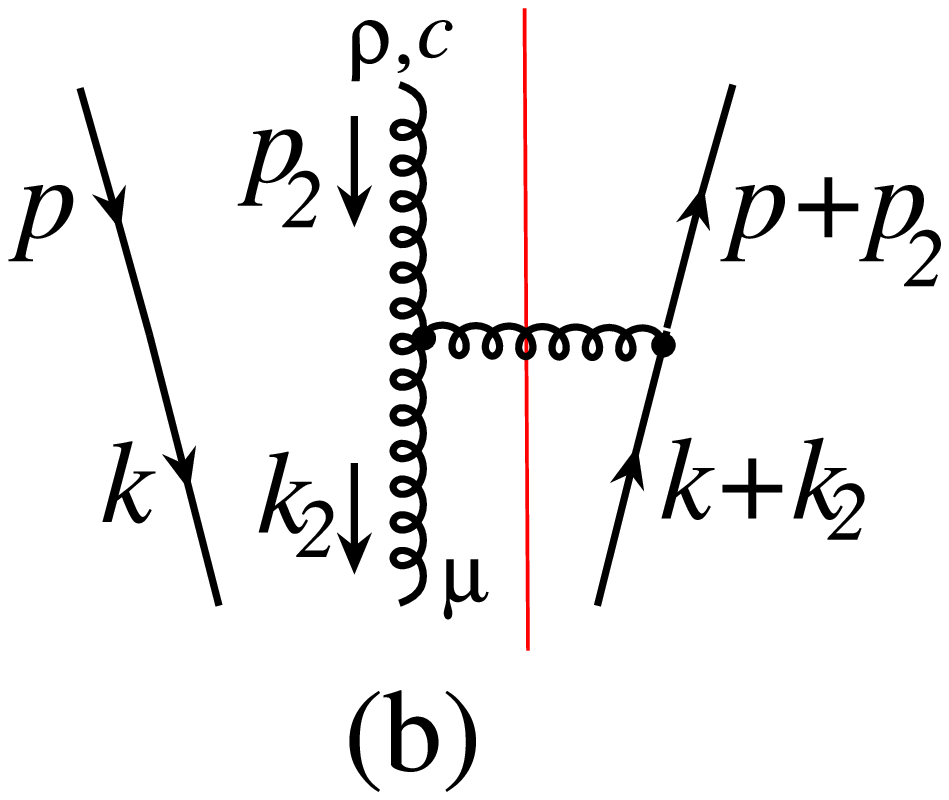,width=1.0in}
\hskip 0.3in
\psfig{file=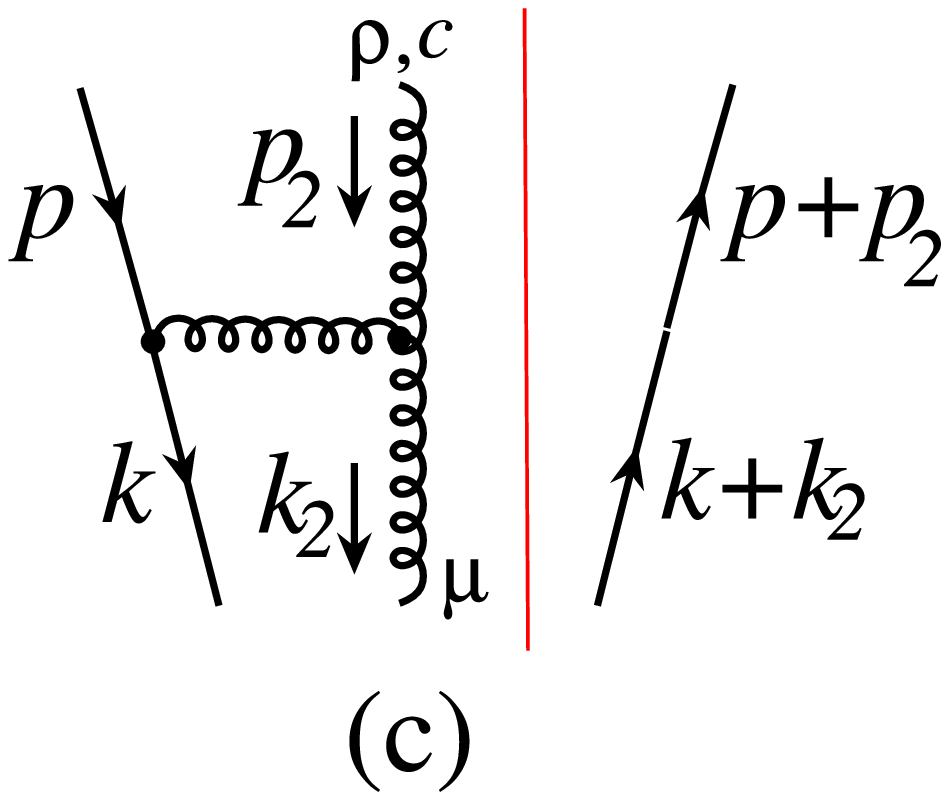,width=1.0in}
\hskip 0.2in
\psfig{file=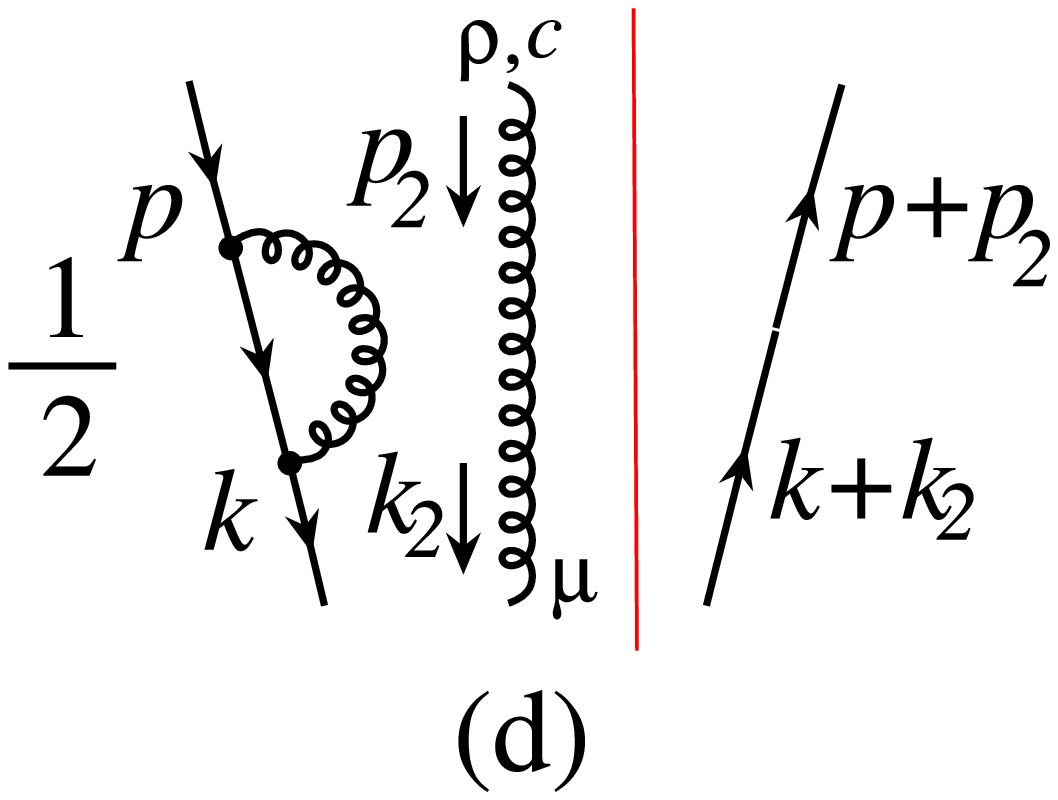,width=1.1in}
\hskip 0.2in
\psfig{file=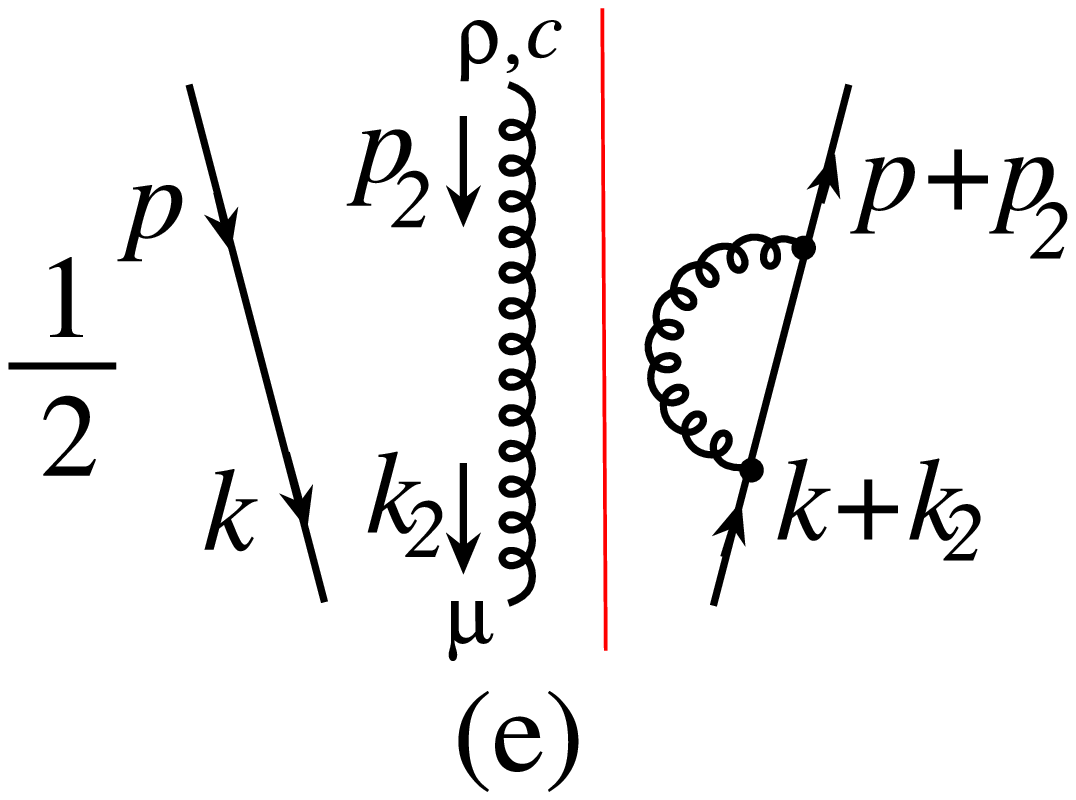,width=1.1in}
\\
\psfig{file=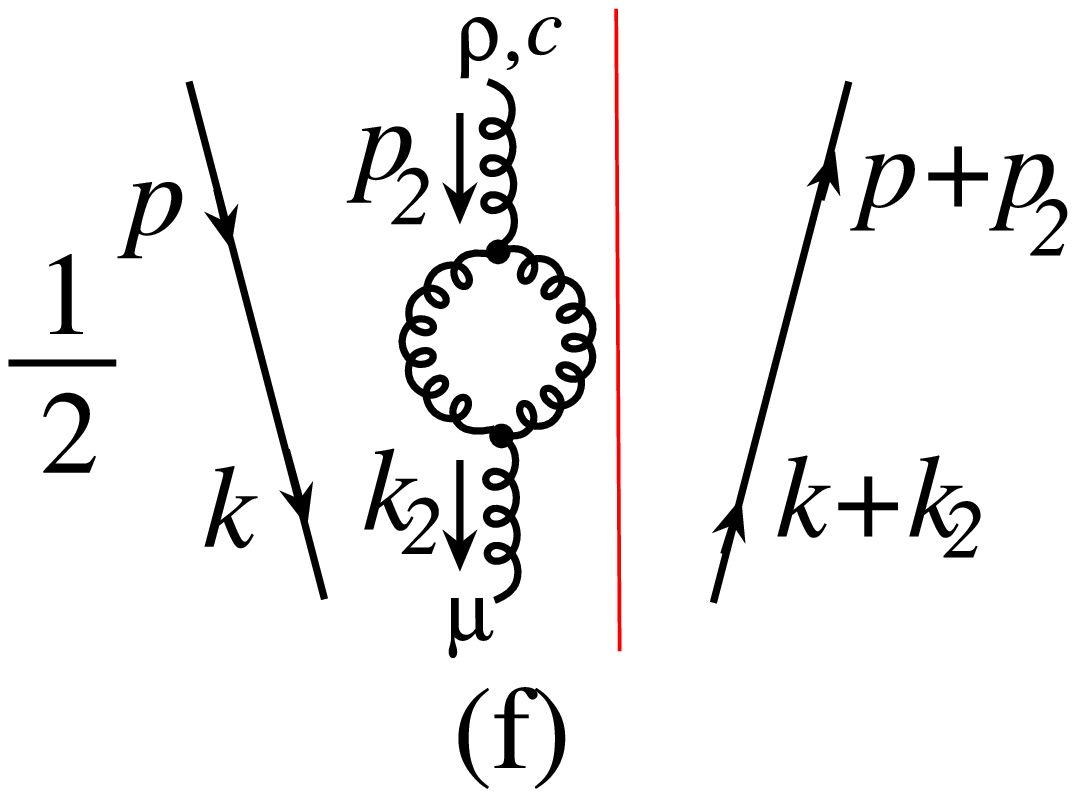,width=1.1in}
\hskip 0.2in
\psfig{file=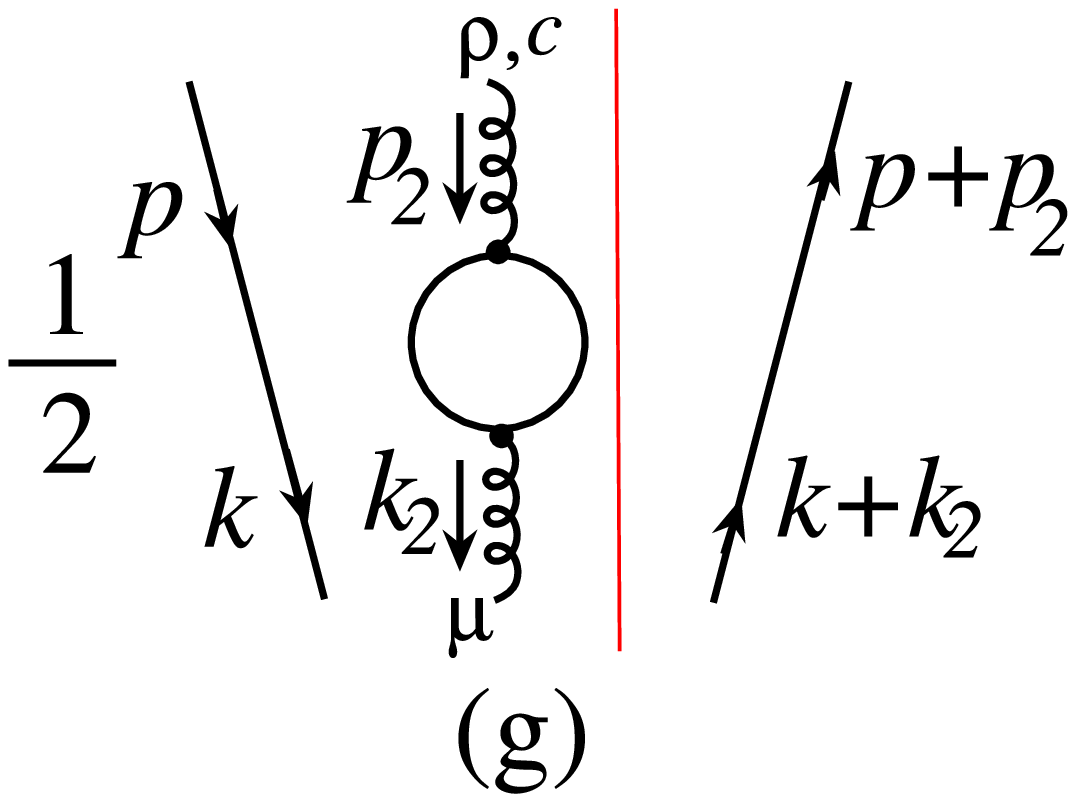,width=1.1in}
\hskip 0.2in
\psfig{file=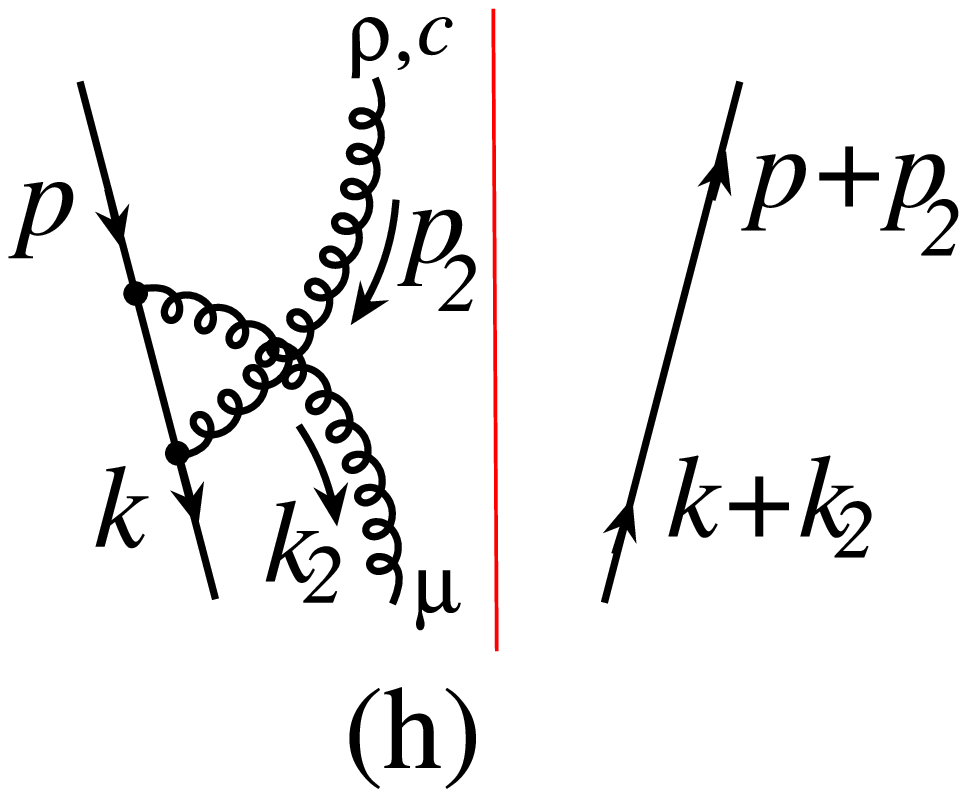,width=1.0in}
\hskip 0.3in
\psfig{file=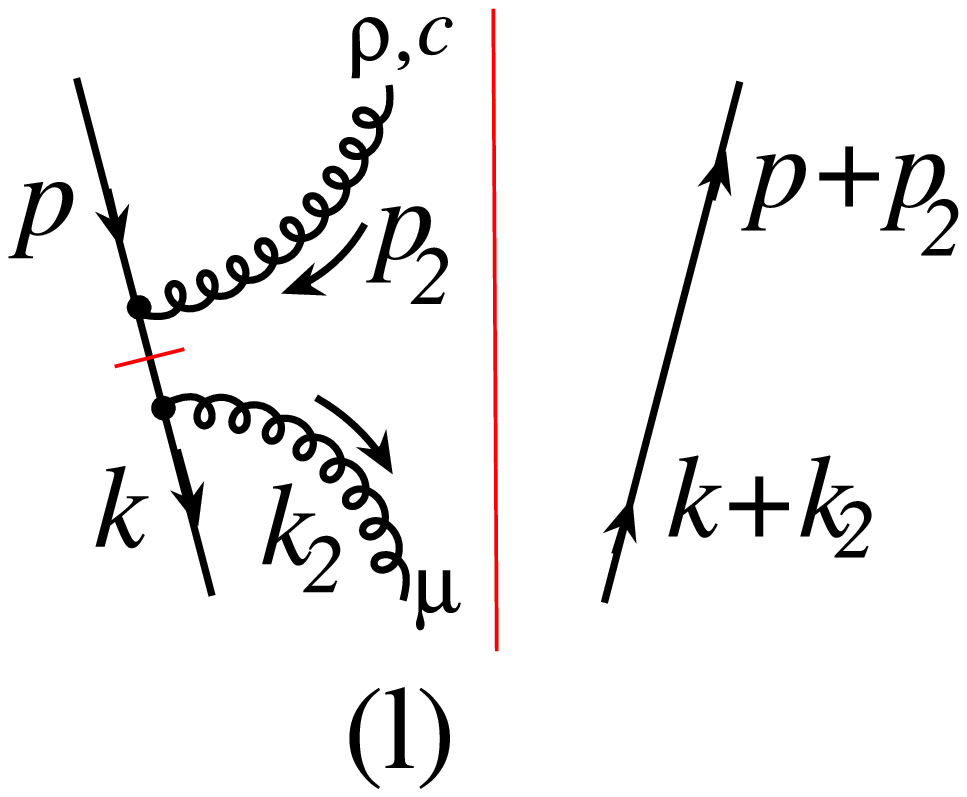,width=1.0in}
\hskip 0.3in
\psfig{file=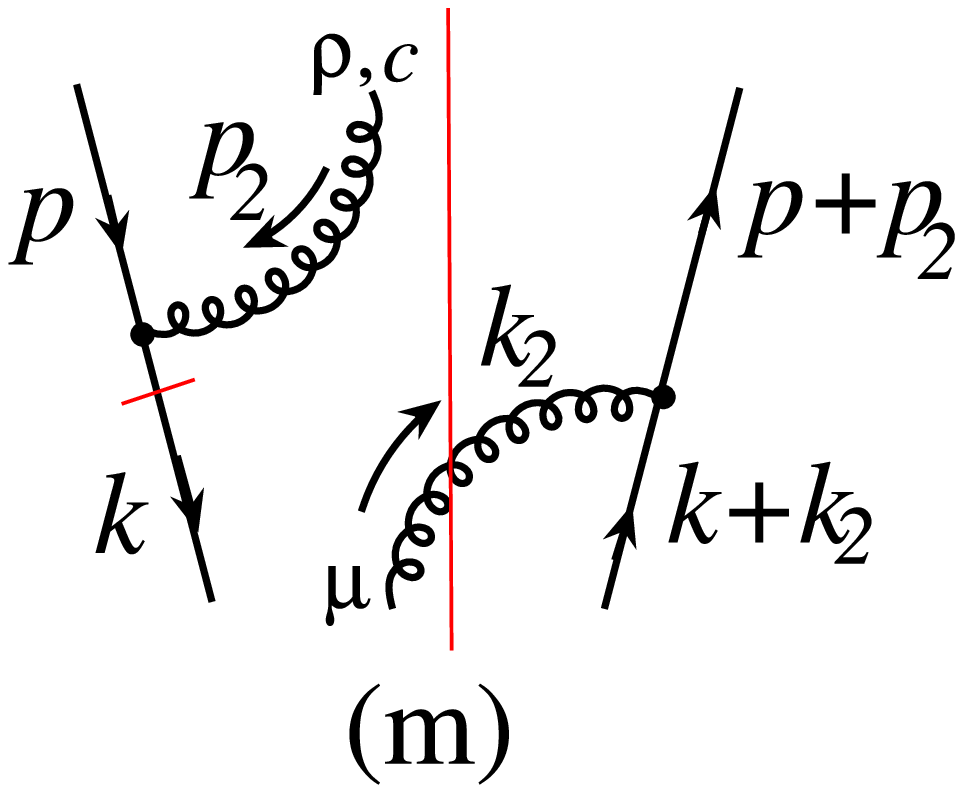,width=1.0in}
\\
\psfig{file=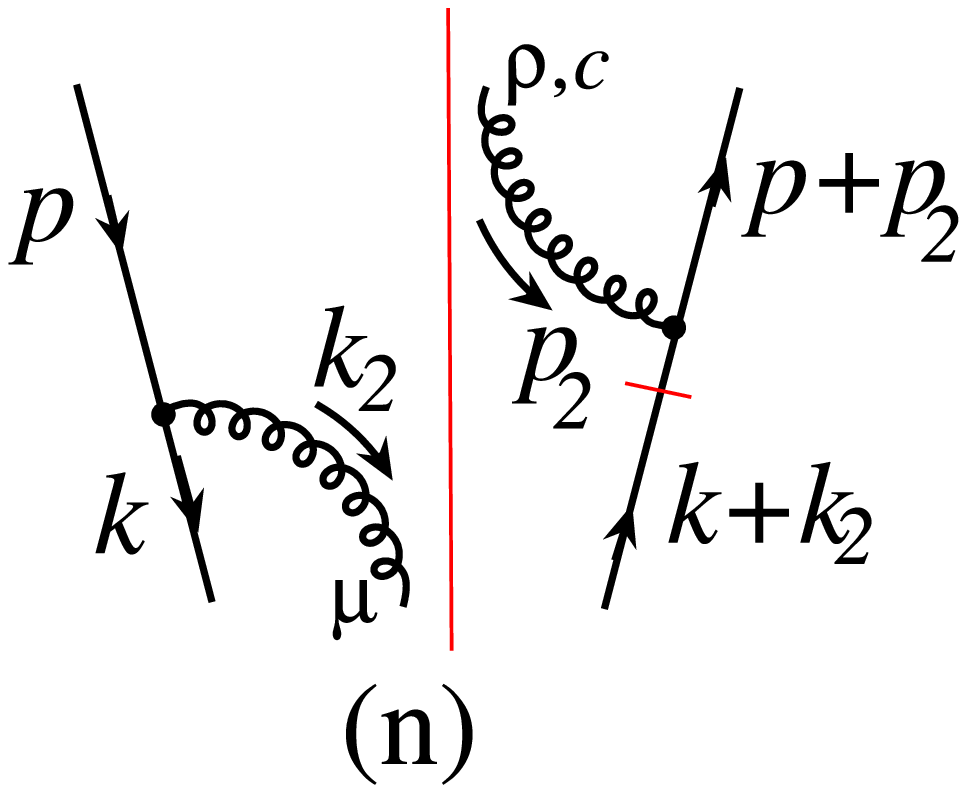,width=1.0in}
\hskip 0.3in
\psfig{file=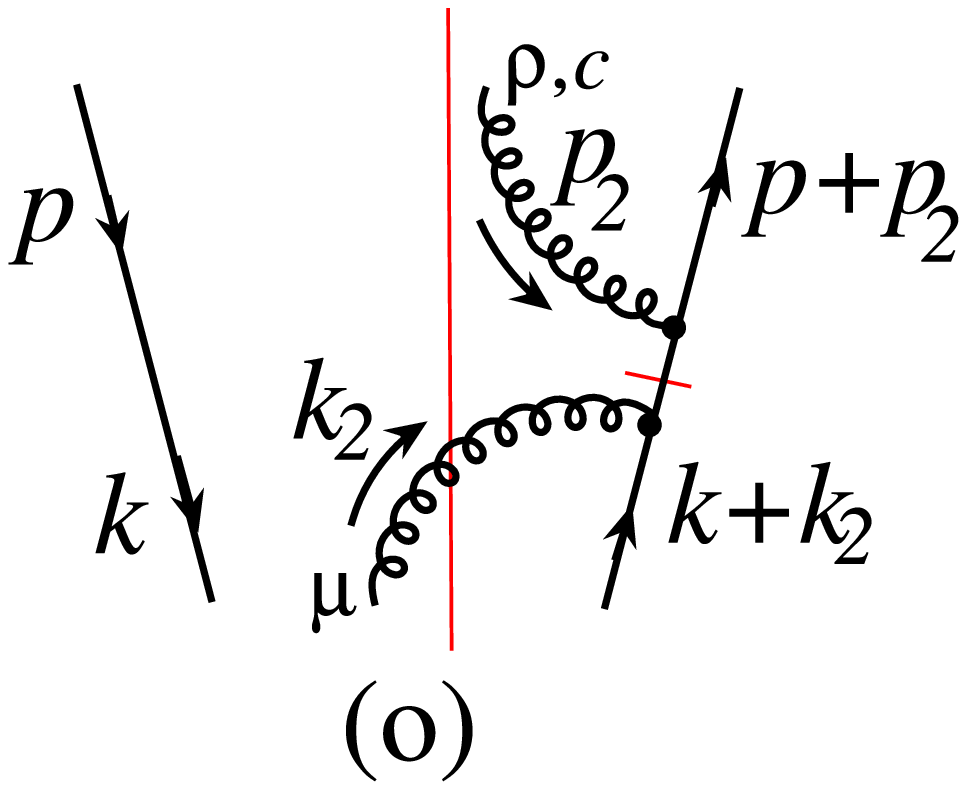,width=1.0in}
\hskip 0.3in
\psfig{file=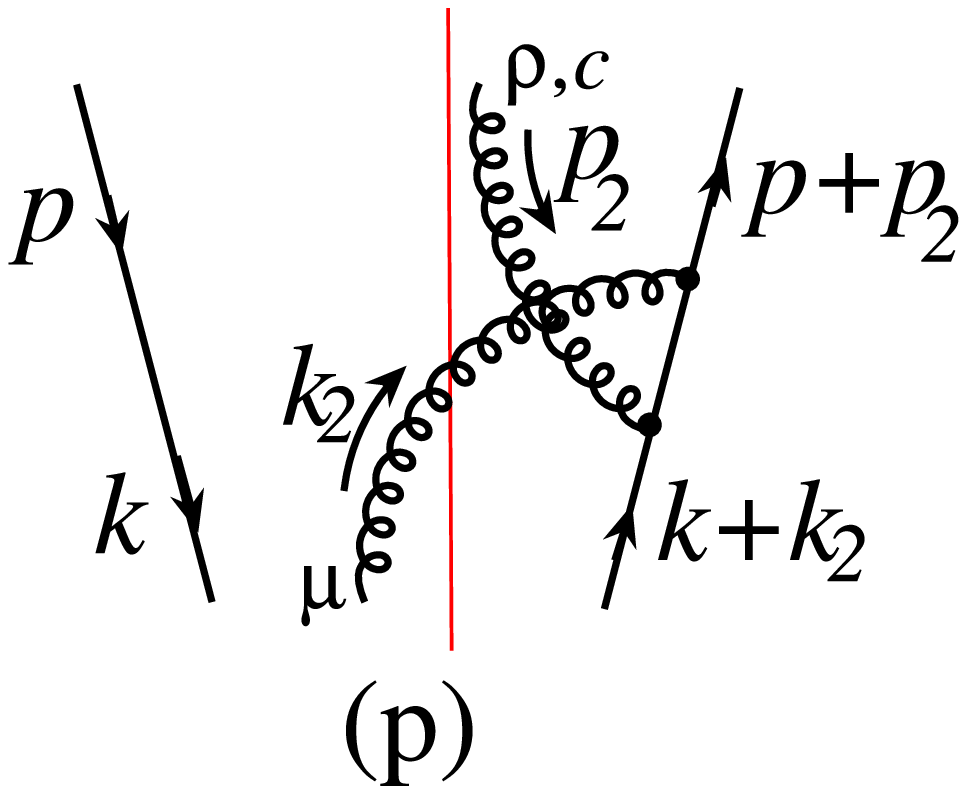,width=1.0in}
\hskip 0.3in
\psfig{file=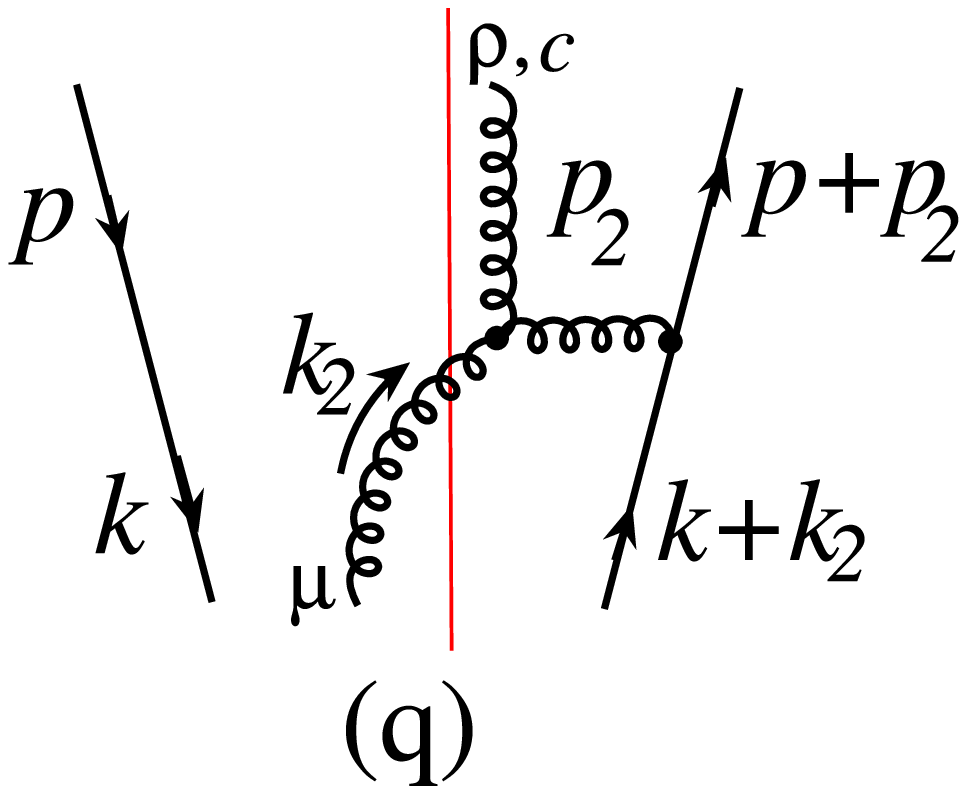,width=1.0in}
\caption{Feynman diagrams that contribute to 
the leading order flavor non-singlet evolution kernel 
of the twist-3 quark-gluon correlation function. }
\label{fig7}
\eef

We start with a detailed calculation of the order of $\alpha_s$
evolution kernels for the evolution equations of 
$\widetilde{\cal T}_{q,F}(x,x+x_2,\mu_F,s_T)$ and 
$\widetilde{\cal T}_{q,F}(x+x_2,x,\mu_F,s_T)$, and then, 
we construct the evolution equation for
${\cal T}_{q,F}(x,x,\mu_F)$ from Eq.~(\ref{ssa_q}).  
Finally, from Eq.~(\ref{T_diag}), 
we have the diagonal correlation function,
$T_{q,F}(x,x,\mu_F)=2\pi{\cal T}_{q,F}(x,x,\mu_F)$.  
We define
\ben
d{\cal I}_{qq}\equiv
\int dx_2\, \delta(x_2)\, 
dK_{qq}(\xi,\xi+\xi_2,x,x+x_2,\alpha_s)\, ,
\label{dIqq}
\een
where $dK_{qq}$ is given by the diagrams in Fig.~\ref{fig3}(a)
with the cut vertex in Eq.~(\ref{cv_q_lc}) and the projection 
operator in Eq.~(\ref{proj_q_lc}). 
We list in Fig.~\ref{fig7} all cut Feynman diagrams at order of 
$\alpha_s$ with the gluon at the cut vertex in the LHS of the 
cut.  Diagrams labeled from (a) to (m) have the top middle gluon 
in the LHS of the cut while the diagrams from (n) to (q) have the 
top middle gluon in the RHS of the cut.  The quark propagator 
with a short bar for the diagrams labeled by (l), (m), (n), and 
(o) is the special propagator introduced in Ref.~\cite{qiu_t4} to
represent the contact interaction.  These diagrams represent
the contribution from the diagram in Fig.~\ref{fig2}(a) that is 
necessary to make the full twist-3 contribution gauge invariant.
In the $n\cdot A=0$ light-cone gauge, the Feynman rule for the
special quark propagator of momentum $k$ is \cite{qiu_t4}
\ben
 \frac{i\gamma\cdot n}{2k\cdot n}\, 
 \frac{k^2}{k^2+i\epsilon}\, .
\label{sp_q}
\een
Having the cut vertex and the projection operator for the bottom 
and top quark and gluon lines, respectively, calculation of 
these Feynman diagrams in Fig.~\ref{fig7} is straightforward.  
In particular, after setting $x_2=0$ or integrating over $x_2$ 
weighted by the $\delta(x_2)$, all diagrams labeled from 
(f) to (q) give the vanishing contribution to the diagonal 
evolution kernel, $K_{qq}(\xi,\xi+\xi_2,x,x,\alpha_s)$.  
Using the technique introduced in Ref.~\cite{Collins:1988wj},
we find the following results for the rest of diagrams,
\ben
d{\cal I}_{qq}^{(a)}
&=&
\delta(\xi_2)\, \frac{1}{\xi}
\int^{\mu_F^2} \frac{dk_T^2}{k_T^2}\,
\left[C_F-\frac{C_A}{2}\right]
\frac{\alpha_s}{2\pi}
\left(\frac{1+z^2}{1-z}\right)\, ,
\label{dKqq_a}
\\
d{\cal I}_{qq}^{(b)}
&=&
\delta(\xi-x)\, \frac{1}{\xi_2}
\int^{\mu_F^2} \frac{dk_T^2}{k_T^2}\,
\left[\frac{C_A}{2}\right]
\frac{\alpha_s}{2\pi}
\left(\frac{1}{2}\,\frac{2x+\xi_2}{x+\xi_2}\right)\, ,
\label{dKqq_b}
\\
d{\cal I}_{qq}^{(c)}
&=&
\delta(\xi+\xi_2-x)\, \frac{1}{\xi}
\int^{\mu_F^2} \frac{dk_T^2}{k_T^2}\,
\left[\frac{C_A}{2}\right]
\frac{\alpha_s}{2\pi}
\left(\frac{1}{2}\,\frac{1+z}{1-z}\right)\, ,
\label{dKqq_c}
\\
d{\cal I}_{qq}^{(d+e)}
&=&
- \delta(\xi_2)\, \delta(\xi-x)\,
\int^{\mu_F^2} \frac{dk_T^2}{k_T^2}\,
\int_0^1 dz\,
\left[C_F\right]
\frac{\alpha_s}{2\pi}
\left(\frac{1+z^2}{1-z}\right)\, .
\label{dKqq_de}
\een
In above equations, $z=x/\xi$ and the color factor 
for each diagram is explicitly shown in the square brackets 
with $C_F=(N_c^2-1)/2N_c$, $C_A=N_c$ and $N_c=3$, 
the number of color.  
We notice that the RHS of the last equation
from the diagrams (d) and (e) is infrared divergent, and
the divergence is needed to cancel the infrared divergence from
the term proportional to $C_F$ in Eq.~(\ref{dKqq_a}) 
when $z\to 1$.  This cancellation of infrared divergence
between the real and virtual diagrams is the same as 
that takes place in the evolution kernel of 
normal PDFs \cite{Collins:1988wj}.  The remaining infrared
divergence as $z\to 1$ in Eq.~(\ref{dKqq_a}) is proportional 
to a different color factor, $C_A/2$, and 
is cancelled by the contribution from diagrams (b) and (c).

From the same Feynman diagrams in Fig.~\ref{fig7}, we can
also calculate the contribution from 
$\widetilde{\cal T}_{\Delta q,F}$,
\ben
d{\cal I}_{q\Delta q}\equiv
\int dx_2\, \delta(x_2)\, 
dK_{q\Delta q}(\xi,\xi+\xi_2,x,x+x_2,\alpha_s)
\label{dIqDq}
\een
by using the projection operator in Eq.~(\ref{proj_Dq_lc}).
In this case, only diagrams (b) and (c) give nonvanishing 
results,
\ben
d{\cal I}_{q\Delta{q}}^{(b)}
&=&
\delta(\xi-x)
\int^{\mu_F^2} \frac{dk_T^2}{k_T^2}\,
\left[\frac{C_A}{2}\right]
\frac{\alpha_s}{2\pi}
\left(\frac{1}{2}\,\frac{1}{x+\xi_2}\right)\, ,
\label{dKqDq_b}
\\
d{\cal I}_{q\Delta{q}}^{(c)}
&=&-
\delta(\xi+\xi_2-x)\, \frac{1}{\xi}
\int^{\mu_F^2} \frac{dk_T^2}{k_T^2}\,
\left[\frac{C_A}{2}\right]
\frac{\alpha_s}{2\pi}
\left(\frac{1}{2}\right)\, .
\label{dKqDq_c}
\een
By comparing above calculated results with Eq.~(\ref{dk_ij}),
we extract evolution kernels, $K_{qq}(\xi,\xi+\xi_2,x,x)$ 
and $K_{q\Delta q}(\xi,\xi+\xi_2,x,x)$.  By calculating the 
same diagrams in Fig.~\ref{fig3} with momentum fractions 
$\xi$ and $x$ switched with $\xi+\xi_2$ and $x+x_2$, 
respectively, we derive evolution kernels, 
$K_{qq}(\xi+\xi_2,\xi,x,x)$ 
and $K_{q\Delta q}(\xi+\xi_2,\xi,x,x)$.  
By integrating Eq.~(\ref{ssa_q}) over $x_2$ weighted by 
$\delta(x_2)$ or simply setting $x_2=0$, we obtain
the order of $\alpha_s$ evolution equation for 
${\cal T}_{q,F}(x,x,\mu_F)$ from flavor non-singlet 
interactions,
\ben
\frac{\partial {\cal T}_{q,F}(x,x,\mu_F)}
     {\partial{\ln \mu_F^2}}
&=&
\frac{\alpha_s}{2\pi}
\int_x^1\frac{d\xi}{\xi}
\bigg\{
P_{qq}(z)\, {\cal T}_{q,F}(\xi,\xi,\mu_F)
\nonumber\\
&\ & \hskip 0.6in
+\frac{C_A}{2}
\left[\frac{1+z^2}{1-z}
\left[{\cal T}_{q,F}(\xi,x,\mu_F)
     -{\cal T}_{q,F}(\xi,\xi,\mu_F)\right]
     +z\, {\cal T}_{q,F}(\xi,x,\mu_F) \right]
\nonumber\\
&\ & \hskip 0.6in
+\frac{C_A}{2}\,
\bigg[{\cal T}_{\Delta{q}, F}(x,\xi,\mu_F) \bigg] \bigg\}\, ,
\label{evo_Tq_ns}
\een
where 
\ben
P_{qq}(z)  = C_F \left[
\frac{1+z^2}{(1-z)_+} + \frac{3}{2}\,\delta(1-z)
\right]
\label{Pqq}
\een
is the LO quark-to-quark splitting function 
for the normal PDFs.  The standard definition of ``+'' 
distribution is 
\ben
\int_x^1\, dz\, \frac{f(z)}{(1-z)_+} =
\int_x^1\, dz\, \frac{f(z)-f(1)}{1-z} + f(1)\,\ln(1-x)
\een
for a smooth function $f(z)$.  
In deriving Eq.~(\ref{evo_Tq_ns}), Eqs.~(\ref{Tqasy})
and (\ref{TDqasy}) were used.  It is clear from 
Eq.~(\ref{evo_Tq_ns}) that the flavor non-singlet evolution 
kernels for the diagonal twist-3 quark-gluon correlation 
function $T_{q,F}(x,x,\mu_F)=2\pi{\cal T}_{q,F}(x,x,\mu_F)$ 
are all infrared safe.  The evolution equation for the 
diagonal correlation function ${\cal T}_{q,F}(x,x,\mu_F)$
is not a closed one since it gets contribution not only from 
the same diagonal function ${\cal T}_{q,F}(\xi,\xi,\mu_F)$
but also from the off-diagonal part of the same function 
as well as gets the contribution from a different function 
${\cal T}_{\Delta q,F}(x,\xi,\mu_F)$.
 
\bef
\psfig{file=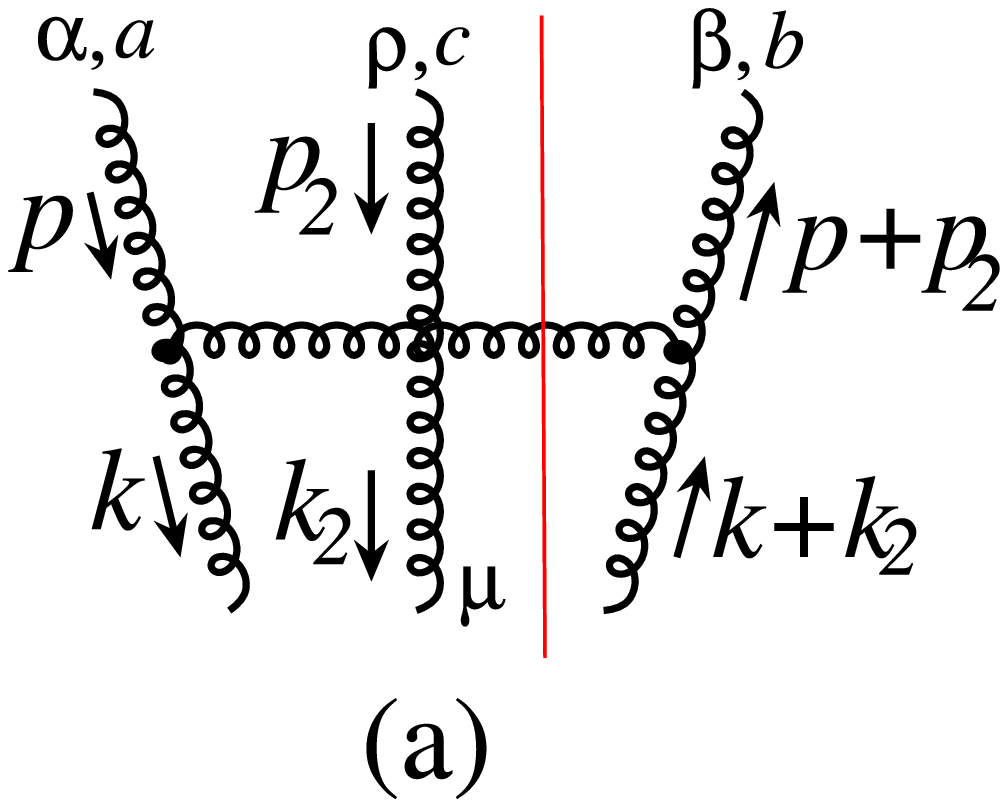,width=1.0in}
\hskip 0.3in
\psfig{file=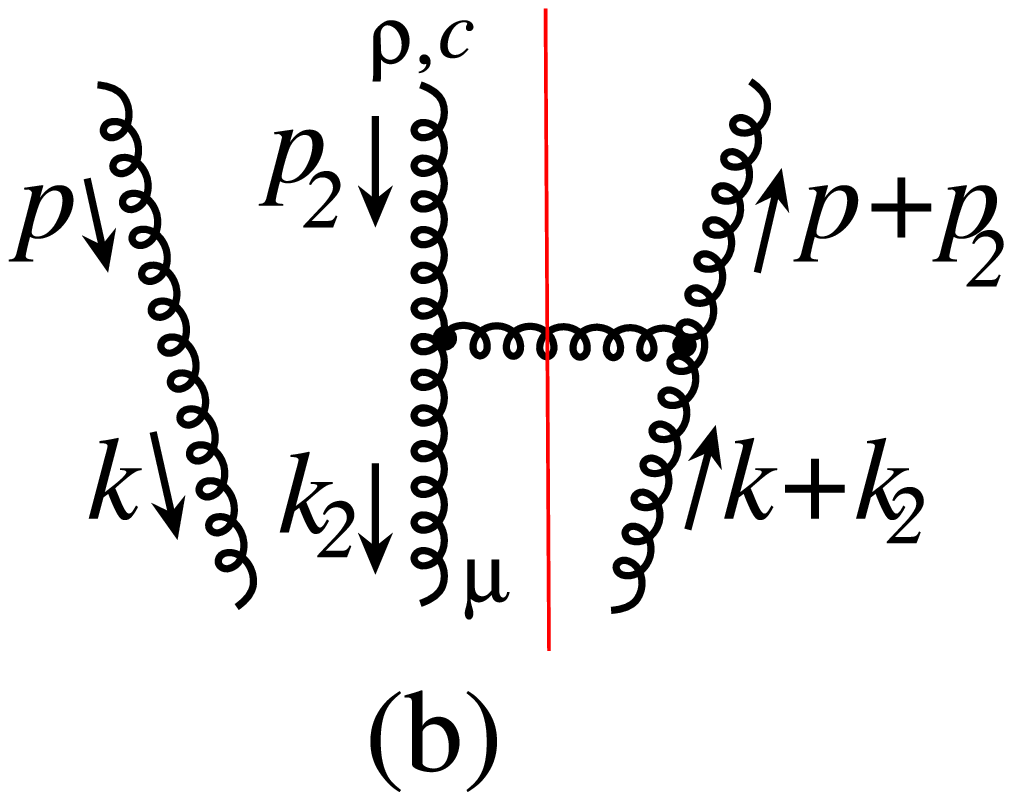,width=1.0in}
\hskip 0.3in
\psfig{file=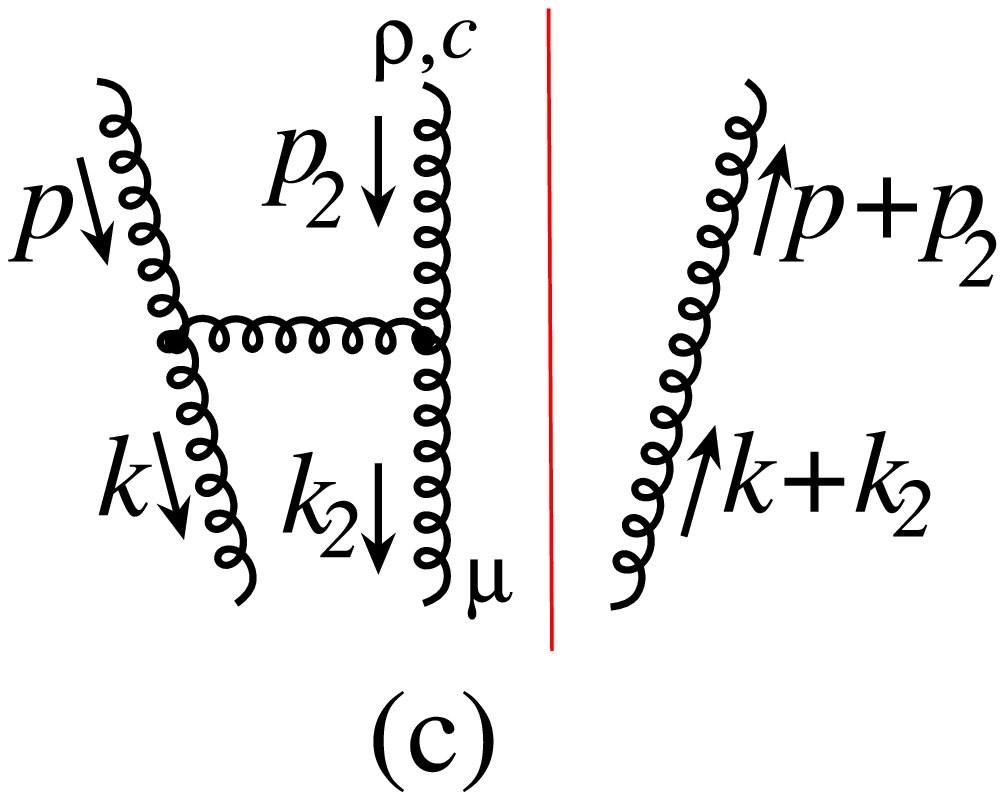,width=1.0in}
\hskip 0.2in
\psfig{file=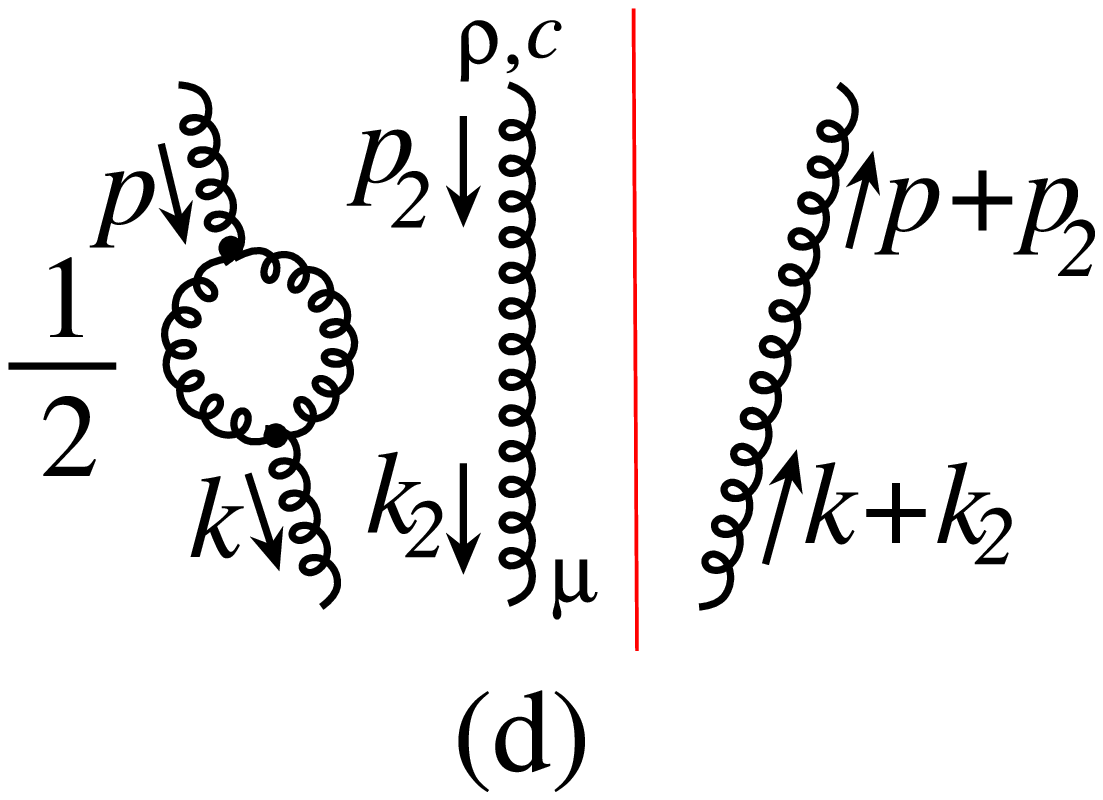,width=1.1in}
\hskip 0.2in
\psfig{file=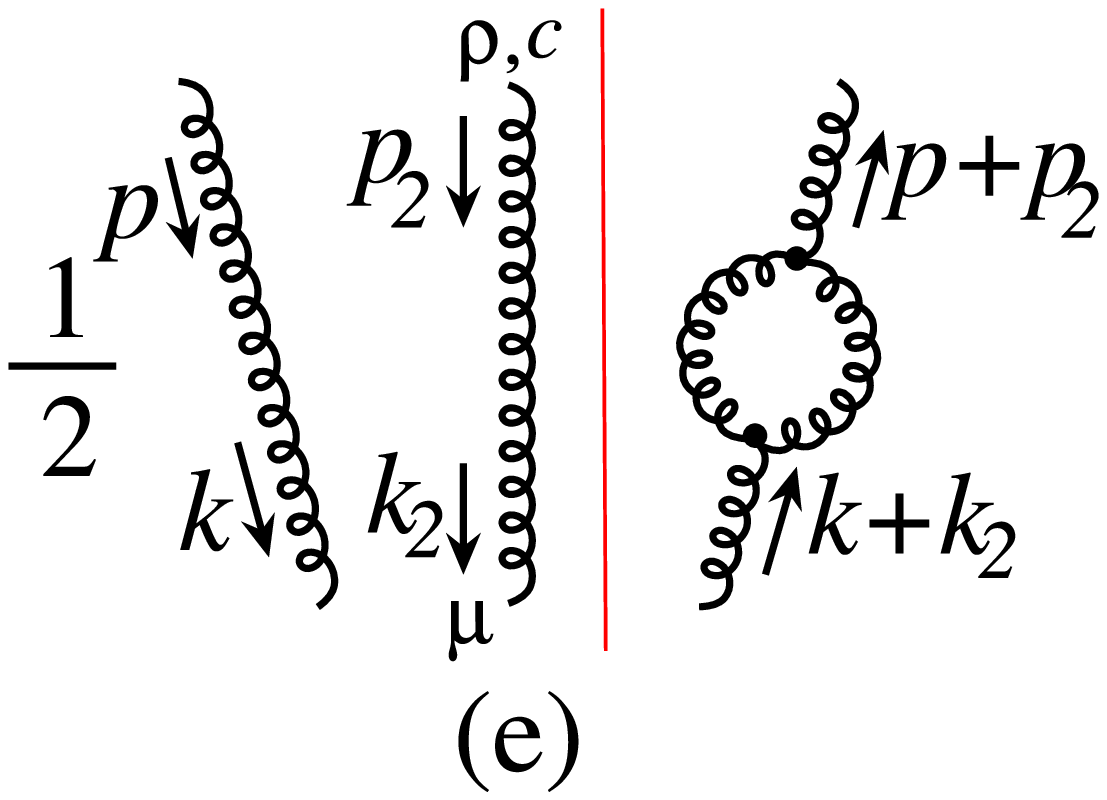,width=1.1in}
\\
\psfig{file=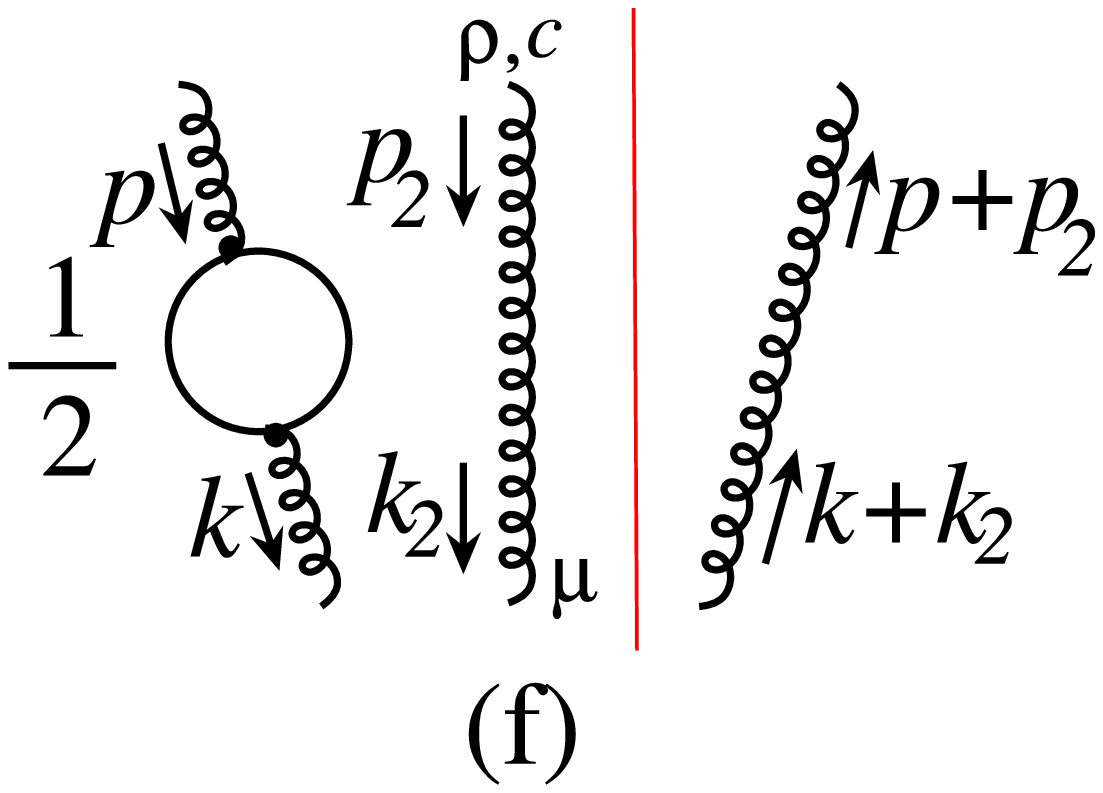,width=1.1in}
\hskip 0.2in
\psfig{file=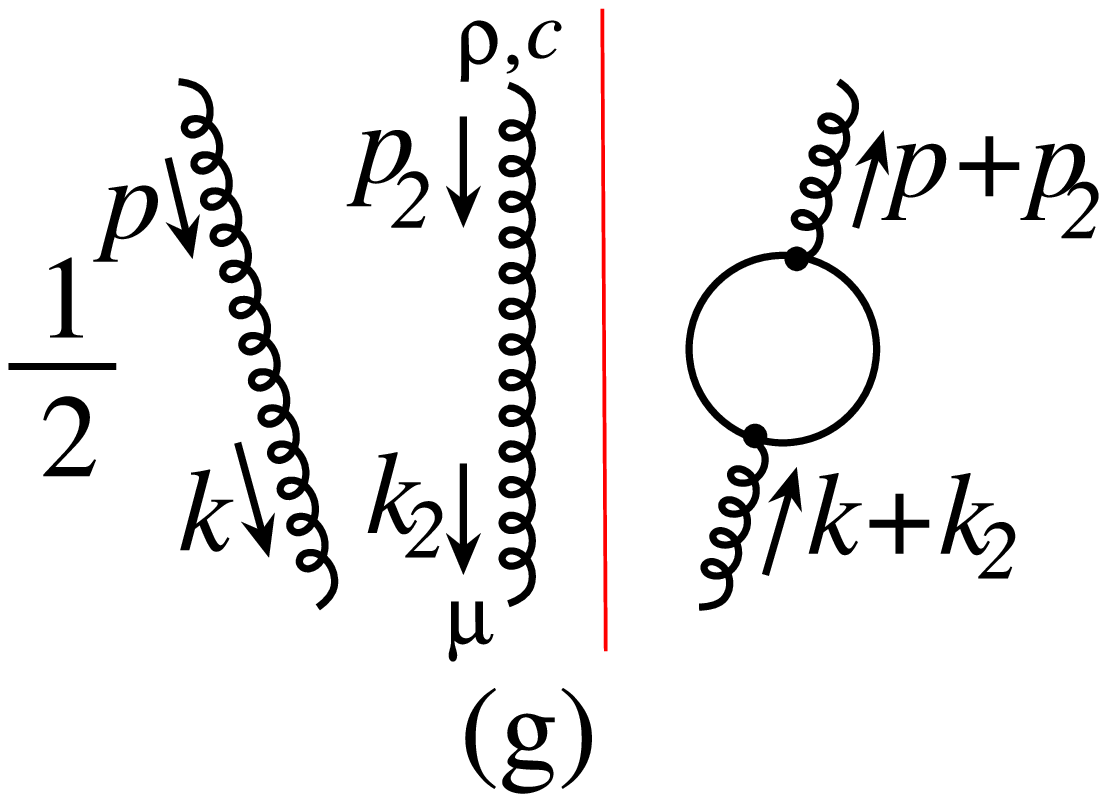,width=1.1in}
\hskip 0.2in
\psfig{file=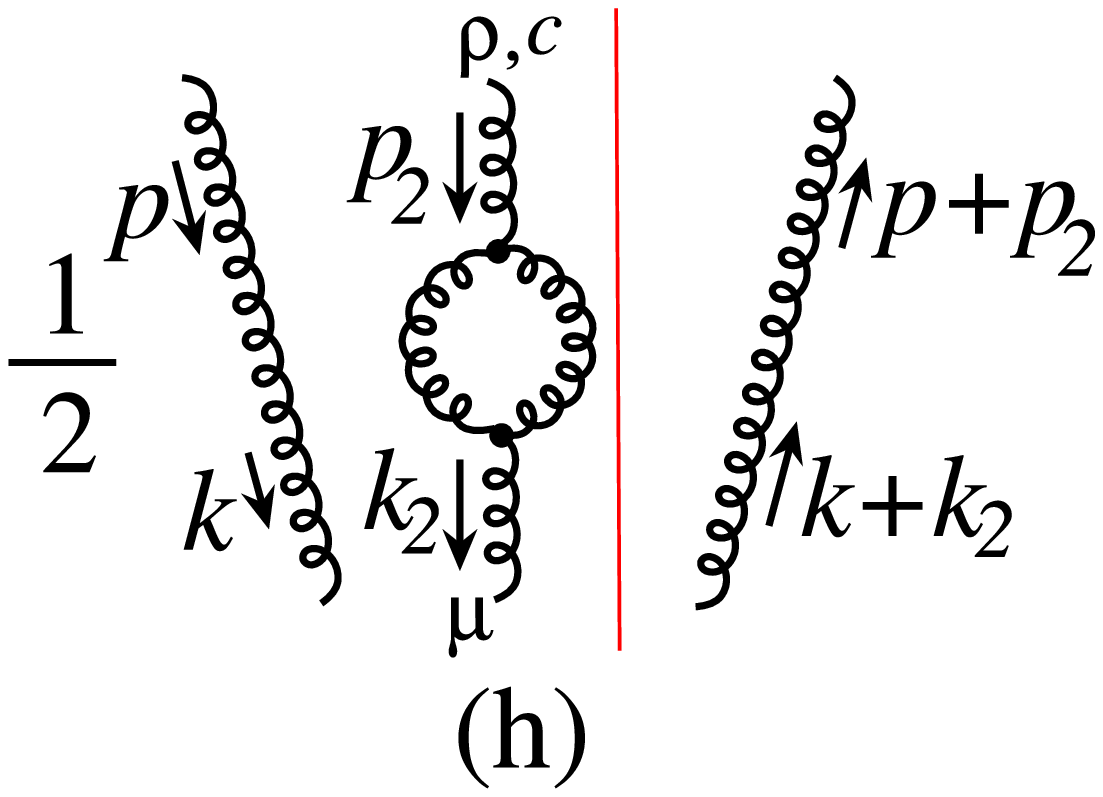,width=1.1in}
\hskip 0.2in
\psfig{file=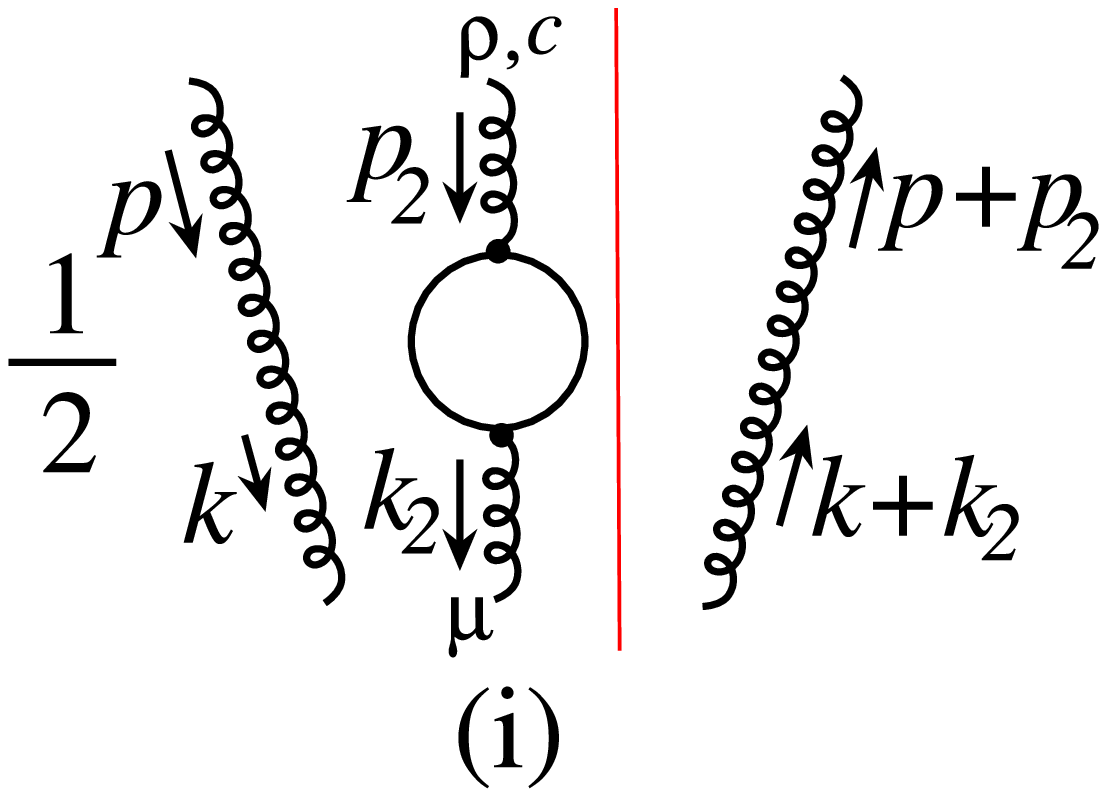,width=1.1in}
\\
\psfig{file=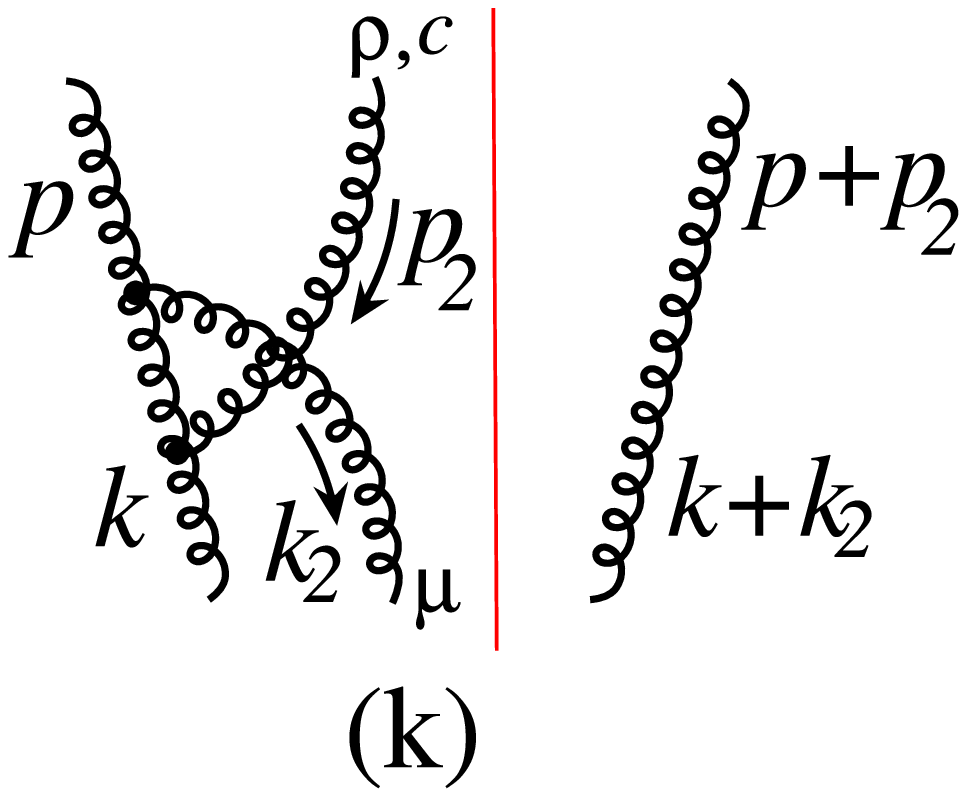,width=1.0in}
\hskip 0.3in
\psfig{file=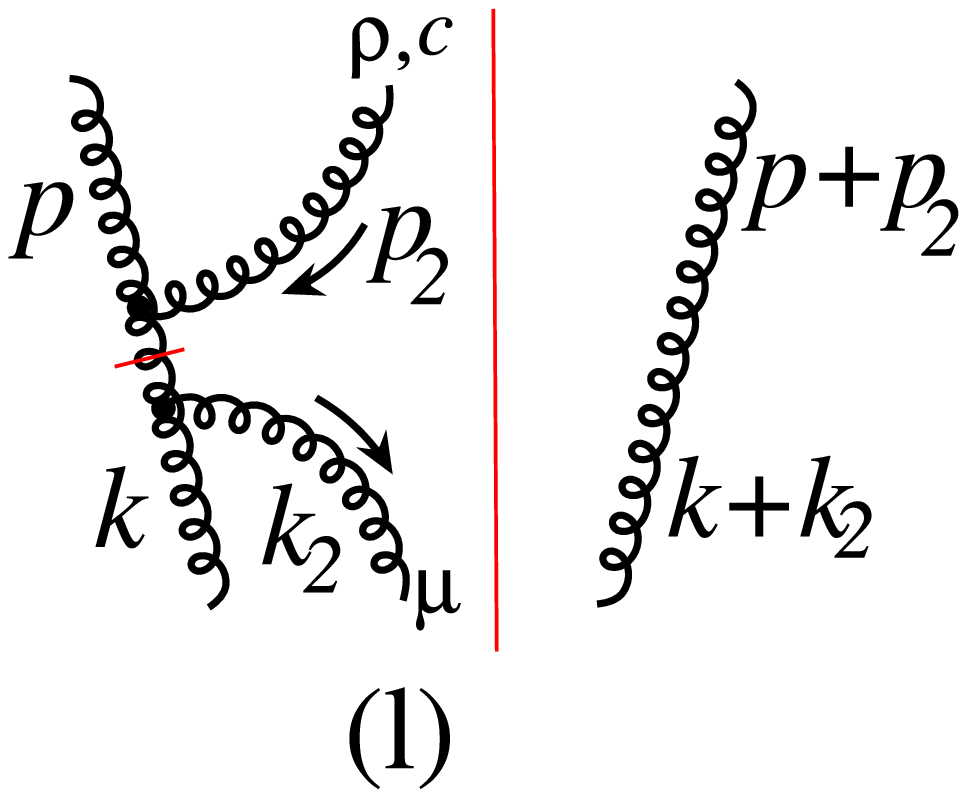,width=1.0in}
\hskip 0.3in
\psfig{file=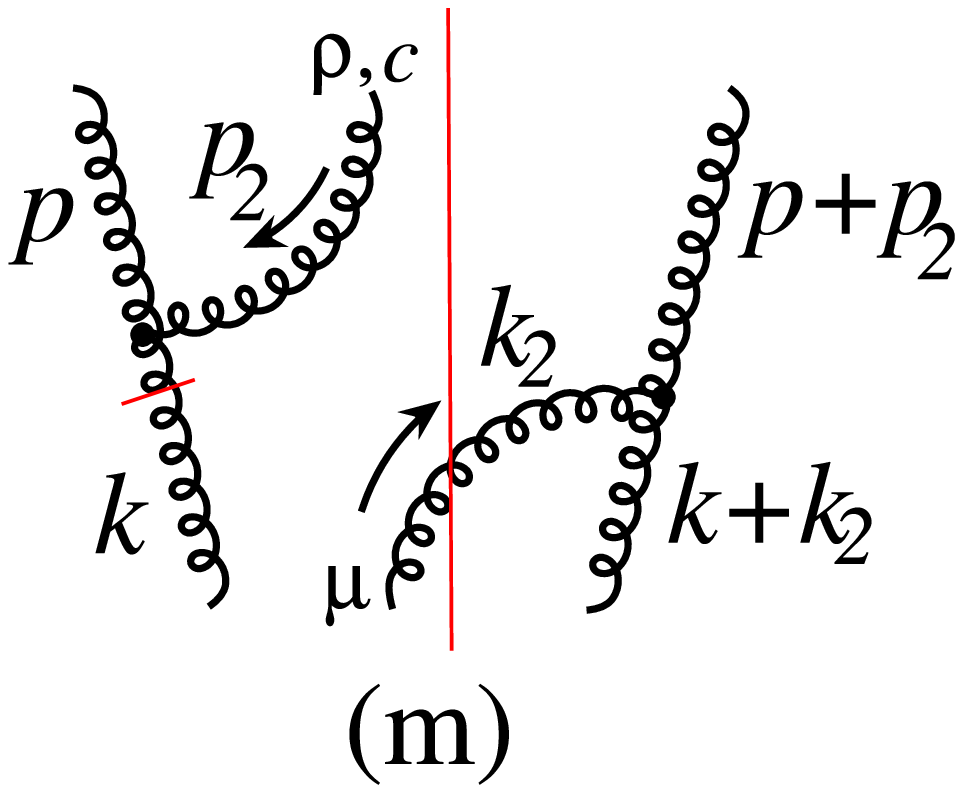,width=1.0in}
\\
\psfig{file=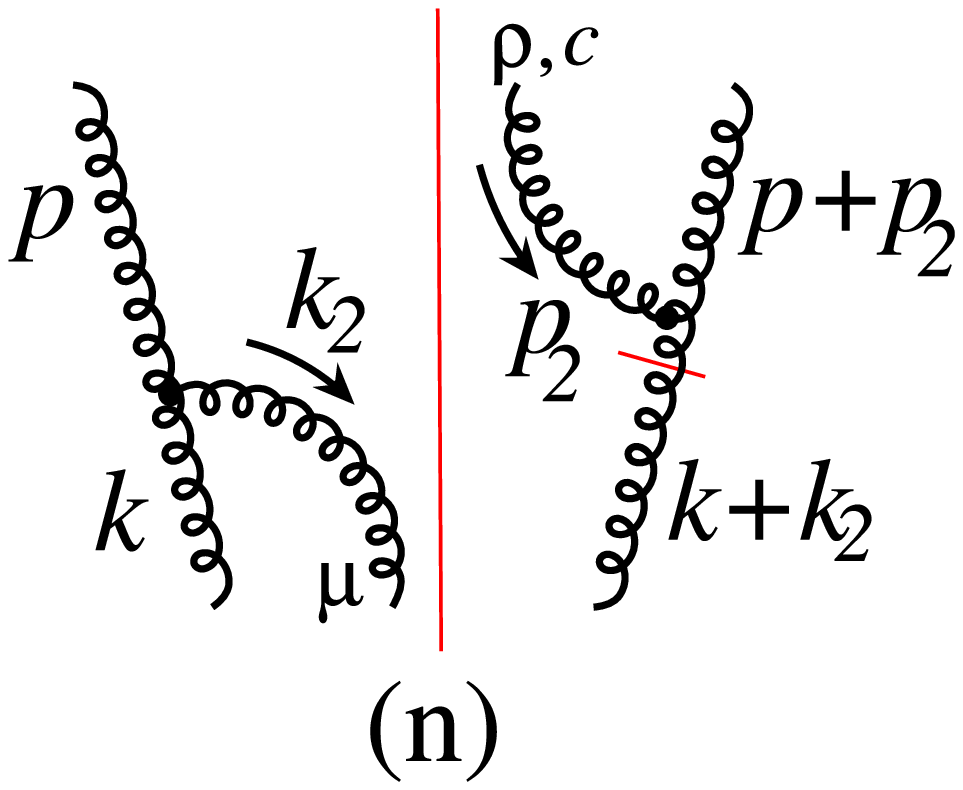,width=1.0in}
\hskip 0.3in
\psfig{file=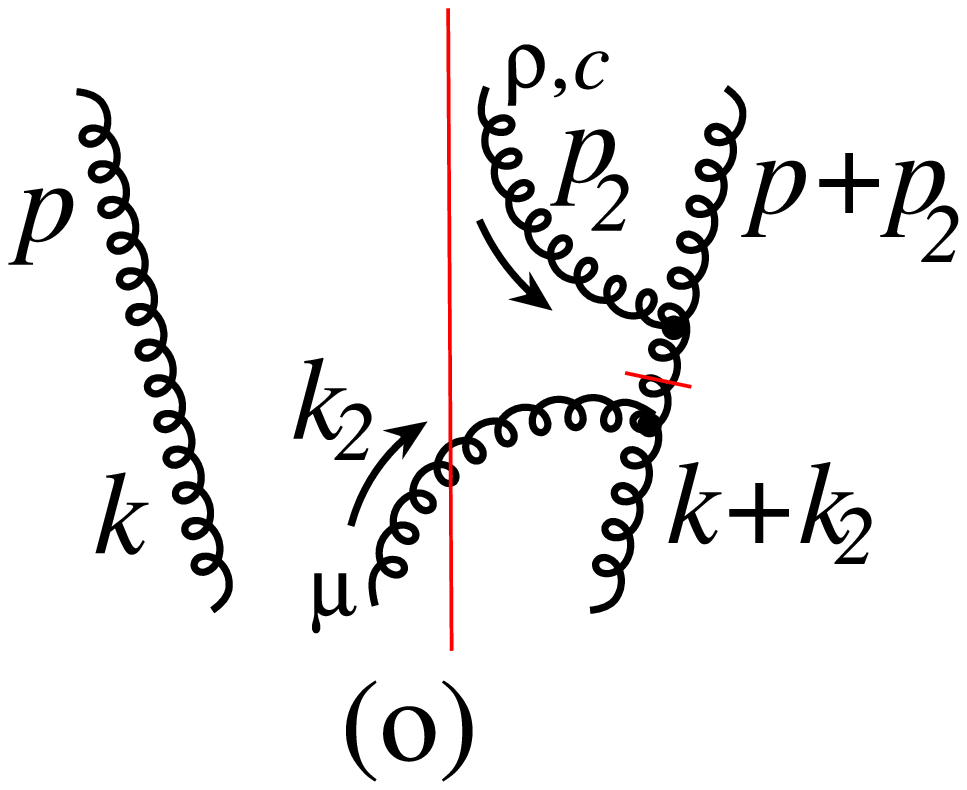,width=1.0in}
\hskip 0.3in
\psfig{file=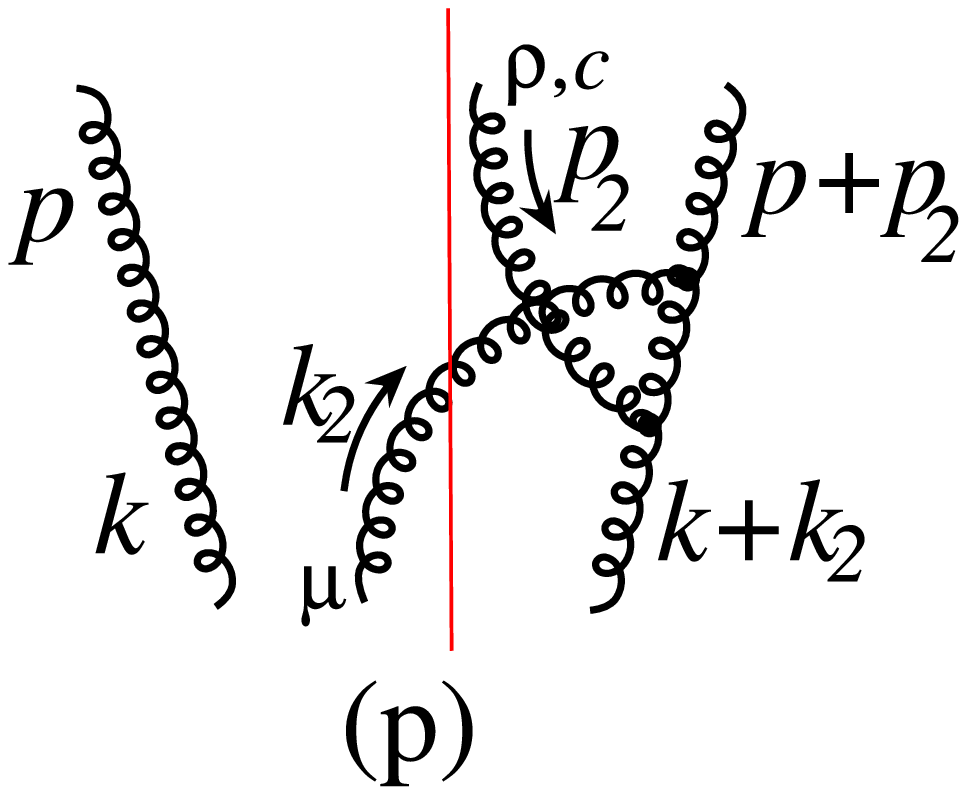,width=1.0in}
\hskip 0.3in
\psfig{file=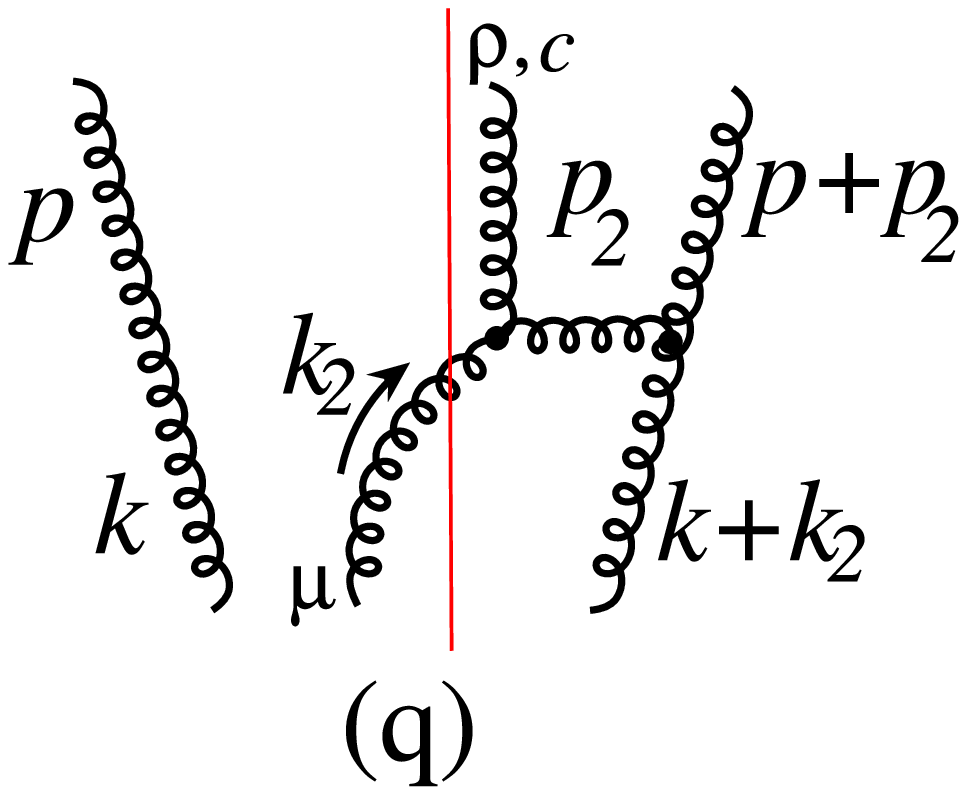,width=1.0in}
\caption{Feynman diagrams that contribute to 
the leading order evolution kernel from the tri-gluon
correlation functions to the tri-gluon correlation function. }
\label{fig8}
\eef

In the rest of this section, we derive the order of $\alpha_s$ 
evolution kernels involving gluons as well as those with 
the flavor change.
In Fig.~\ref{fig8}, we list all cut Feynman diagrams at the order 
of $\alpha_s$ that could contribute to the evolution kernels,
$K_{gg}^{(ij)}$ and $K_{\Delta g\Delta g}^{(ij)}$ with
$i,j=f,d$, when
proper cut vertices and projection operators are used.  
The gluon propagator with a short bar in the diagrams 
(l), (m), (n), and (o) is the gluonic special propagator 
defined in Ref.~\cite{qiu_t4}, which represents the 
contact interaction.  The diagrams with the contact
interaction are responsible for the twist-3 contribution 
from the diagram in Fig.~\ref{fig4}(a).
We calculate all diagrams with the cut vertices and projection
operators derived in this section and setting $x_2=0$.
We find that after taking $x_2=0$ or integrating over
$x_2$ weighted with $\delta(x_2)$,  
only diagrams (a), (b), (c), (d), (e), (f), and (g) give 
the nonvanishing contribution to the evolution kernel,
$K_{gg}^{(i,j)}$,
\ben
d{\cal I}_{gg}^{(a)}
&=&
2\pi\,\delta(\xi_2)\, \frac{1}{\xi}
\int^{\mu_F^2} \frac{dk_T^2}{k_T^2}\,
\left[C_A-\frac{C_A}{2}\right]
\frac{\alpha_s}{2\pi}\
2z\left(\frac{z}{1-z}+\frac{1-z}{z}+z(1-z)\right)\, ;
\label{dKgg_a}
\\
d{\cal I}_{gg}^{(b)}
&=&
2\pi\,\delta(\xi-x)\, \frac{1}{\xi_2}
\int^{\mu_F^2} \frac{dk_T^2}{k_T^2}\,
\left[\frac{C_A}{2}\right]
\frac{\alpha_s}{2\pi}
\left(\frac{1}{2}\,\frac{x^2+(x+\xi_2)^2}{(x+\xi_2)^2}\right)\, ;
\label{dKgg_b}
\\
d{\cal I}_{gg}^{(c)}
&=&
2\pi\,\delta(\xi+\xi_2-x)\, \frac{1}{\xi}
\int^{\mu_F^2} \frac{dk_T^2}{k_T^2}\,
\left[\frac{C_A}{2}\right]
\frac{\alpha_s}{2\pi}
\left(\frac{1}{2}\,\frac{1+z^2}{1-z}\right)\, ;
\label{dKgg_c}
\\
d{\cal I}_{gg}^{(d+e)}
&=&
- 2\pi\,\delta(\xi_2)\, \delta(\xi-x)\,
\int^{\mu_F^2} \frac{dk_T^2}{k_T^2}\,
\int_0^1 dz\
\frac{1}{2}\,\left[C_A\right]
\frac{\alpha_s}{2\pi}\
2\left(\frac{z}{1-z}+\frac{1-z}{z}+z(1-z)\right)\, ;
\label{dKgg_de}
\\
d{\cal I}_{gg}^{(f+g)}
&=&
- 2\pi\,\delta(\xi_2)\, \delta(\xi-x)\,
\int^{\mu_F^2} \frac{dk_T^2}{k_T^2}\,
\int_0^1 dz\
\frac{1}{2}\,(2\,n_f)\left[\frac{1}{2}\right]
\frac{\alpha_s}{2\pi}\,
\left((1-z)^2+z^2\right)\, ,
\label{dKgg_fg}
\een
where $n_f=1,2,...$ the number of active quark flavors
and the color factors are shown in the square brackets.
We find that the results from Eq.~(\ref{dKgg_a}) to 
(\ref{dKgg_fg}) are the same for the evolution kernel
$K_{gg}^{(ff)}$ and $K_{gg}^{(dd)}$.  All evolution kernels
of the crossing contribution $K_{gg}^{(fd)}=K_{gg}^{(df)}=0$.

\bef
\psfig{file=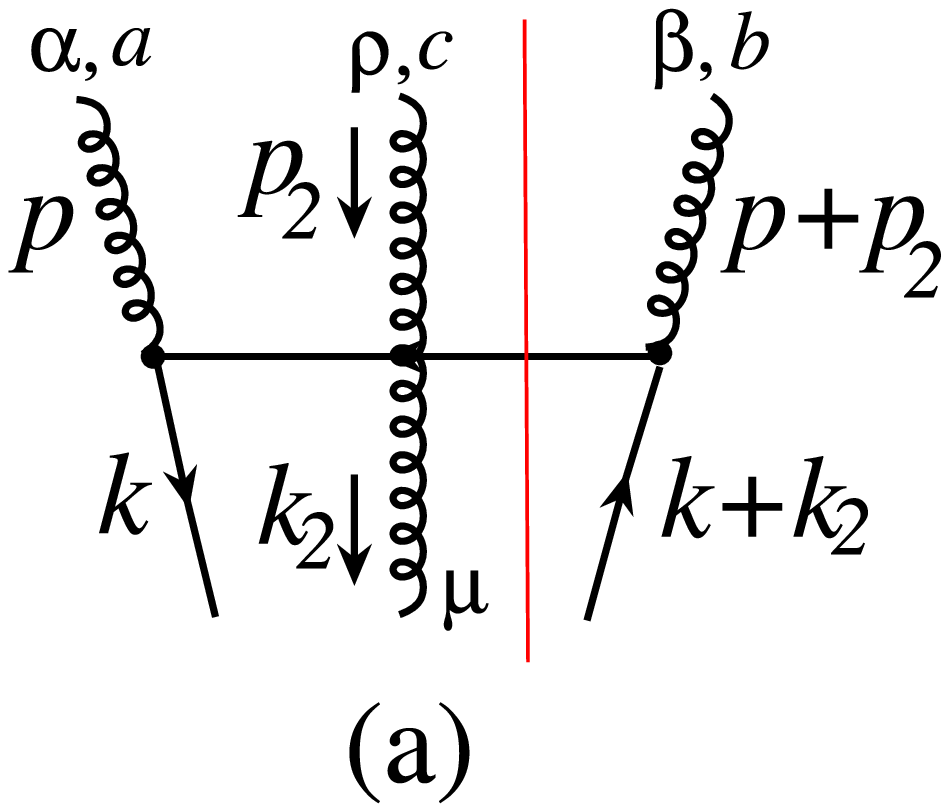,width=1.0in}
\hskip 0.3in
\psfig{file=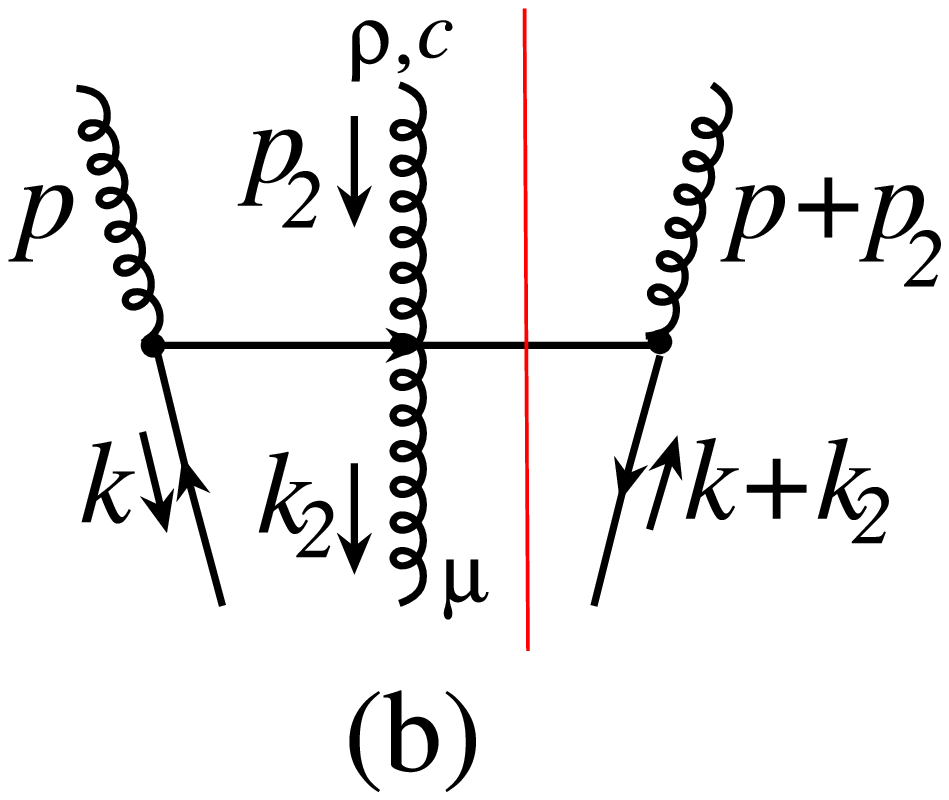,width=1.0in}
\caption{Feynman diagrams that contribute to 
the leading order evolution kernel from the tri-gluon
correlation functions to the quark-gluon correlation function. }
\label{fig9}
\eef

\bef
\psfig{file=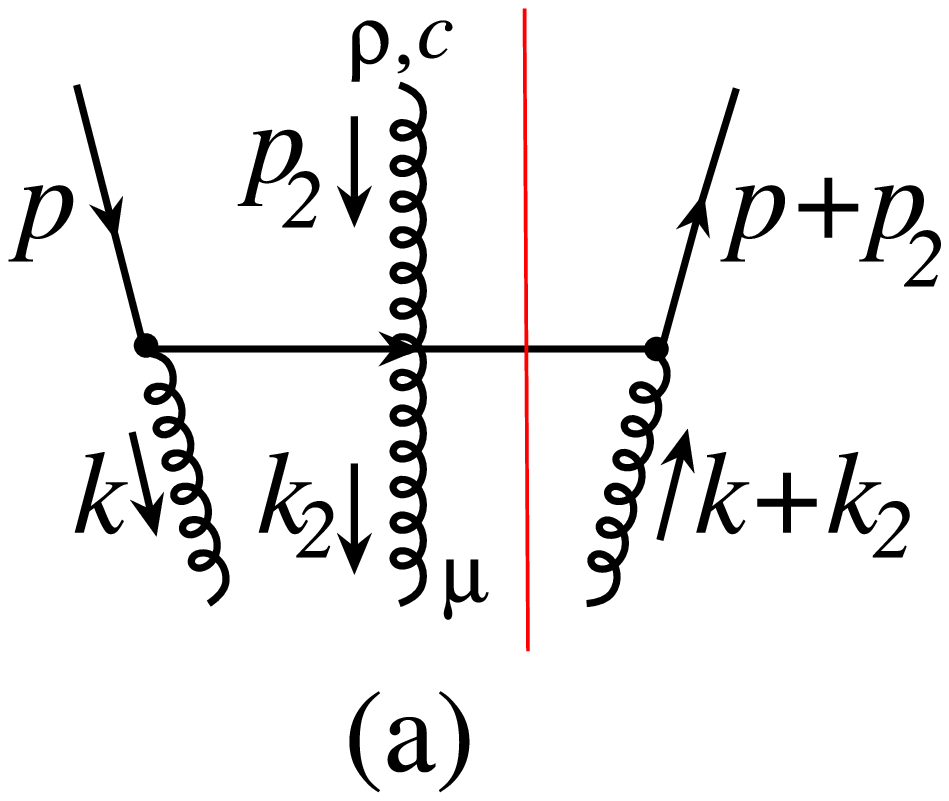,width=1.0in}
\hskip 0.3in
\psfig{file=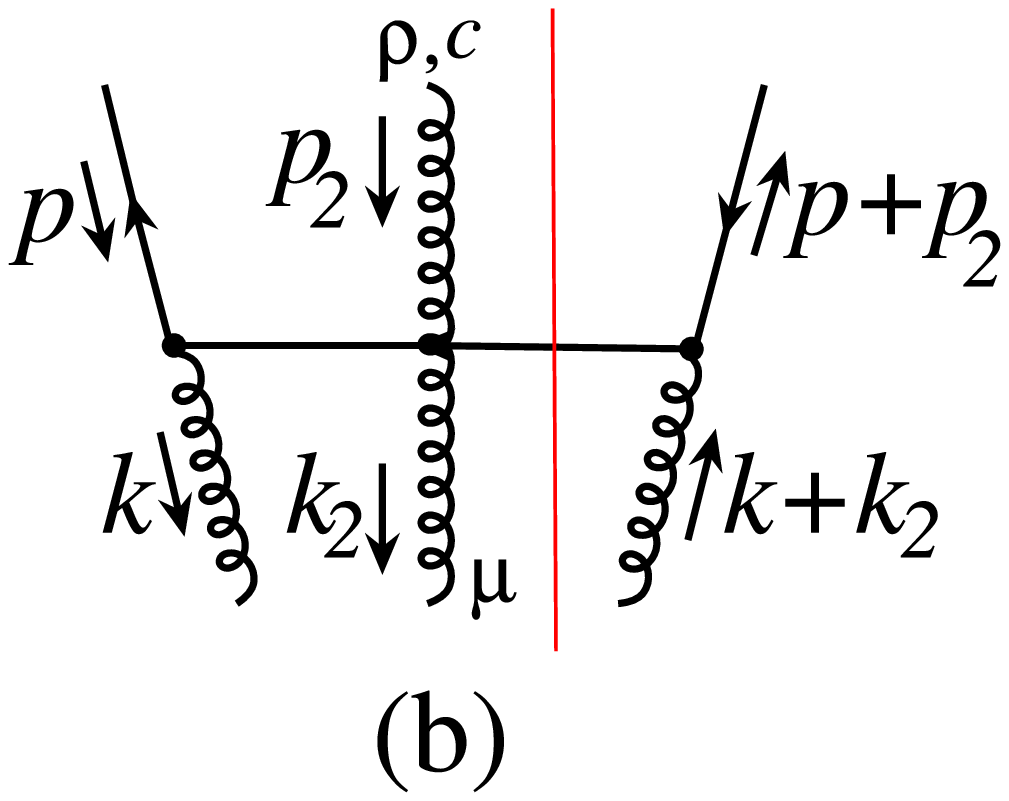,width=1.0in}
\caption{Feynman diagrams that contribute to 
the leading order evolution kernel from the quark-gluon
correlation functions to the tri-gluon correlation functions. }
\label{fig10}
\eef

By adding the flavor changing contribution to the evolution
kernels from Figs.~\ref{fig9} and \ref{fig10}, and adding
the contributions from the same diagrams but with their parton 
momentum fractions $\xi$ and $\xi+\xi_2$ switched,
we derive the order of $\alpha_s$ evolution equations 
for the diagonal twist-3 quark-gluon and tri-gluon correlation 
functions defined in Eqs.~(\ref{Tq}) and (\ref{Tg}),
\ben
\frac{\partial {T}_{q,F}(x,x,\mu_F)}{\partial{\ln \mu_F^2}}
&=&
\frac{\alpha_s}{2\pi}
\int_x^1\frac{d\xi}{\xi}
\bigg\{
P_{qq}(z)\, {T}_{q,F}(\xi,\xi,\mu_F)
\nonumber\\
&\ & \hskip 0.6in
+\frac{C_A}{2}
\left[\frac{1+z^2}{1-z}
\left[{T}_{q,F}(\xi,x,\mu_F)
     -{T}_{q,F}(\xi,\xi,\mu_F)\right]
     +z\, {T}_{q,F}(\xi,x,\mu_F)
\right]
\nnu
&\ & \hskip 0.6in
+\frac{C_A}{2}
\bigg[{T}_{\Delta q,F}(x,\xi,\mu_F)\bigg]
\nnu
&\ & \hskip 0.6in
+ P_{qg}(z)
\left(\frac{1}{2}\right)
\left[{T}_{G,F}^{(d)}(\xi,\xi,\mu_F)
     +{T}_{G,F}^{(f)}(\xi,\xi,\mu_F)\right]
\bigg\}\, ;
\label{Fevo_q}
\\
\frac{\partial {T}_{\bar{q},F}(x,x,\mu_F)}
     {\partial{\ln \mu_F^2}}
&=&
\frac{\alpha_s}{2\pi}
\int_x^1\frac{d\xi}{\xi}\
\bigg\{P_{qq}(z)\, {T}_{\bar{q},F}(\xi,\xi,\mu_F)
\nnu
&\ & \hskip 0.6in
+\frac{C_A}{2}
\left[\frac{1+z^2}{1-z}
\left[{T}_{\bar{q},F}(\xi,x,\mu_F)
     -{T}_{\bar{q},F}(\xi,\xi,\mu_F)\right]
     +z\, {T}_{\bar{q},F}(\xi,x,\mu_F)
\right]
\nnu
&\ & \hskip 0.6in
+\frac{C_A}{2}
\bigg[{T}_{\Delta\bar{q},F}(x,\xi,\mu_F)\bigg]
\nnu
&\ & \hskip 0.6in
+ P_{qg}(z)
\left(\frac{1}{2}\right)
\left[{T}_{G,F}^{(d)}(\xi,\xi,\mu_F)
     -{T}_{G,F}^{(f)}(\xi,\xi,\mu_F)\right]
\bigg\} \, ;
\label{Fevo_qb}
\\
\frac{\partial { T}_{G,F}^{(f)}(x,x,\mu_F)}
     {\partial{\ln \mu_F^2}}
&=&
\frac{\alpha_s}{2\pi}
\int_x^1\frac{d\xi}{\xi}
\bigg\{P_{gg}(z)\, {T}_G^{(f)}(\xi,\xi,\mu_F)
\nnu
&\ & \hskip 0.6in
+\frac{C_A}{2}
\left[2\left(\frac{z}{1-z}+\frac{1-z}{z}+z(1-z)\right)
\left[{T}_{G,F}^{(f)}(\xi,x,\mu_F)
     -{T}_{G,F}^{(f)}(\xi,\xi,\mu_F)\right]\right.
\nnu
&\ & \hskip 0.9in \left.
+2\left(1-\frac{1-z}{2z}-z(1-z)\right)
  {T}_{G,F}^{(f)}(\xi,x,\mu_F)
\right]
\nnu
&\ & \hskip 0.6in
+\frac{C_A}{2}
\bigg[(1+z)\,{T}_{\Delta G,F}^{(f)}(x,\xi,\mu_F)\bigg]
\nnu
&\ & \hskip 0.6in
+P_{gq}(z)\left(\frac{N_c^2}{N_c^2-1}\right)\sum_{q}
\left[{T}_{q,F}(\xi,\xi,\mu_F)
     -{T}_{\bar{q},F}(\xi,\xi,\mu_F)\right]
\bigg\} \, ;
\label{Fevo_gf} 
\\
\frac{\partial {T}_{G,F}^{(d)}(x,x,\mu_F)}
     {\partial{\ln \mu_F^2}}
&=&
\frac{\alpha_s}{2\pi}
\int_x^1\frac{d\xi}{\xi}
\bigg\{P_{gg}(z)\, {T}_{G,F}^{(d)}(\xi,\xi,\mu_F)
\nnu
&\ & \hskip 0.6in
+\frac{C_A}{2}
\left[2\left(\frac{z}{1-z}+\frac{1-z}{z}+z(1-z)\right)
\left[{T}_{G,F}^{(d)}(\xi,x,\mu_F)
     -{T}_{G,F}^{(d)}(\xi,\xi,\mu_F)\right]\right.
\nnu
&\ & \hskip 0.6in \left.
+2\left(1-\frac{1-z}{2z}-z(1-z)\right)
{T}_{G,F}^{(d)}(\xi,x,\mu_F)
\right]
\nnu
&\ & \hskip 0.6in
+\frac{C_A}{2}
\bigg[(1+z)\,{T}_{\Delta G,F}^{(d)}(x,\xi,\mu_F)\bigg]
\nnu
&\ & \hskip 0.6in
+P_{gq}(z)\left(\frac{N_c^2-4}{N_c^2-1}\right)\sum_{q}
\left[{T}_{q,F}(\xi,\xi,\mu_F)
     +{T}_{\bar{q},F}(\xi,\xi,\mu_F)\right]
\bigg\} \, .
\label{Fevo_gd} 
\een
In above evolution equations, $z=x/\xi$,
the LO quark-to-quark splitting function 
is given in Eq.~(\ref{Pqq}), and the rest LO parton-to-parton
splitting functions for the normal PDFs are given by
\ben
P_{qg}(z) &=&
\frac{1}{2}\, \left[(1-z)^2 + z^2\right]\, ,
\nnu
P_{gg}(z) &=& 
2\,C_A \left[
\frac{z}{(1-z)_+} + \frac{1-z}{z} + z(1-z)\right]
+ \left(C_A \frac{11}{6} - \frac{n_f}{3} \right)
\delta(1-z)\, ,
\nnu
P_{gq}(z) &=&
C_F\, \left[\frac{(1-z)^2 + 1}{z}\right]\, .
\label{Pij}
\een
Our explicit calculation also verifies that evolution equations
for the {\it diagonal} parts of the second set of twist-3 
correlation functions vanish,
\ben
\frac{\partial {T}_{\Delta q(\bar{q}),F}(x,x,\mu_F)}
     {\partial{\ln \mu_F^2}}
= 0\, ,
\ \ \
\frac{\partial {T}_{\Delta G,F}^{(f,d)}(x,x,\mu_F)}
     {\partial{\ln \mu_F^2}}
= 0\, ,
\label{Fevo_D}
\een
which are consistent with the antisymmetric nature of the 
second set of twist-3 correlation functions.
The {\it off-diagonal} correlation functions in 
Eqs.~(\ref{Fevo_q}) to (\ref{Fevo_gd}) are defined as 
\ben
{T}_{q(\bar{q}),F}(\xi,x,\mu_F) 
&\equiv& 
2\pi\, {\cal T}_{q(\bar{q}),F}(\xi,x,\mu_F)\, ,
\ \ \
{T}_{\Delta q(\bar{q}),F}(\xi,x,\mu_F) 
\equiv 
2\pi\, {\cal T}_{\Delta q(\bar{q}),F}(\xi,x,\mu_F)\, ,
\nnu
{T}_{G,F}^{(f,d)}(\xi,x,\mu_F) 
&\equiv & 
2\pi\, \frac{{\cal T}_{G,F}^{(f,d)}(\xi,x,\mu_F)}{\xi}\, ,
\ \ \
{T}_{\Delta G,F}^{(f,d)}(\xi,x,\mu_F) 
\equiv
2\pi\, \frac{{\cal T}_{\Delta G,F}^{(f,d)}(\xi,x,\mu_F)}{\xi}\, ,
\label{off_TqTg}
\een
where the more general correlation functions
${\cal T}_{q,F}(\xi,x,\mu_F)$, 
${\cal T}_{G,F}^{(f,d)}(\xi,x,\mu_F)$,
${\cal T}_{\Delta q,F}(\xi,x,\mu_F)$, 
and ${\cal T}_{\Delta G,F}^{(f,d)}(\xi,x,\mu_F)$ 
are defined in Eqs.~(\ref{Tqasy}), (\ref{Tgasy}),
(\ref{TDqasy}), and (\ref{TDgasy}), respectively
and the extra $2\pi$ is due to the fact that the 
$\int dy_2^-$ in Eqs.~(\ref{Tq}) and (\ref{Tg}) does not
have $1/2\pi$.

Equations from (\ref{Fevo_q}) to (\ref{Fevo_gd}) 
plus Eq.~(\ref{Fevo_D}) form a complete
set of evolution equations of the {\it diagonal}\ twist-3 
correlation functions relevant to the leading 
gluonic pole contribution to the SSAs.
All evolution kernels at the order of $\alpha_s$ 
are infrared safe and perturbatively calculated. However,
unlike the evolution equations for the full twist-3 correlation
functions from Eq.~(\ref{ssa_q}) to Eq.~(\ref{ssa_Dg}), these 
evolution equations do not form a closed equation set. 

The evolution equations of the diagonal twist-3 correlation
functions have a lot in common with the DGLAP evolution equations
of normal PDFs.  Every channel of parton-to-parton evolution 
kernel is led by a term that is proportional to the DGLAP
splitting function and the diagonal twist-3 correlation functions.
All other terms are either proportional to the difference 
between the diagonal and the off-diagonal correlation functions 
or proportional to the off-diagonal correlation functions.
Therefore, we expect that the scale dependence of the 
diagonal part of the twist-3 correlation functions 
is more close to the scale dependence of spin-averaged PDFs, 
not the spin-dependent helicity distributions.

Unlike the normal PDFs, the quark-gluon and antiquark-gluon 
correlation functions could have a different evolution equation 
unless the tri-gluon correlation function 
${T}_{G,F}^{(f)}=0$.
The difference was caused by the difference in color contraction 
for ${T}_{G,F}^{(f)}$ and ${T}_{G,F}^{(d)}$.  
As pointed out in Ref.~\cite{Kang:2008qh}, the production of 
open charm mesons in semi-inclusive deep inelastic scattering
or hadron-hadron collisions can provide the excellent information
on the tri-gluon correlation functions.  
If ${T}_{G,F}^{(f)}\neq 0$,
the difference between the quark-gluon and antiquark-gluon 
correlation functions could be enhanced as the scale evolves.
The difference should lead to interesting measurable consequences
when we compare the SSAs generated by the quark-gluon correlation
with that by the antiquark-gluon correlation.

It was argued in Ref.~\cite{Kang:2008qh} that one of the two tri-gluon 
correlation functions ${T}_{G,F}^{(f)}$ 
can be related to the moment 
of a TMD gluon distribution, known as the gluonic Sivers function, 
in terms of their operator definitions.  
However, the other tri-gluon correlation function 
${T}_{G,F}^{(d)}$ 
does not have a direct operator connection to the TMD gluon
distribution.  Equation (\ref{Fevo_gd}) indicates that 
within QCD collinear factorization formalism, 
the ${T}_{G,F}^{(d)}$ 
can be generated perturbatively from the quark-gluon and 
antiquark-gluon correlation functions as long as 
${T}_{q,F}(x,x,\mu_F)+{T}_{\bar{q},F}(x,x,\mu_F)\neq 0$
or ${T}_{\Delta G,F}^{(d)}(x,x,\mu_F)\neq 0$,
even if ${T}_{G,F}^{(d)}$ vanishes at one scale.

To complete this subsection, we state that we also 
examined the infrared sensitivity of the order of 
$\alpha_s$ evolution kernels for correlation functions 
that give the leading fermionic pole contribution to 
the SSAs.   
The fermionic pole contribution is generated by
the off-diagonal correlation functions, 
${\cal T}_{q,F}(0,x,\mu_F)$,
${\cal T}_{G,F}(0,x,\mu_F)$,
${\cal T}_{\Delta q,F}(0,x,\mu_F)$,
and ${\cal T}_{\Delta G,F}(0,x,\mu_F)$
(or equivalently from 
${\cal T}_{q,F}(x,0,\mu_F)$,
${\cal T}_{G,F}(x,0,\mu_F)$,
${\cal T}_{\Delta q,F}(x,0,\mu_F)$,
and ${\cal T}_{\Delta G,F}(x,0,\mu_F)$
for the diagrams in which the gluon at the cut vertex is 
in the RHS of the cut).  
At the order of $\alpha_s$, all evolution kernels are also 
infrared safe. 
For example, the flavor non-singlet evolution kernel 
for the quark-gluon correlation function 
$\widetilde{\cal T}_{q,F}$ can be calculated from the 
diagrams in Fig.~\ref{fig7} by setting $x=0$.
We find that after setting $x=0$, only diagrams
(a), (b), (c), (e), (f), and (g) in Fig.~\ref{fig7} give 
nonvanishing contribution to the evolution kernel.  Again,
the evolution kernel is infrared safe and all infrared 
divergences cancel between diagrams.  We will present
the full evolution equations for the off-diagonal twist-3
correlation functions in a future publication.


\section{Scale dependence}
\label{phenomenology}

In this section, we study the scale dependence of the diagonal
twist-3 quark-gluon and tri-gluon correlation functions relevant 
to SSAs by solving the evolution equations derived 
in the last section.

Since the evolution equations for the diagonal twist-3 correlation 
functions from Eq.~(\ref{Fevo_q}) to (\ref{Fevo_gd}) do not form 
a closed set of differential equations, we need to make a model for
off-diagonal correlation functions before we can study the scale
dependence of the diagonal correlation functions.  
For the following numerical study, we introduce the following model
to express the {\it symmetric} off-diagonal correlation functions 
in terms of diagonal correlation functions and a universal width,
\ben
{T}_{q,F}(x_1,x_2,\mu_F)
&=&
\frac{1}{2}\,
\left[{T}_{q,F}(x_1,x_1,\mu_F)
     +{T}_{q,F}(x_2,x_2,\mu_F)\right]\, 
e^{-\frac{(x_1-x_2)^2}{2\sigma^2}}\, ,
\nnu
{\cal T}_{G,F}^{(f,d)}(x_1,x_2,\mu_F)
&=&
\frac{1}{2}\,
\left[{\cal T}_{G,F}^{(f,d)}(x_1,x_1,\mu_F)
+{\cal T}_{G,F}^{(f,d)}(x_2,x_2,\mu_F)\right]\,
e^{-\frac{(x_1-x_2)^2}{2\sigma^2}}\, ,
\label{model_off}
\een
where both $T_{q,F}$ and ${\cal T}_{G,F}^{(f,d)}$ are
symmetric in exchange of $x_1$ and $x_2$ and the $\sigma$ is 
a width for the strength of the off-diagonal correlation.
From Eq.~(\ref{off_TqTg}), the off-diagonal correlation
function $T_{G,F}^{(f,d)}(x_1,x_2,\mu_F)$ defined in 
Eq.~(\ref{off_TqTg}) is not symmetric.
From Eq.~(\ref{model_off}), we have  
\ben
T_{G,F}^{(f,d)}(x_1,x_2,\mu_F)
= \frac{1}{2}
\left[T_{G,F}^{(f,d)}(x_1,x_1,\mu_F)
+\frac{x_2}{x_1}\,T_{G,F}^{(f,d)}(x_2,x_2,\mu_F)\right]\,
e^{-\frac{(x_1-x_2)^2}{2\sigma^2}}\, .
\label{model_off2}
\een
We choose the width $\sigma$ such that 
\ben
e^{-\frac{(x_1-x_2)^2}{2\sigma^2}}\sim 0 
\een
when $|x_1-x_2|\to 1$.  In Fig.~\ref{width}, we 
plot the factor $e^{-\frac{x^2}{2\sigma^2}}$ as
a function of $x$ for $\sigma=1/4$ (solid line) and $1/8$
(dashed line).

\bef
\psfig{file=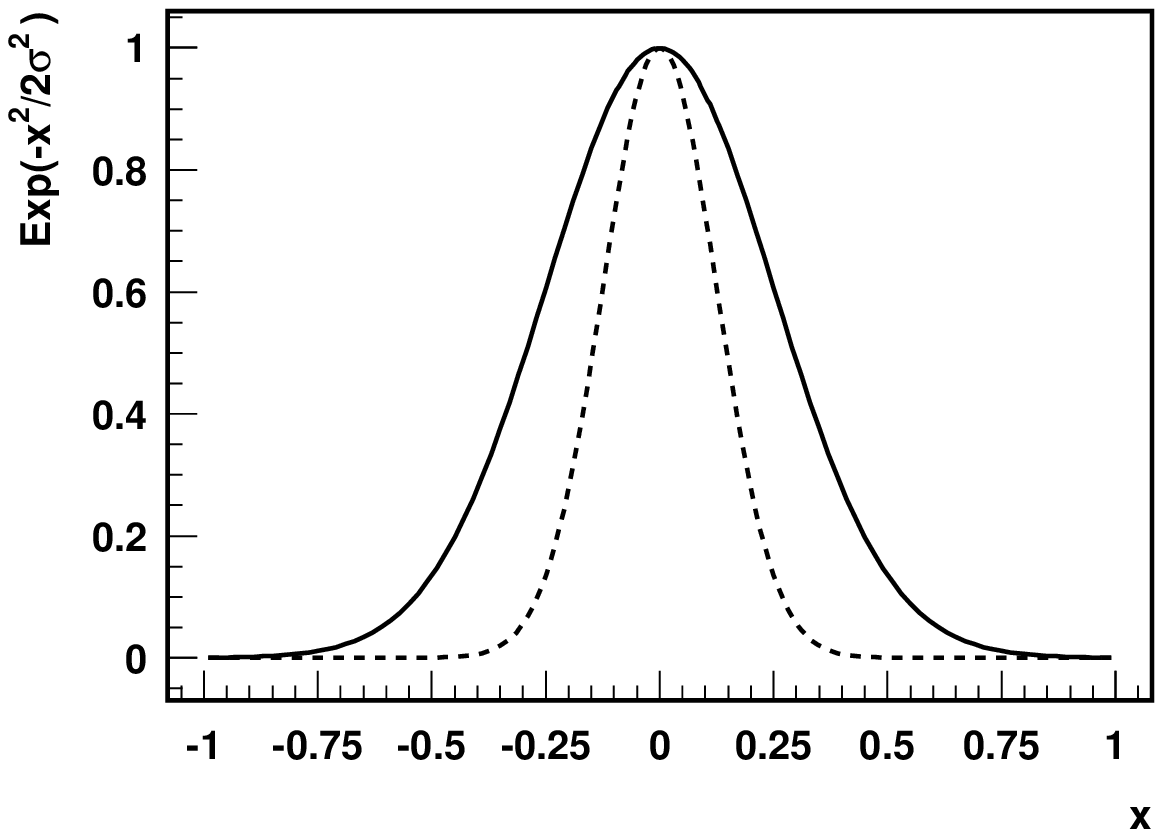,width=2.5in}
\caption{The factor $e^{-\frac{x^2}{2\sigma^2}}$ 
as a function of $x$ for $\sigma=1/4$ (solid) and 
$\sigma=1/8$ (dashed).}
\label{width}
\eef

To numerically solve the evolution equations in Eqs.~(\ref{Fevo_q}) 
to (\ref{Fevo_gd}), we choose the following input correlation
functions at $\mu_{F0}=2$~GeV.  For the quark-gluon correlation
function $T_{q,F}(x,x,\mu_{F0})$, we choose the Fit.~II of the 
quark-gluon correlation function $T_{q,F}(x,x,\mu_F)$ 
from Ref.~\cite{Kouvaris:2006zy}.  For the tri-gluon correlation
functions, we choose the model introduced in Ref.~\cite{Kang:2008qh},
\ben
T_{G,F}^{(f)}(x,x,\mu_{F0})=\lambda_f\, G(x,\mu_{F0})
\qquad
T_{G,F}^{(d)}(x,x,\mu_{F0})=\lambda_d\, G(x,\mu_{F0})
\een
with $\lambda_f=\lambda_d=0.07$ GeV at $\mu_{F0}=2$ GeV and
CTEQ6L unpolarized gluon distribution \cite{CTEQ6}.  
As an approximation, we also set 
$T_{\Delta q,F}(x,\xi,\mu_F)=0$ and 
$T_{\Delta G,F}(x,\xi,\mu_F)=0$ for less parameters since
they have vanishing diagonal contribution and the size of 
the off-diagonal part could be smaller than that of set
one correlation functions.  For a better
convergence of the numerical solution, we 
use the linear combination of the two tri-gluon correlation 
functions, $T_{G,F}^{(\pm)}=T_{G,F}^{(d)}\pm T_{G,F}^{(f)}$, 
when we solve for the evolution equations. 

\bef
\psfig{file=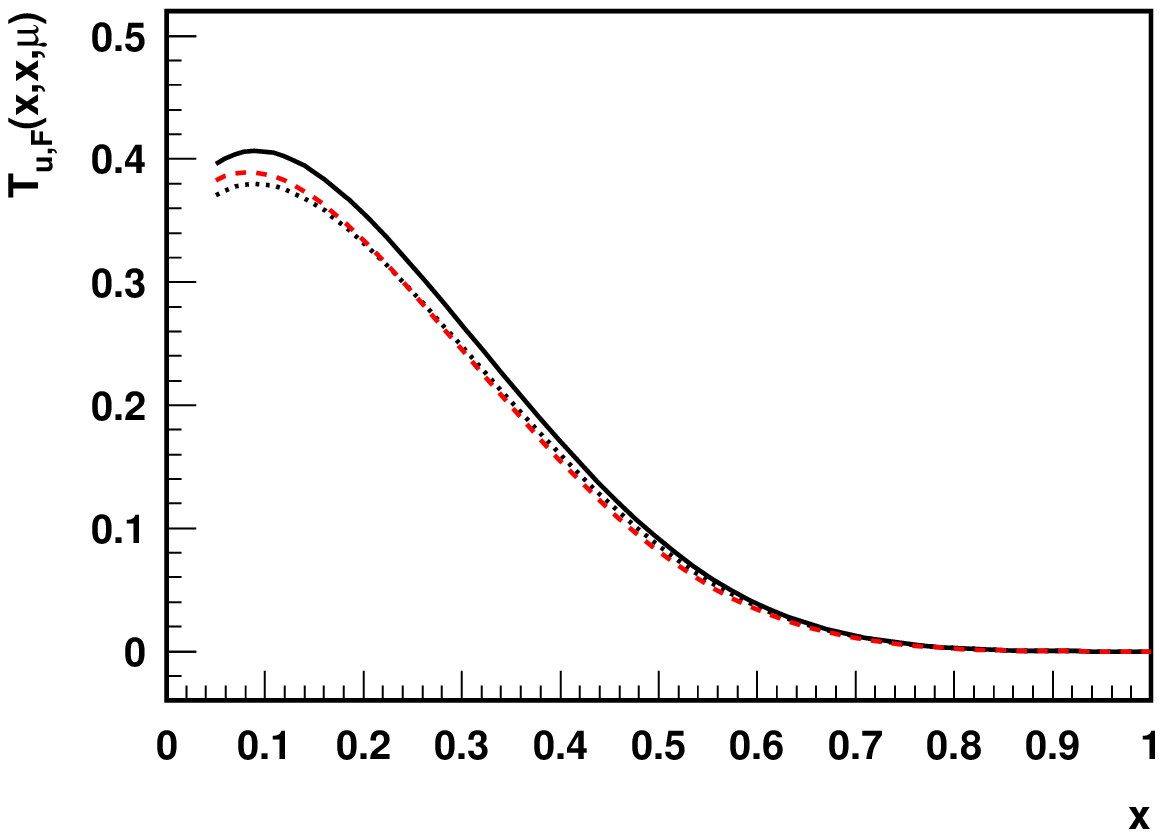,width=2.5in}
\hskip 0.2in
\psfig{file=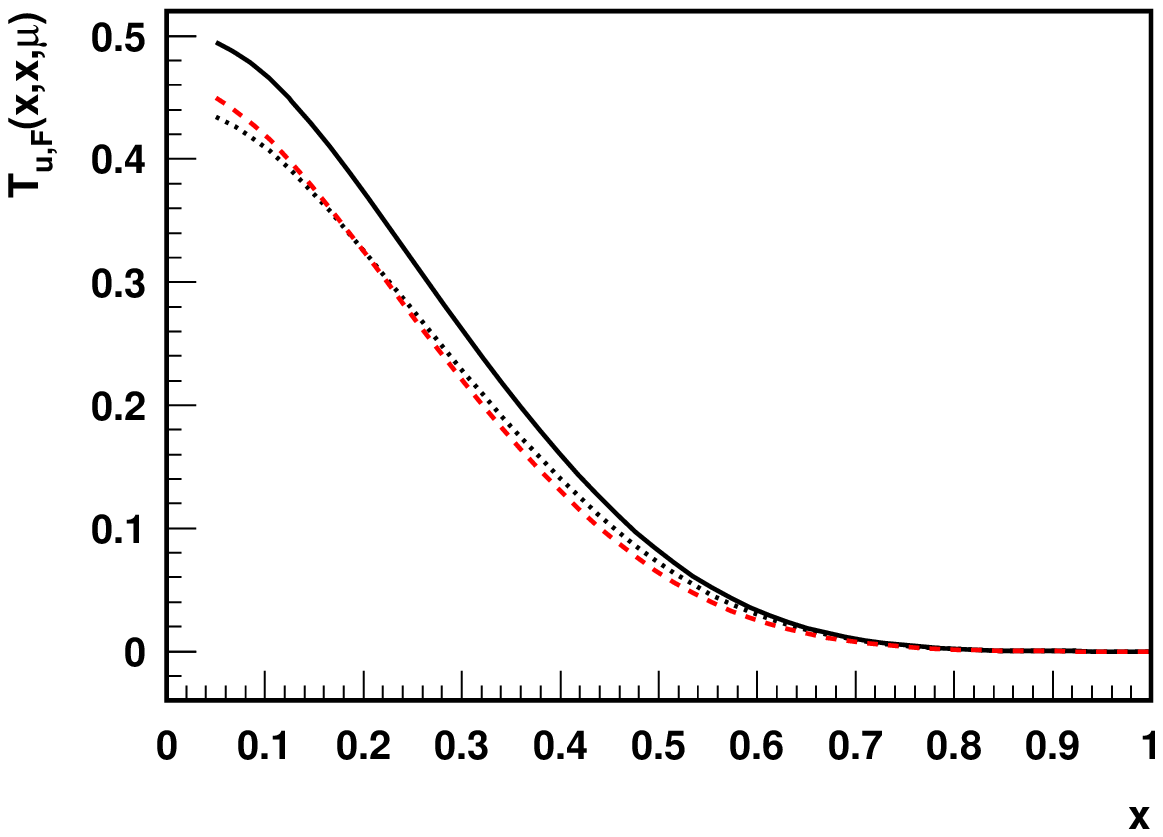,width=2.5in}
\caption{Twist-3 up-quark-gluon correlation $T_{u,F}(x,x,\mu_F)$ 
as a function of $x$ at $\mu_F=4$ GeV (left) and $\mu_F=10$ GeV (right).  
The factorization scale dependence is a solution of the 
flavor non-singlet evolution equation in Eq.~(\ref{evo_Tq_ns}). 
Solid and dotted curves correspond to $\sigma=1/4$ and 1/8, 
while the dashed curve is obtained by keeping only the DGLAP 
evolution kernel $P_{qq}(z)$ in Eq.~(\ref{evo_Tq_ns}).}
\label{ug_ns}
\eef

\bef
\psfig{file=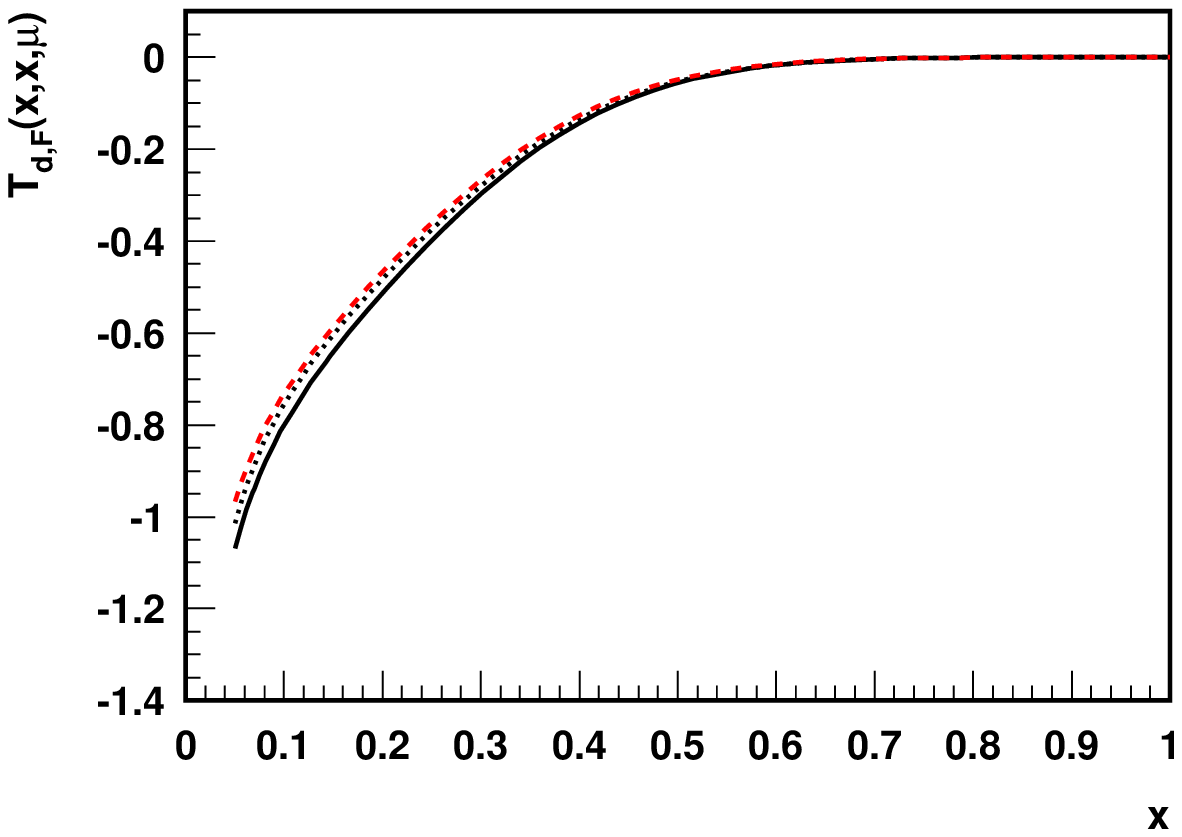,width=2.5in}
\hskip 0.2in
\psfig{file=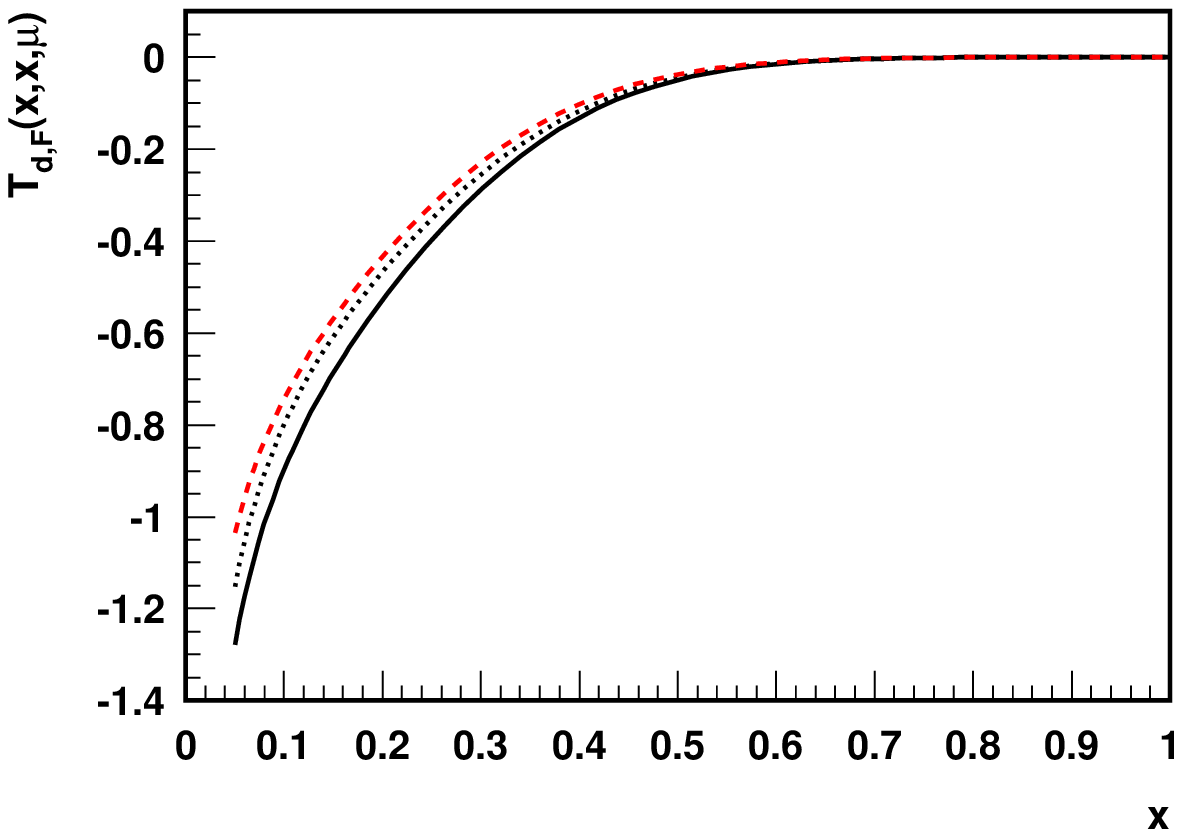,width=2.5in}
\caption{Twist-3 down-quark-gluon correlation $T_{d,F}(x,x,\mu_F)$ 
as a function of $x$ at $\mu_F=4$ GeV (left) and $\mu_F=10$ GeV (right).  
Solid and dotted curves correspond to $\sigma=1/4$ and 1/8, 
while the dashed curve is obtained by keeping only the DGLAP 
evolution kernel $P_{qq}(z)$ in Eq.~(\ref{evo_Tq_ns}).}
\label{dg_ns}
\eef

We first solve the flavor non-singlet evolution equation for 
the quark-gluon correlation function in Eq.~(\ref{evo_Tq_ns})
to test the relative role of the normal DGLAP evolution term
that is proportional to $P_{qq}(z)$ and the new piece that depends 
on the off-diagonal correlation function.  In Fig.~\ref{ug_ns},
we plot the twist-3 up-quark-gluon correlation $T_{u,F}(x,x,\mu_F)$ 
as a function of $x$ at the factorization scale $\mu_F=4$ GeV 
(left) and $\mu_F=10$ GeV (right).  The difference between 
the left figure and the one on the right indicates 
the evolution of the twist-3 correlation functions.
The factorization scale dependence is a solution of the 
flavor non-singlet evolution equation in Eq.~(\ref{evo_Tq_ns}). 
Solid and dotted curves correspond to two different choices
of the width for the off-diagonal input correlation function 
at $\sigma=1/4$ and 1/8, respectively. 
The dashed curve is obtained by keeping only the DGLAP 
evolution kernel $P_{qq}(z)$  when we
solve the flavor non-singlet evolution equation 
in Eq.~(\ref{evo_Tq_ns}).
Similarly, we plot the twist-3 down-quark-gluon correlation 
$T_{d,F}(x,x,\mu_F)$ as a function of $x$ 
at the factorization scale $\mu_F=4$ GeV (left) and 
$\mu_F=10$ GeV (right) in Fig.~\ref{dg_ns}.  Unlike the
up-quark-gluon correlation function $T_{u,F}$, the down-quark-gluon 
correlation function $T_{d,F}$ is negative \cite{qiu}.
Figures~\ref{ug_ns} and \ref{dg_ns} clearly show that 
the scale dependence of the diagonal twist-3 quark-gluon 
correlation function does follow the evolution of 
the unpolarized quark distribution.  The difference between
the solid and the dashed curves indicates that the effect of 
non-DGLAP type contribution from the off-diagonal correlation 
function could be very important at small $x$ 
if the width of the off-diagonal correlation function is large.  

\bef
\psfig{file=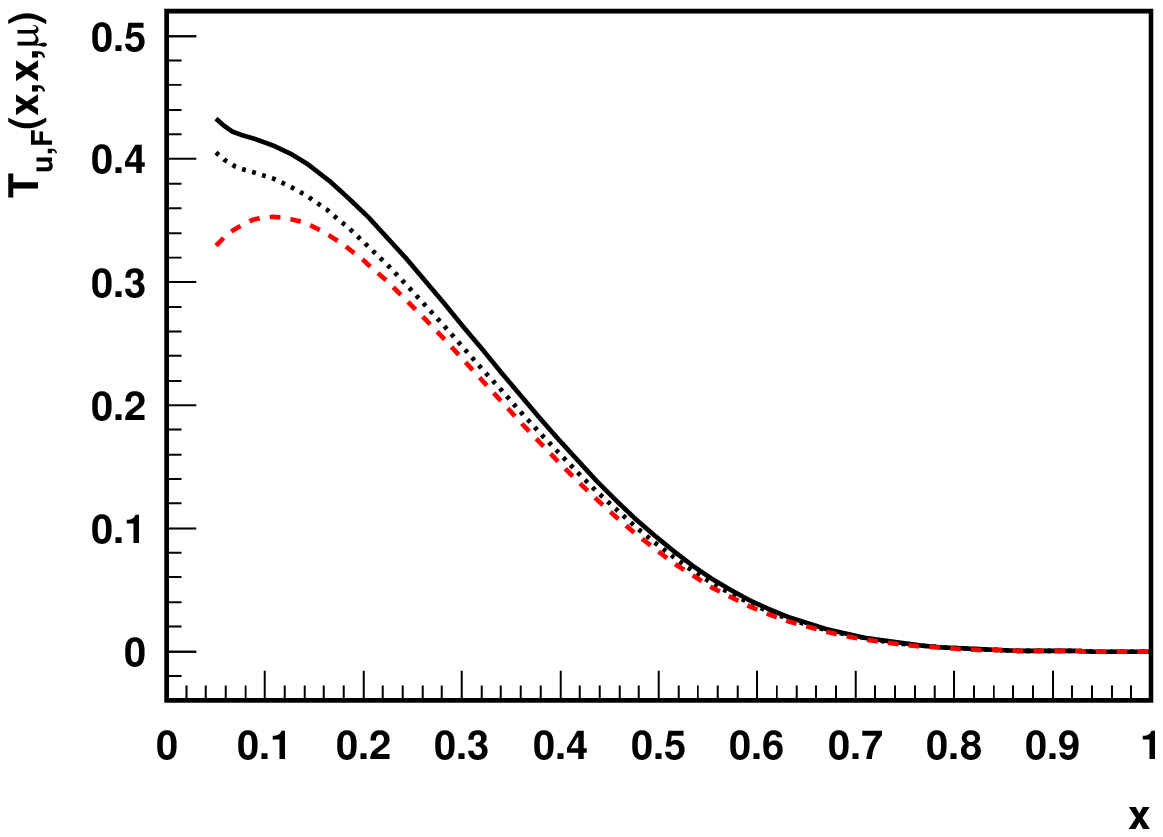,width=2.5in}
\hskip 0.2in
\psfig{file=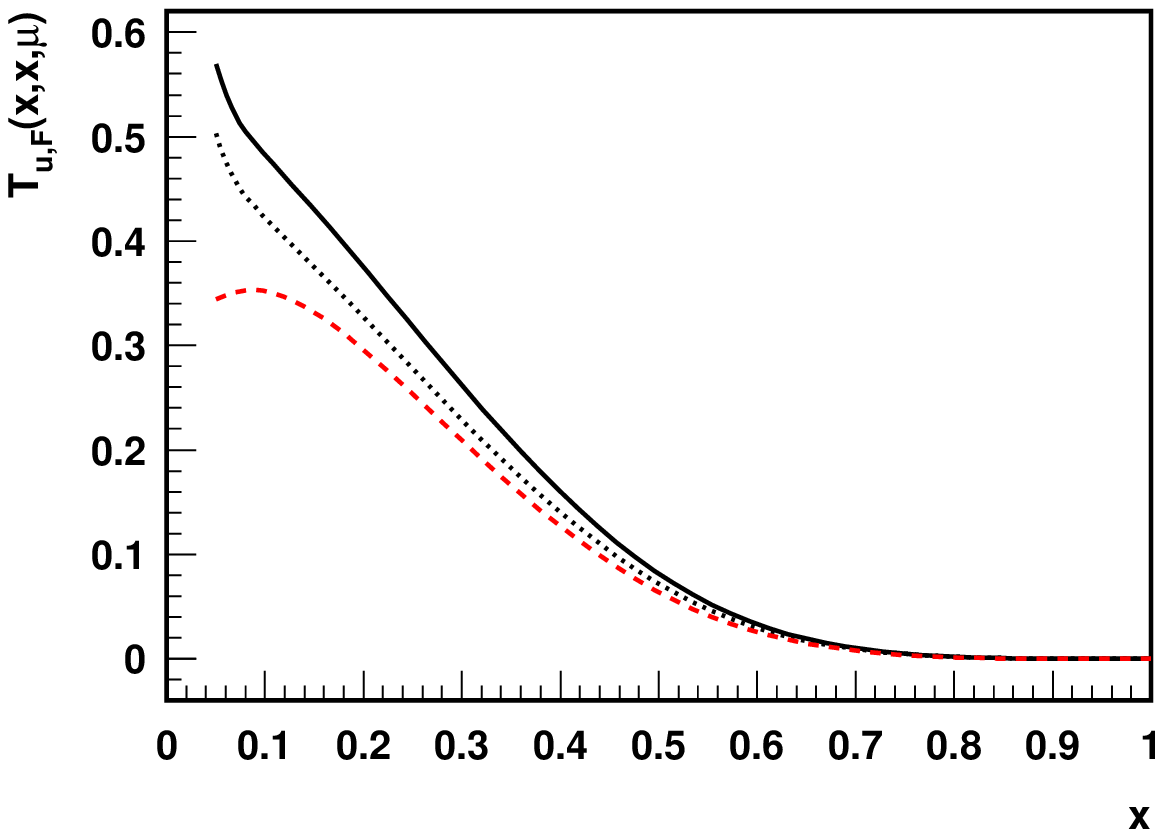,width=2.5in}
\caption{Twist-3 up-quark-gluon correlation $T_{u,F}(x,x,\mu_F)$ 
as a function of $x$ at $\mu_F=4$ GeV (left) and $\mu_F=10$ GeV (right).  
The factorization scale dependence is obtained by solving the
full set of evolution equations in Eq.~(\ref{Fevo_q}) through 
(\ref{Fevo_gd}).
Solid and dotted curves correspond to $\sigma=1/4$ and 1/8 for
the width of input off-diagonal correlation functions.  
The dashed curves represent the quark-gluon correlation functions 
obtained from the parametrization of Fit~II in Ref.~\cite{Kouvaris:2006zy}
by assuming all quark-gluon and tri-gluon correlation functions 
obey the DGLAP.}
\label{ug_s}
\eef

\bef
\psfig{file=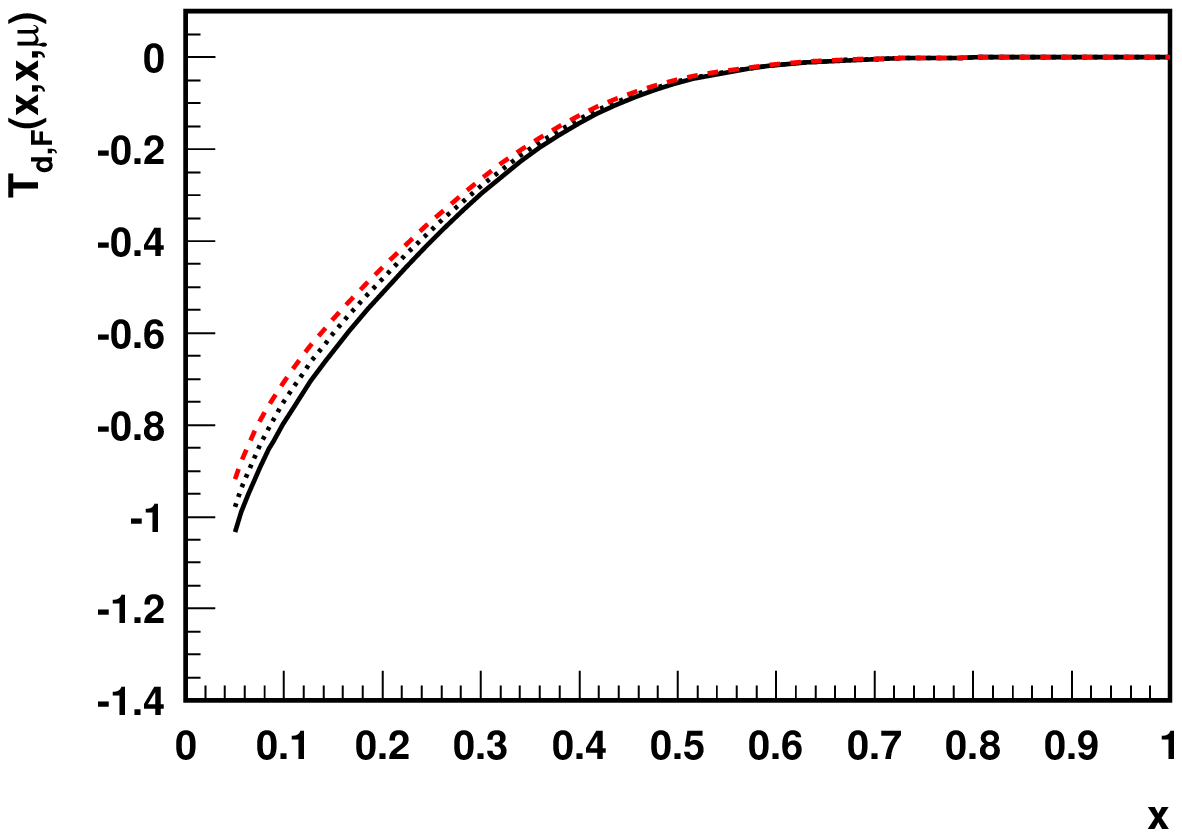,width=2.5in}\hskip 0.2in
\psfig{file=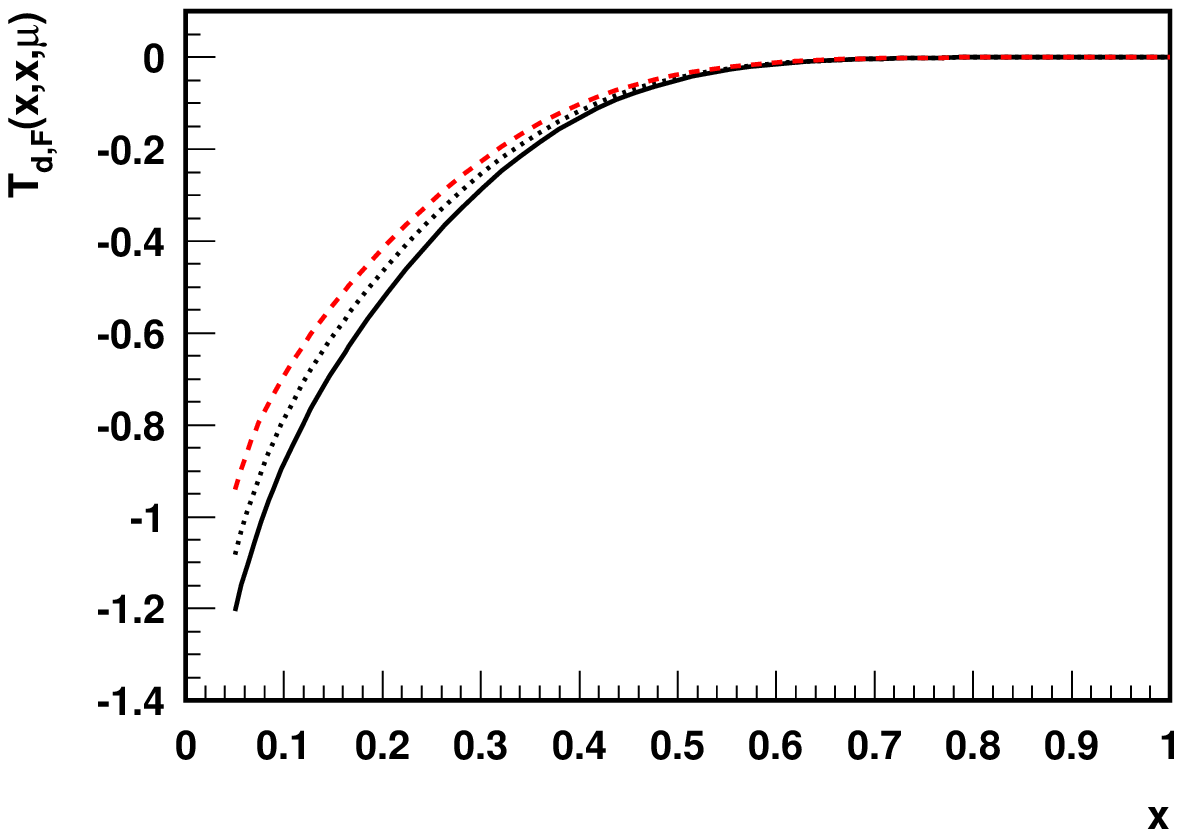,width=2.5in}
\caption{Twist-3 down-quark-gluon correlation $T_{u,F}(x,x,\mu_F)$ 
as a function of $x$ at $\mu_F=4$ GeV (left) and $\mu_F=10$ GeV (right).
all curves are defined in the same way as those in Fig.~\ref{ug_s}.}
\label{dg_s}
\eef
In Figs.~\ref{ug_s} and \ref{dg_s}, we plot the twist-3 
up-quark-gluon and down-quark-gluon correlation functions,
$T_{u,F}(x,x,\mu_F)$ and $T_{d,F}(x,x,\mu_F)$, as a function of $x$ 
at $\mu_F=4$ GeV (left) and $\mu_F=10$ GeV (right).  
Only difference between the solid and dotted curves in these figures 
and those in Figs.~\ref{ug_ns} and \ref{dg_ns} is that we use the
full set of evolution equations in Eq.~(\ref{Fevo_q}) through 
(\ref{Fevo_gd}) to solve for the factorization scale dependence
of these correlation functions. 
The dashed curves represent the quark-gluon correlation functions 
obtained from the parametrization of Fit~II in Ref.~\cite{Kouvaris:2006zy}
by assuming all quark-gluon and tri-gluon correlation functions 
obey the DGLAP evolution.
We find that non-DGLAP terms in the full evolution equations 
for the diagonal twist-3 correlation functions play a significant role
in modifying the evolution of these correlation functions at small
$x$, where the role of the off-diagonal correlation functions is 
enhanced due to a larger available phase space for the evolution 
kernels.  The extra enhancement of the solid and dotted curves
over the dashed curves in Figs.~\ref{ug_s} and \ref{dg_s} is mainly
from the term proportional to the sum of both tri-gluon correlation 
functions $T_{G,F}^{(f)}$ and $T_{G,F}^{(d)}$ that we assumed to have the 
same sign.

\bef
\psfig{file=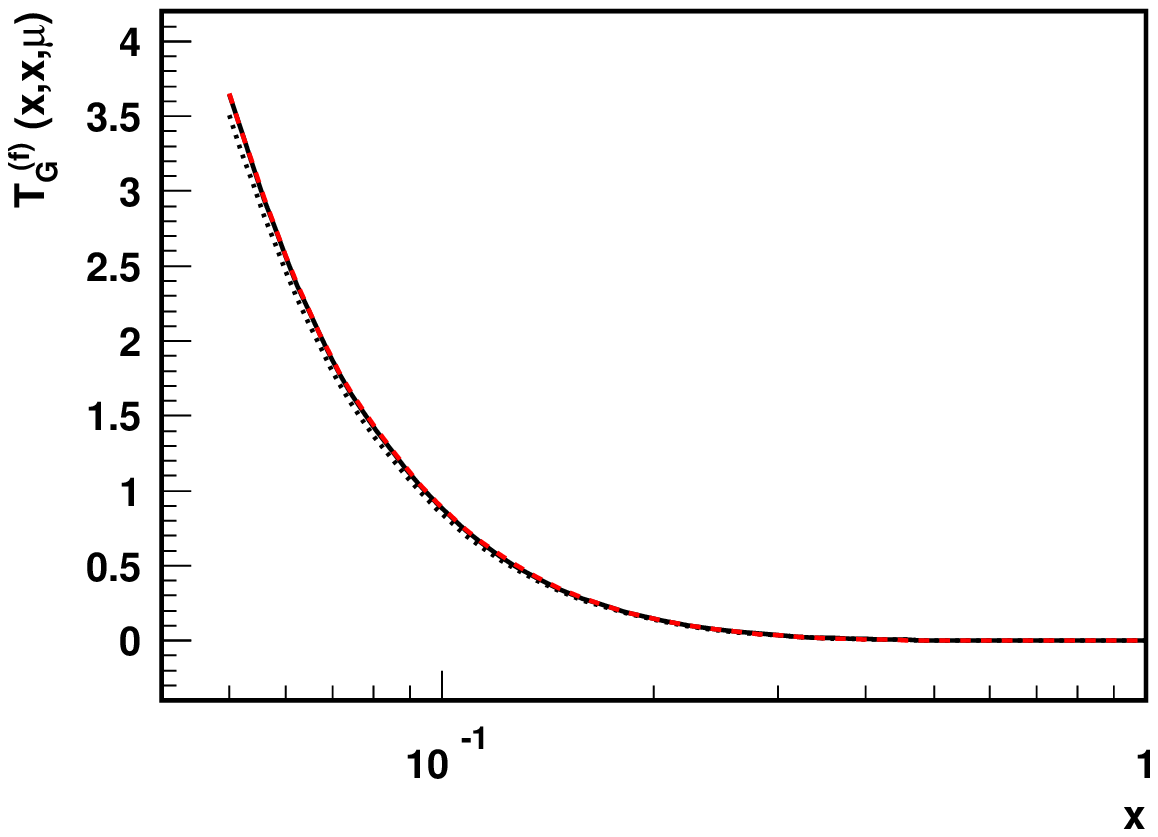,width=2.5in}\hskip 0.2in
\psfig{file=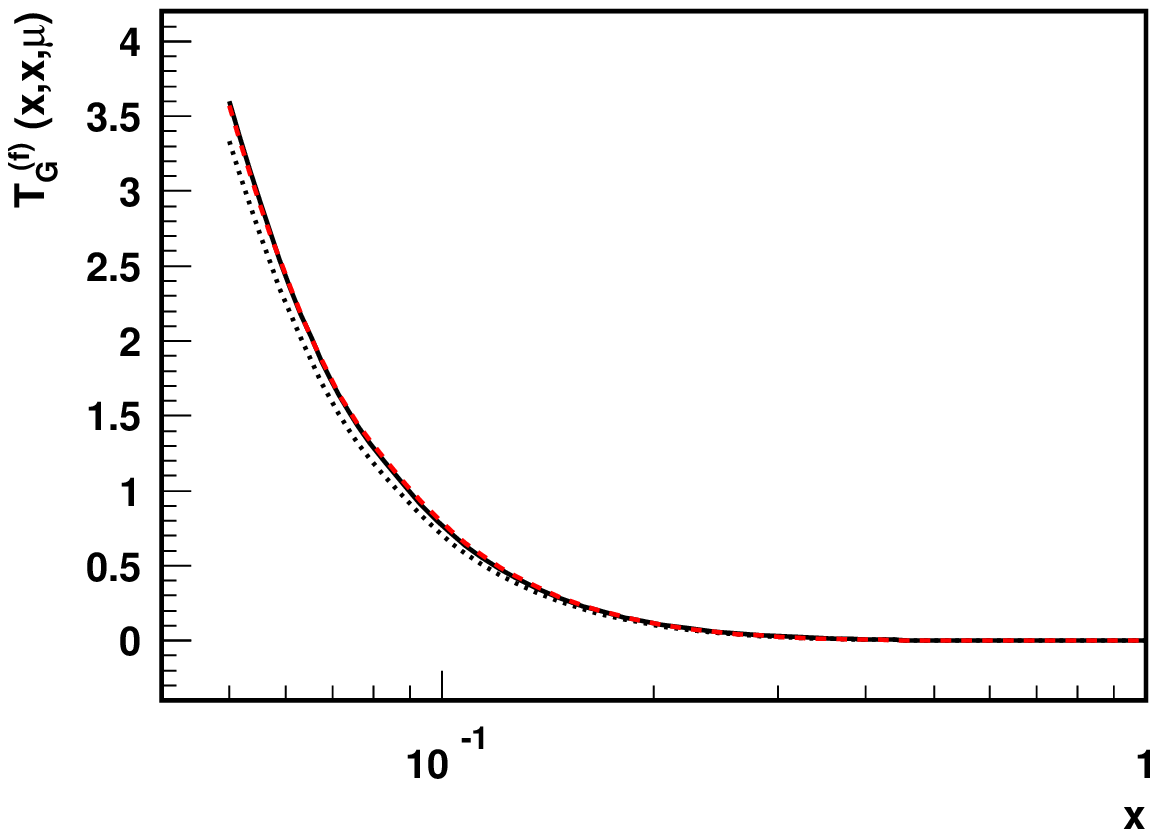,width=2.5in}
\caption{Twist-3 tri-gluon correlation function $T_{G,F}^{(f)}(x,x,\mu_F)$ 
as a function of $x$ at $\mu_F=4$ GeV (left) and $\mu_F=10$ GeV (right). 
Dashed curves are from $T_{G,F}^{(f)}(x,x,\mu_F)=\lambda_f\, G(x,\mu_F)$,
and solid and dotted curves are from solving the full evolution
equations with $\sigma=1/4$ and 1/8 for the input correlation functions,
respectively.}
\label{trig_f}
\eef

\bef
\psfig{file=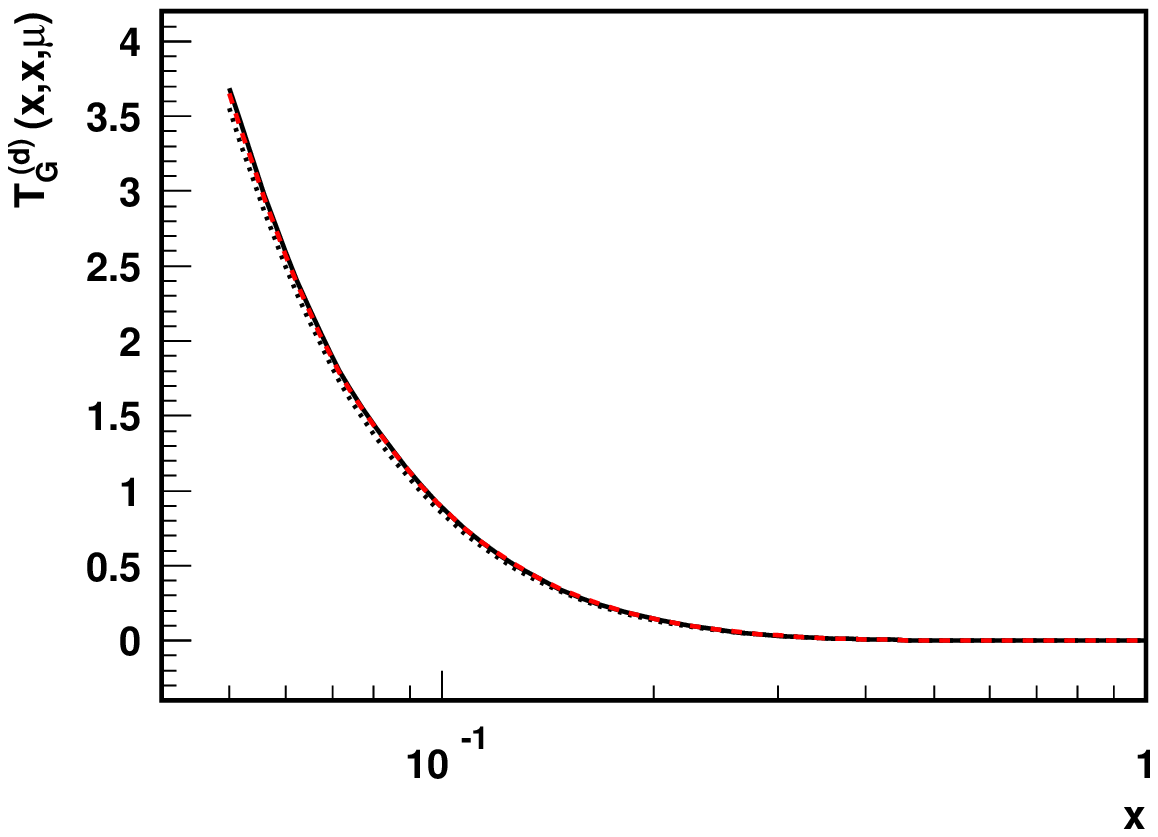,width=2.5in}\hskip 0.2in
\psfig{file=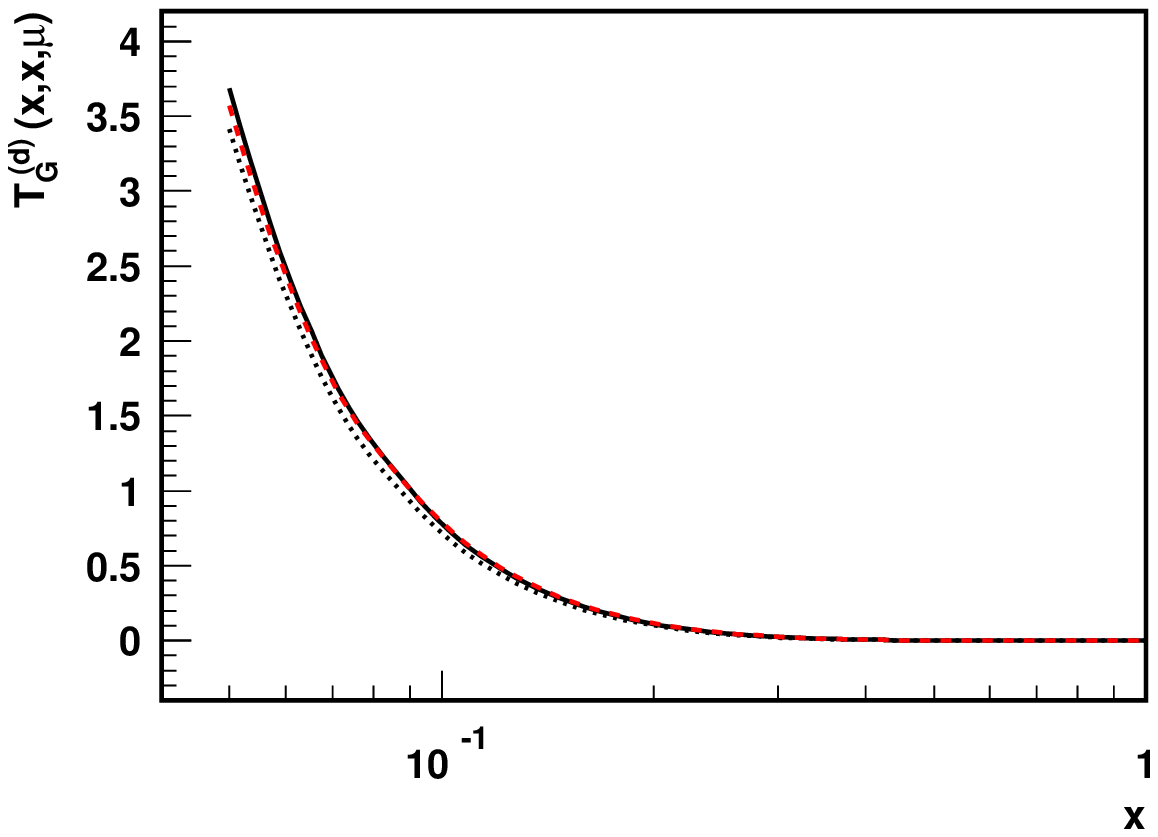,width=2.5in}
\caption{Twist-3 tri-gluon correlation function $T_{G,F}^{(d)}(x,x,\mu_F)$ 
as a function of $x$ at $\mu_F=4$ GeV (left) and $\mu_F=10$ GeV (right). 
Dashed curves are from $T_{G,F}^{(d)}(x,x,\mu_F)=\lambda_d\, G(x,\mu_F)$,
and solid and dotted curves are from solving the full evolution
equations with $\sigma=1/4$ and 1/8 for the input correlation functions,
respectively.}
\label{trig_d}
\eef

In Figs.~\ref{trig_f} and \ref{trig_d}, we plot the twist-3
tri-gluon correlation functions, $T_{G,F}^{(f)}(x,x,\mu_F)$ and
$T_{G,F}^{(d)}(x,x,\mu_F)$, as a function of $x$ 
at $\mu_F=4$ GeV (left) and $\mu_F=10$ GeV (right), respectively.
Solid and dotted curves are from solving the full evolution
equations with the input correlation functions evaluated at
$\sigma=1/4$ and 1/8, respectively.  Dashed curves are given
by the normal CTEQ6L gluon distribution multiplied by
the normalization constant $\lambda_f$ (or $\lambda_d$), which
corresponds to making an assumption that all twist-3 correlation
functions obey the DGLAP evolution, like the normal unpolarized
PDFs.  We notice that for the evolution of tri-gluon correlation 
functions, the difference in color factor for the DGLAP-type 
terms in the full evolution equations tends to compensate 
the contribution from the terms proportional to the off-diagonal 
correlation functions, so that the evolution of the tri-gluon
correlation functions follow more closely to the DGLAP evolution  
as shown in Figs.~\ref{trig_f} and \ref{trig_d}.

We complete this section by stressing that the scale dependence
presented in this section is sensitive to our assumption 
to neglect the role of the second set of twist-3 correlation
functions and our model for the input tri-gluon correlation 
functions (equal and positive at the input scale).
Although the overall features found here should be valid, 
the precise numerical values of these correlation functions 
should be extracted from a consistent global QCD analysis 
by comparing experimental data on SSAs and corresponding 
theoretical calculations, like what have been done to
test the leading power QCD factorization formalism 
\cite{CTEQPDF,MRSTPDF}.  
The new evolution equation derived in this paper is the necessary
step to make such a consistent global QCD analysis possible 
for twist-3 correlation functions relevant to SSAs.

\section{Summary and Conclusions}
\label{summary}

We constructed two sets of twist-3 correlation functions that are
responsible for generating the SSAs in the 
QCD collinear factorization approach.  These correlation functions
do not contribute to the long-distance correlation functions 
extracted from any measurement of parity conserving double-spin
asymmetries, such as the DIS structure function $g_2$.
We introduced the Feynman diagram representation for the twist-3
quark-gluon and tri-gluon correlation functions and 
derived the cutvertices to connect the hadronic matrix 
elements of these correlation functions to 
the forward scattering Feynman diagrams.  
In terms of the Feynman diagram representation, we derived for 
the first time a closed set of evolution equations in 
Eqs.~(\ref{ssa_q}) and (\ref{ssa_Dg}) for these 
quark-gluon and tri-gluon correlation functions.  We also 
provide the explicit prescription for calculating the corresponding
evolution kernels.

We calculated in the light-cone gauge the order of $\alpha_s$ 
evolution kernels for the scale dependence of the diagonal 
quark-gluon and tri-gluon correlation functions 
that are responsible for the leading gluonic pole contribution 
to the SSAs.  We also provided in this paper the cut vertices and 
corresponding projection operators necessary for performing 
the calculation of evolution kernels in a covariant
gauge.  A covariant gauge calculation requires additional Feynman 
diagrams that have eikonal lines connecting quark and gluon lines
at the cut vertices.  These additional diagrams are from the 
expansion of the gauge links in the matrix element definition of 
the twist-3 correlation functions \cite{qiu}.
Our calculation explicitly shows that all evolution
kernels are infrared safe as they should be.  In addition, we also
examined infrared behavior of the evolution kernels that are 
more relevant to the fermionic pole contribution to the SSAs and 
found that these kernels are also infrared safe.  We leave the
details on the scale dependence of more general off-diagonal 
twist-3 correlation functions to a future publication. 

We found that the evolution equations for the scale dependence
of the {\it diagonal} twsit-3 correlation functions do not form a 
closed set of differential equations.  The variation of the 
diagonal correlation functions gets contribution from the 
off-diagonal correlation functions.  At the order of $\alpha_s$,
all contributions to the evolution kernels from the diagonal 
correlation functions are proportional to corresponding 
DGLAP evolution kernels of unpolarized PDFs.  In order to test
the role of the off-diagonal correlation functions to the 
scale dependence of the diagonal correlation functions, we 
introduced a model for the off-diagonal correlation functions
in Eq.~(\ref{model_off}) in terms of the diagonal functions 
plus a correlation factor characterized by a parameter $\sigma$.
With our model, we were able to solve the evolution equations
and derive the scale dependence of the diagonal twist-3 
correlation functions, and test the relative strength of 
different sources of contribution to the scale dependence.
We found within a reasonable parameter space of our model that
the scale dependence of the diagonal twist-3 quark-gluon and
tri-gluon correlation functions follows closely the 
DGLAP evolution of unpolarized PDFs, and the contribution 
from the off-diagonal correlation functions becomes more
significant at small $x$.  

Although our observations are dependent of the model for 
the off-diagonal correlation functions, we believe that 
the overall feature found here should be valid.  
The precise behavior of these twist-3 correlation functions 
should be extracted from a consistent global QCD analysis, just 
like what have been done for the leading power QCD dynamics
and the scale dependence of the universal PDFs 
\cite{CTEQPDF,MRSTPDF}.  The evolution equations derived 
in this paper enable us to evaluate the NLO corrections 
to the SSAs systematically, 
which represents a necessary step moving toward the 
goal of global analysis of QCD dynamics beyond what have been 
explored by the very successful leading power QCD collinear
factorization formalism.  With such a global analysis, we 
will be able to explore the rich field of quantum correlation
of multiple fields inside a polarized hadron.

\section*{Acknowledgments}

We thank G.~Sterman, W.~Vogelsang and F.~Yuan 
for helpful discussion.  We also thank 
A.~Belitsky and P.~Ratcliffe for correspondence on 
references of evolution equations of twist-3 correlation functions.
J.Q. thanks the Institute of 
High Energy Physics, Chinese Academy of Science 
for its hospitality during the completion of this work. 
This work was supported in part by the U. S. Department of Energy 
under Grant No.~DE-FG02-87ER40371.


\end{document}